\documentclass[12pt,twoside,english]{article}
\pdfoutput=1    % flag for arXiv compilation using gnuplot diagrams
\usepackage{ifthen}                % if then else Abfragen
\newboolean{boldeqnitalic}
\setboolean{boldeqnitalic}{true} 

\usepackage{caption}
\usepackage[intlimits] {amsmath}   % muss for defjs# stehen 
\usepackage{defjs2}                % eigene Definitionen    
\usepackage{epsfig}
\usepackage{psfig}
\usepackage{graphicx}
\usepackage{psfrag}
\usepackage[usenames]{color}
\usepackage{colortbl}
\usepackage{ifthen}
\usepackage[cp1252]{inputenc}
\usepackage{fancyhdr}
\usepackage{rotating}
\usepackage{multirow}
\usepackage{booktabs}              % Dicke Tabellenlinien
\usepackage{amssymb}
\usepackage{amsbsy}
\usepackage{bm}                    % fette Schrift in Formeln
\usepackage{babel}
\usepackage{theorem}
\usepackage{upgreek}  % gerade (roman) griechische Buchstaben z.B. \upalpha
\usepackage{pifont}   % zusaetzliche Schriften, z.B. eingekreiste Zahlen
\usepackage{mathrsfs}
\usepackage{datetime}
\usepackage{tikz}
\usetikzlibrary{matrix}
\usepackage[all]{xy}
\usepackage{rotating}
\usepackage{mathtools}   % e.g. text over arrow
\usepackage{cancel}
\usepackage{subfig}
\usepackage[labelfont=bf, justification=justified]{caption}
\usepackage{algpseudocode}
\usepackage{algorithm}
\algnewcommand{\algorithmicor}{\textbf{ or }}
\algnewcommand{\OR}{\algorithmicor}

\definecolor{dunkelgrau}{rgb}{0.8,0.8,0.8}
\definecolor{hellgrau}{rgb}{0.90,0.90,0.90} %... slightly darker 
\fancypagestyle{plain}{%
  \fancyhead{}%
  \fancyfoot[c]{\sffamily\thepage}%
}
\makeatletter                           % Leere Fuellseiten
\def\cleardoublepage{\clearpage\if@twoside \ifodd\c@page\else
  \hbox{}
  \vspace*{\fill}
  \thispagestyle{empty}
  \newpage
  \if@twocolumn\hbox{}\newpage\fi\fi\fi}
\makeatother
\begin{document}
\unitlength1.0cm
\frenchspacing

\thispagestyle{empty}
 
\vspace{-3mm}
  
%\begin{center}
%%  {\bf \large Meso-to-Macro Scale Transition for 3d Concrete} 
%  {\bf \large Image-based Modeling of 3d Concrete} 
%  \\[2mm]
%  {\bf \large Part I: Accuracy Analysis for Elastic Deformations}
%\end{center}
  
\vspace{2mm}

%\begin{center}
%%  {\bf \large Meso-to-Macro Scale Transition for 3d Concrete} 
%  {\bf \large Towards the Digital Twin of a Concrete Specimen --} 
%  \\[2mm]
%  {\bf \large Part I: Elastic Analyses for Size, Resolution, Discretization}
%\end{center}
%
%\vspace{2mm}

\begin{center}
%  {\bf \large Meso-to-Macro Scale Transition for 3d Concrete} 
  {\bf \large Computational Homogenization of Concrete in the} 
  \\[2mm]
  {\bf \large Cyber Size-Resolution-Discretization (SRD) Parameter Space}
\end{center}

\vspace{4mm}

\ce{Ajinkya Gote$^{1}$, Andreas Fischer$^{1}$, Chuanzeng Zhang$^{2}$, Bernhard Eidel$^{1\ast}$}
 
\vspace{4mm}

\ce{\small $^{1}$DFG-Heisenberg-Group, Institute of Mechanics, Department of Mechanical Engineering} 
\ce{\small $^{2}$Chair of Structural Mechanics, Department of Civil Engineering} 
\vspace{2mm}
\ce{\small University Siegen, 57068 Siegen, Paul-Bonatz-Str. 9-11, Germany} 
\vspace{2mm}
\ce{\small $^{\ast}$e-mail: bernhard.eidel@uni-siegen.de, phone: +49 271 740 2224, fax: +49 271 740 2436} 
\vspace{2mm}

\bigskip

\begin{center}
{\bf \large Highlights}

\bigskip

{\footnotesize
\begin{minipage}{15.5cm} 

\vspace*{-2mm}

\begin{itemize}
 \item Numerical homogenization analysis for effective elastic properties of concrete specimen  
 \\[-6mm]
 \item Comprehensive sampling of the Size-Resolution-Discretization (SRD) parameter space 
 \\[-6mm]
 \item Transition from apparent to effective properties invariant to RVE boundary conditions 
 \\[-6mm] 
  \item Quality of error estimation transferable from 2d to 3d and from small to large volumes 
 \\[-6mm] 
 \item Considerable efficiency gain in SRD parameter space at controlled errors 
% \\[-6mm] 
% \item ...  
\end{itemize}

\end{minipage}
}
\end{center}

\bigskip

\begin{center}
{\bf \large Abstract}

\bigskip

{\footnotesize
\begin{minipage}{14.5cm}
\noindent
Micro- and mesostructures of multiphase materials obtained from tomography and image acquisition are an ever more important database for simulation analyses. Huge data sets for reconstructed 3d volumes typically as voxel grids call for criteria and measures to find an affordable balance of accuracy and efficiency. The present work shows for a 3d mesostructure of concrete in the elastic deformation range, how the computational complexity in analyses of numerical homogenization can be reduced at controlled errors. Reduction is systematically applied to specimen size S, resolution R, and discretization D, which span the newly introduced SRD parameter space. Key indicators for accuracy are (i) the phase fractions, (ii) the homogenized elasticity tensor, (iii) its invariance with respect to the applied boundary conditions and (iv) the total error as well as spatial error distributions, which are computed and estimated. Pre-analyses in the 2d SRD parameter sub-space explore the transferability to the 3d case. Beyond the concrete specimen undergoing elastic deformations in the present work, the proposed concept enables accuracy-efficiency balances for various classes of heterogeneous materials in different deformation regimes and thus contributes to build comprehensive digital twins of materials with validated attributes.
\end{minipage}
}
\end{center}

{\bf Keywords:}
Homogenization; Elasticity; Concrete; Error Analysis; Adaptivity; Finite elements 
%\hfill  vers.\,\today %\, at \currenttime\\
%\hspace*{6.3cm} 
%\\[2mm]
%\hfill vers.\,\today \, at \currenttime\\

% {\bf Content}: include results for distinction of apparent to effective properties in 2d and 3d. homogenization eqs., {\bf Formal}: ... 
 
\section{Introduction} 
\label{sec:intro} 
Concrete is a multiphase material consisting of a mortar phase and an aggregate phase, frequently accompanied by distributed pores. For the modeling and simulation of real lab-experiments such as e.g. a Brazilian-type compression test Huang et al. \cite{Huang.2015} there is no ambiguity about model dimensions and boundary conditions (BC); the simulation model must coincide with the experiment. For simulations on larger length scales in the context of computational homogenization, the appropriate size is closely related to the definition of a representative volume element (RVE), see Hill \cite{Hill.1963}, Drugan and Willis \cite{Drugan.1996}, Kanit et al. \cite{Kanit.2006}, Ostoja-Starzewski \cite{OstojaStarzewski.2006}, Gitman et al. \cite{Gitman.2007}, and for a discussion with further references see Geers and Kouznetsova \cite{Geers.2010} and Schr\"oder \cite{Schroder.2014}. 

The criteria for an RVE can directly refer to the statistical representativeness in the composition of a microdomain, for concrete in terms of the mortar-aggregate-pore phase fractions, aggregate shapes and orientations if present, the interface morphologies etc., which then result in a corresponding size of the microdomain. Another perspective points to the observation in simulations that if the considered microdomain is not sufficiently large, its overall response at large will depend on the applied BCs; periodic (PBC), kinematical uniform (KUBC) or stress uniform (SUBC) which all fulfill Hill's condition of micro-macro equivalence of energy densities. Following Huet \cite{Huet.1990} the corresponding homogenized properties are referred to as ''apparent'', which converge into ''effective'' ones, if the microdomain is large enough. In that case the influence of the applied BCs vanishes, the corresponding microdomain achieves the attribute ''representative'' and turns into an RVE, see also Ostoja-Starzewski \cite{OstojaStarzewski.2006}, Kanit et al. \cite{Kanit.2006}. 

%In the perspective of numerical homogenization realized by two-scale finite element methods, both the microscale as well as the macroscale analysis matter, the latter for effective properties and deformations on the length scale of structures at large. 
The vast majority of concrete homogenization analyses so far was carried out rather for apparent properties without testing the required size to achieve invariance with respect to the applied BCs in the above mentioned sense; for elastic properties in Das et al. \cite{Das.2015}, Wimmer et al. \cite{Wimmer.2016}, Youssef et al. \cite{Youssef.2018} and Luo et al. \cite{Luo.2020}; for mechanical, diffusive, and chemo-expansive properties in Bosco et al. \cite{Bosco.2020}; and similarly, in homogenization investigations for effective thermal conductivity in El Moumen et al. \cite{ElMoumen.2015} and in Wei et al. \cite{Wei.2013}, as well as in homogenization with a hydro-thermo-chemo-mechanical coupling in Wu et al. \cite{Wu.2013}.  

Highly resolved FIB-Tomography and X-ray Computed Tomography Technique (XCT) in Pennycoock and Nellist \cite{Pennycook.2011}, Holzer and Cantoni \cite{Holzer.2011}, and in Ketcham and Carlson \cite{Ketcham.2001} have become standard in the characterization of multiphase composites not only for concrete but also for a great variety of materials. 

The obtained ordered stack of pixelized images is used for the reconstruction of 3d micro- and mesostructures. 
Resultant voxel-grids can directly be used as 3d finite element discretizations as shown in Keyak et al. \cite{Keyak.1990}, Hollister et al. \cite{Hollister.1994}, Mishnaevsky \cite{Mishnaevsky.2005}, Young et al. \cite{Young.2008}, with applications to an open-cell foam in Michailidis et al. \cite{Michailidis.2010}, to reinforced composites in Sencu et al. \cite{sencu.2016}, to concrete and asphalt construction materials in the review of Du Plessis and Boshoff \cite{duplessis.2019} and to cement-based materials in the overview work of Chung et al. \cite{chung.2019} to name but a few. For an overview of various techniques for the (re)construction of microstructures and RVEs see Bargmann et al. \cite{Bargmann.2018}.
 
{\color{black} The impact of pixel and voxel resolution on simulation results was studied in Nguyen et al. \cite{Nguyen.2015} for foamed concrete, in Shah et al. \cite{Shah.2016} for conductivity in porous media and for applications in digital rock physics see Berg et al. \cite{Berg.2017}. Huang et al. \cite{Huang.2015} carried out a resolution compression by factor 4 in each direction of space and monitored the phase fraction ratios, which were virtually constant.} 
 
Adaptive mesh coarsening of an initially uniform pixel- or voxel-grid employing quadtree- and octree-type coarsening is instrumental to reduce the computational complexity, interfaces preserve their high resolution for accuracy and phase interiors are subject to coarsening for efficiency, Legrain et al. \cite{Legrain.2011}, 
%Lian et al. \cite{Lian.2013}, 
Huang et al. \cite{Huang.2015}, Ren et al. \cite{Ren.2015}, Saputra et al. \cite{Saputra.2017}, Gravenkamp and Duczek \cite{Gravenkamp.2017}, Saputra et al. \cite{Saputra.2018}. Only very recently the adaptive coarsening was carried out along with an accompanying error control \cite{Fischer.2020, Eidel.2020}.

The present work is based on data for a concrete specimen obtained by Huang et al. in \cite{Huang.2015}. The cubic specimen exhibits 371$^3$ voxels which implies more than 150 million unknown degrees of freedom (ndof). The number of unknowns is prohibitive for two-scale simulations using FE$^2$/FE-HMM, where to each quadrature point of the macroscopic finite element an RVE of that problem size is attached. 
Even more important, high resolution and full specimen size can turn out to be wasteful in view of sub-volumes and/or reduced resolutions which are enough to serve as an RVE.

In either case, and beyond the particular example of investigation in this work, there is a strong need for methods to tame an excess of data and thereby reduce the computational complexity while preserving sufficient accuracy and overall reliability of simulations. Three of these methods spanning the newly-introduced ''SRD parameter space'' of size S, resolution R and discretization D are shown in Fig.~\ref{fig:SRD-paraemeter-sapce}. 

\begin{Figure}[htbp]
	\centering
\fbox{
\parbox{15.2cm}{\hspace*{4mm}
\parbox{14.0cm}{
\begin{minipage}{14cm}
\centering  
		{\includegraphics[width=12cm, angle=0]{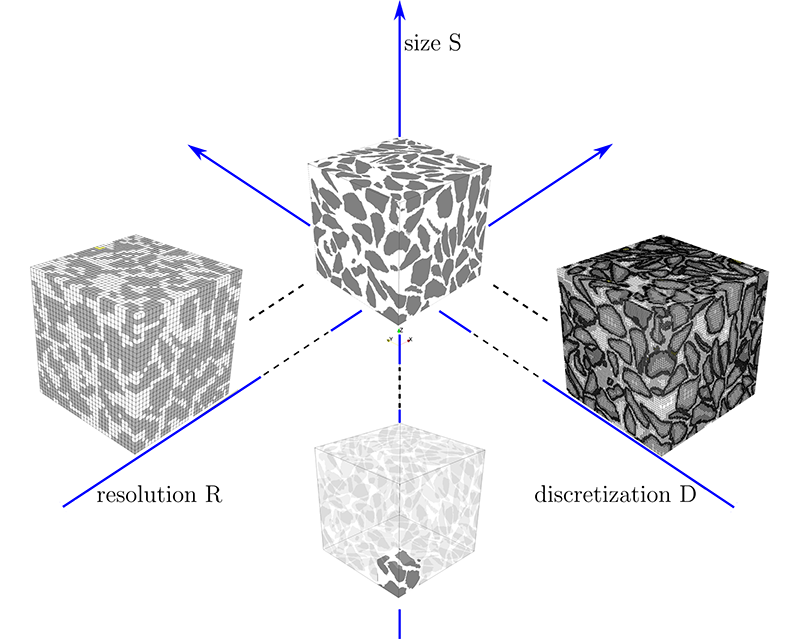}} 
	\caption{\textbf{The SRD parameter space of the digital twin of a concrete specimen.} The computational complexity can be reduced by the reduction of size S, by coarse-graining of resolution R or of discretization D, or a combination thereof. But how about the corresponding losses in accuracy?} 
	\label{fig:SRD-paraemeter-sapce}
\end{minipage}
}}}	
\end{Figure}

\begin{Figure}[htbp]
	\centering
	\includegraphics[width=12cm, angle=0]{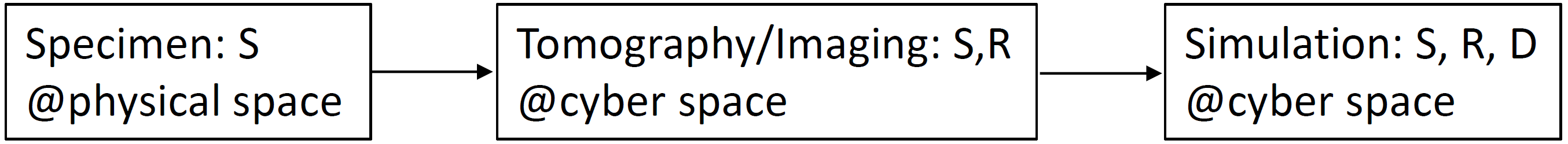} 
	\caption{\textbf{Workflow.} Processing and data transfer from physical to and within cyber space.}
	\label{fig:workflow-SRD}
\end{Figure}
 
In the workflow of Fig.~\ref{fig:workflow-SRD} only size S exists in physical space, whereas resolution R is added by image acquisition in tomography, and discretization D is added in the cyber space of simulation which completes the SRD parameter space. Data transfer exhibits a hierarchy in that the cyber space of tomography and imaging generates data in the physical space of the specimen and transfers them to the cyber agency of simulation and data analysis where it is processed. As a consequence, losses in data generation and transfer with respect to size and resolution can hardly be compensated by the recipient. This circumstance inherently favors image acquisition in size and resolution tending to the edges of feasibility which may turn out luxuriant for particular simulation purposes and therefore calls for adequate reductions. 

For the exploration of the SRD space as illustrated in Fig.~\ref{fig:SRD-paraemeter-sapce} the following questions arise:  
\begin{enumerate}
    \item {\bf Size S:} Which specimen size is sufficiently large to serve as an RVE? 
    \item {\bf Resolution R:} Which image-, hence pixel- and voxel-resolution provides the best balance between accuracy and efficiency? 
    \item {\bf Discretization D:} What can adaptive mesh coarsening contribute to reduce the computational efforts of a discretization which coincides with full resolution? How do resolution errors compare to discretization errors?
    \item {\bf Accuracy:} The complexity reduction in the SRD parameter space must be accompanied by error control. How accurate is error estimation as validated by the computation of actual errors in terms of global errors as well as error-distributions?
    \item {\bf Transfer from 2d to 3d:} To which extent is a transfer from the 2d SRD parameter subspace to the 3d space possible? How can 3d simulations profit from 2d pre-analyses? 
\end{enumerate}   

As the above introduction outlines, considerable progress already has been achieved with respect to the points 1.-3., but separately, whereas a panoramic view combining all three dimensions of the SRD parameter space, most notably with a rigorous error control, is lacking and therefore the topic of this work. 

Conclusions from 2d to 3d addressed in point 4. are of general interest, for elastic properties it was considered in Saxena and Mavko \cite{Saxena.2016}; moreover it has revealed a dimensional effect in stiffness, see Gl\"uge et al. \cite{Gluge.2020} and references therein. Here, we test the transferability of specimen size by monitoring ''convergence'' of homogenized elastic properties in the 2d and 3d SRD parameter space and the transferability of the quality of error estimation. Transferability in this respect would enable cheap 2d pre-analyses and thus have positive consequences for all of the above mentioned aspects 1.--4., since e.g. the accuracy of error estimation had not to be validated by expensive ''overkill'' solutions in 3d, which is for large specimen sizes hardly accessible.  
 
For addressing the above questions we restrict here our attention to elastic analyses being aware that spatial variations in the microstructure of composites may have a minor influence on the elastic properties but a strong impact on the overall inelastic behavior \cite{Geers.2010}. 
  
{\bf Notation.} The starting point in this work is the case of full specimen size S in full resolution, where voxel size $h_{\square}$ coincides with finite element size $h$. Within certain bounds however (e.g. $h\leq h_{\square}$) almost arbitrary combinations in the SRD parameter space are possible; for instance S256-RD128adap2 for specimen size S256 (edge length 256 $\times$ 0.1 mm, i.e. length unit is 0.1mm), resolution R128 (voxels per edge, hence after one step of resolution coarsening) and discretization D128 (elements per edge) with two consecutive adaptive mesh coarsening steps. 

\section{Microstructure} 

\begin{figure}[htbp]
	\centering
	\subfloat[3d microstructure]
	{\includegraphics[height=5.0cm, angle=0]{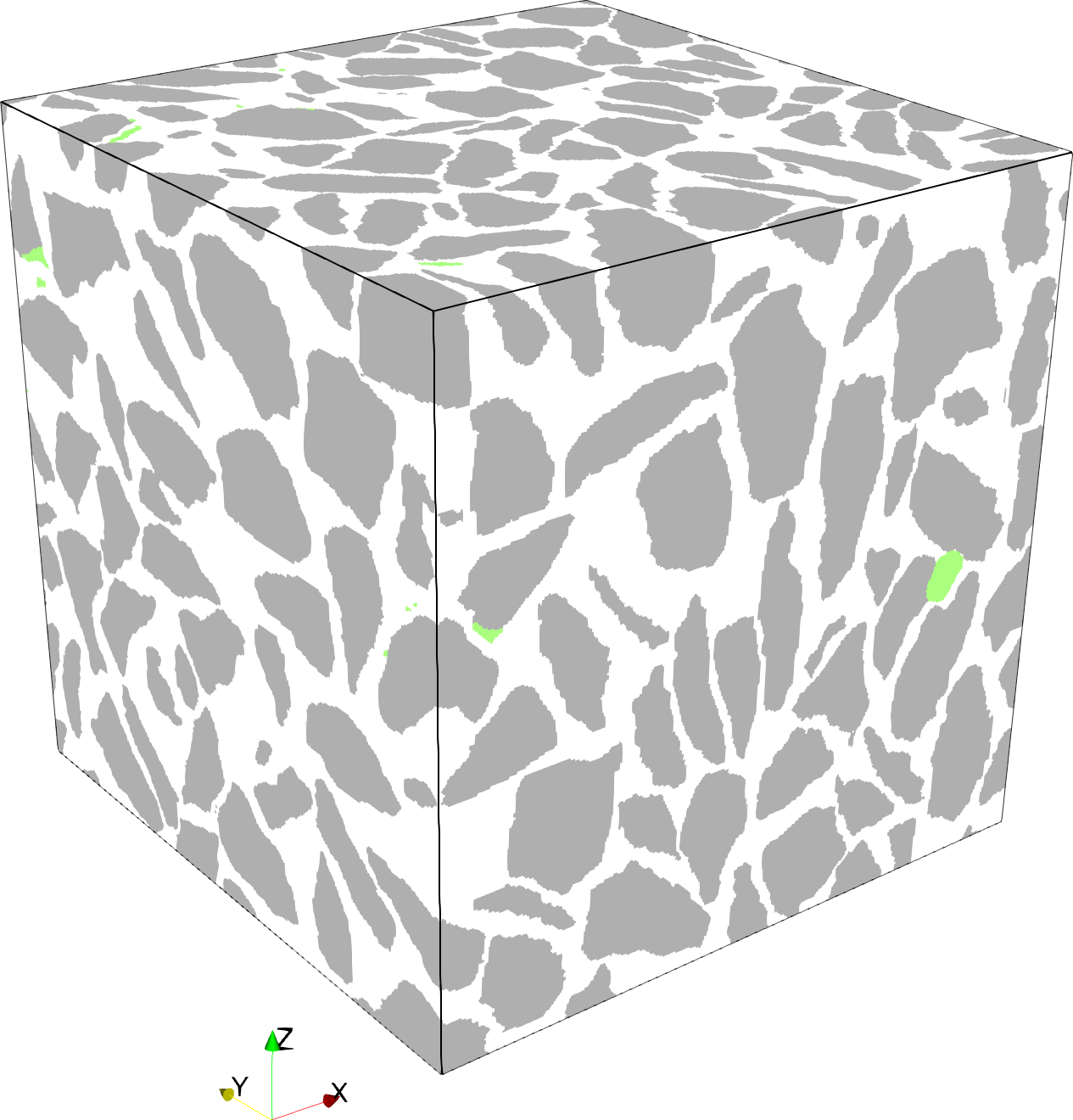}} \hspace*{0.10\linewidth}
	\subfloat[2d slice microstructure]
	{\includegraphics[height=5.0cm, angle=0]{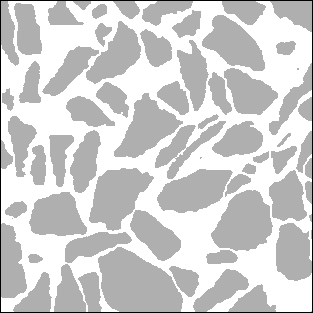}}
	\caption{\textbf{Concrete microstructure.} (a) Original 3d specimen of reference \cite{Huang.2015}, (b) exemplary 2d slice. Three phases, grey for aggregate, white for mortar, and green for pore.} 
	\label{fig:Concrete_structure}
\end{figure}
 
\begin{Table}[htbp]
	\begin{minipage}{16.5cm}  
		\centering
		\renewcommand{\arraystretch}{1.2} 
		\begin{tabular}{r c c c}			
        \hline
		                & Aggregate  & Mortar  & Pore    \\
		\hline
		3d phase fractions  & $0.5580$  & $0.4312$    & $0.0109$    \\
		2d phase fractions  & $0.5629$  & $0.4357$    & $0.0014$    \\
		$E$ (GPa)           & $50$      & $20$        & $-$         \\   % in simulation: 1
		$\nu$               & $0.3$     & $0.3$       & $-$         \\   % in simulation: 0.3
		\hline
		\end{tabular} 
	\end{minipage}
	\caption{\textbf{Phases}. Phase fractions and elastic properties of the constituents. 
	}
	\label{tab:Concrete_constituents} 
\end{Table}

The starting point of the present analysis is a cubic specimen of concrete with a side length of 37.1 mm. The mesoscale multiphase composition referred to as microstructure in the following is displayed in Fig.~\ref{fig:Concrete_structure} (a). It exhibits a resolution of 371$^3$ voxels and is the outcome of reconstruction from in-situ X-ray Computed Tomography (XCT) images provided in Huang et al. \cite{Huang.2015}.  
The specimen consists of three phases; beyond the aggregate and the mortar of a relatively small pore phase. The solid phases each exhibit isotropic, linear elastic material behavior described by $\sigma_{ij} = \mathbb{C}^{\epsilon}_{ijkl} \,\varepsilon_{kl}$, with the Cauchy-stress tensor $\sigma_{ij}$, the fourth order elasticity tensor $\mathbb{C}^{\epsilon}_{ijkl}$, and the infinitesimal strain tensor $\varepsilon_{kl}$. For convenience we use Voigt notation in the following. The volume fractions of the phases and the isotropic elasticity parameters of the solid phases are given in Tab.~\ref{tab:Concrete_constituents}. The aggregate-mortar stiffness contrast is 2.5.
The chosen area element in Fig.~\ref{fig:Concrete_structure} (b) is a slice in the $XZ$-plane. It is considered as representative for its good match of the phase fractions with the 3d specimen.

\section{Computational Homogenization with Adaptivity}
\label{sec:methods}

The methods used in the present work are briefly described with links to further references.

\subsection{Numerical homogenization} 

\begin{Figure}[htbp]
	\centering
\fbox{
\parbox{14.2cm}{\hspace*{4mm}
\parbox{14.0cm}{
\begin{minipage}{14cm}
\centering  
	{\includegraphics[width=12.0cm, angle=0]{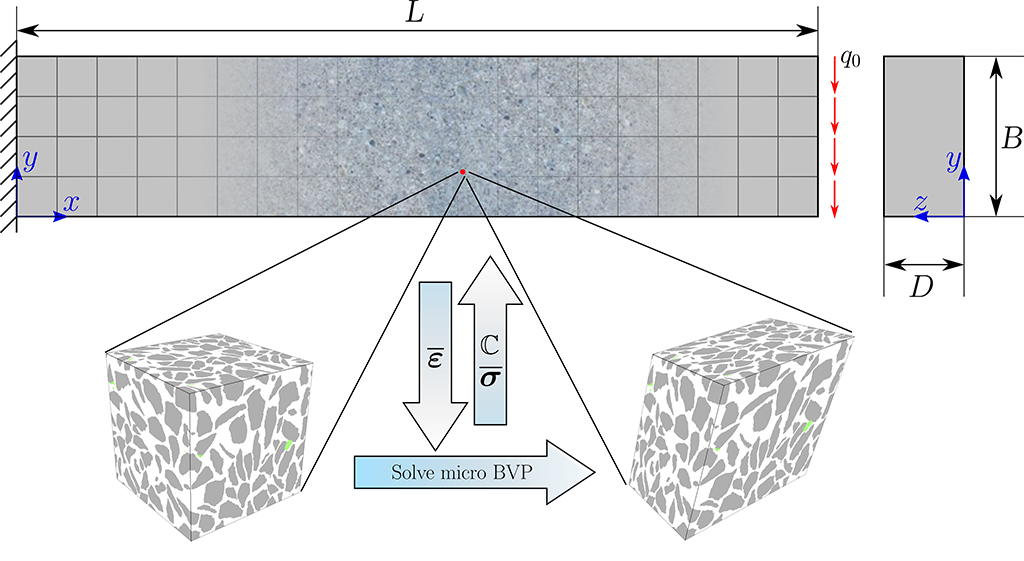}}
	\caption{\textbf{Micro-macro-transition in computational homogenization}. Macro-micro transfer (localization): macro-strain  $\overline{\bm \varepsilon}$, micro-macro transfer (homogenization): averaged stress $\overline{\bm \sigma}$ and homogenized elasticity tensor ${\mathbb{C}}$ from the solution of the micro boundary value problem (BVP).} 
	\label{fig:mic-mac-transition}
\end{minipage}
}}}	
\end{Figure}
The mechanical analysis of the concrete specimen is conducted in the framework of a two-scale finite element method for computational homogenization, the Finite Element Heterogeneous Multiscale Method FE-HMM or FE$^2$ based on the equivalence of macro and micro energy densities 
%$\langle \bm \sigma : \bm \varepsilon \rangle = \langle \bm \sigma \rangle : \langle \bm \varepsilon \rangle$
$\overline{\bm \sigma : \bm \varepsilon} = \overline{\bm \sigma} : \overline{\bm \varepsilon}$
with volume averages over the microdomain $\mathcal{B}_{\epsilon}$ defined as 
%$\langle \bullet \rangle := \frac{1}{|\mathcal{B}_{\epsilon}|} \int_{\mathcal{B}_{\epsilon}} \bullet \, \text{d}V$.
 $\overline{\bullet} := \frac{1}{|\mathcal{B}_{\epsilon}|} \int_{\mathcal{B}_{\epsilon}} \bullet \, \text{d}V$. The application of energetically consistent boundary conditions of KUBC, PBC and SUBC to the microproblem is realized by the method of Lagrange multipliers \cite{Fischer.2019}, and the computation of the homogenized elasticity tensor\footnote{Here we distinguish between the elasticity tensors for single phases $\mathbb{C}^{\epsilon}$ and the homogenized elasticity tensor $\mathbb{C}$, for the latter frequently the symbol $\overline{\mathbb{C}}$ is used.} $\mathbb{C}$ is done according to Abdulle \cite{Abdulle.2006} along with a modification for efficient computation~\cite{Eidel.2018}. Details of FE-HMM for linear elastic solids along with error and convergence analyses can be found in \cite{Eidel.2016, Eidel.2018}, and extensions to geometrical and material nonlinearity in \cite{Eidel.2018b, Eidel.22.08.2019}

\subsection{Resolution coarsening}  
\label{subsec:ResolutionCoarsening}

In resolution coarsening all voxels of the initial, finely resolved microstructure image undergo a uniform coarsening, in the phases and at interfaces. At interfaces rules must be introduced which define on how a cube of $2^3$ fine voxels having different Young's moduli (represented by different color codes of images) merge into one coarser voxel. Since the finite element discretization follows this uniform coarsening, voxels represent hexahedral finite elements with shape functions of polynomial order $q=1$ and pass their properties onto them. The same applies to pixel-coarsening in 2d.

In the following, two different variants of voxel coarsening are applied. The first version obeys a rule of mixtures; the newly created voxel exhibits properties of the volume average of the voxels merging in that coarser voxel, which amounts to taking the arithmetic mean for the same voxel size. For a sketch see the transformation of (a) into (b) in Fig. \ref{fig:coarsening-total}. The newly introduced phases with mixed properties at interfaces are an artefact for the case of distinguished phases with sharp boundaries. As a result, the sharp contrast e.g. in stiffness at interfaces is abraded which similarly reduces the corresponding stress jumps as the material response. The volume average of Young's moduli over the entire microdomain is preserved by the very definition of the mixture rule. This coarsening rule creates microstructures which resemble or even coincide with the outcome of tomography image acquisition in that they are typically raster-graphics showing mixed colors for pixels at interfaces. In a novel approach characterizing digital microstructures by the Minkowski-based quadratic normal tensor proposed by Ernesti et al. \cite{Ernesti.30.07.2020}, blurred interfaces are a feature inherent in the method.

\begin{Figure}[htbp]
	\centering
	\subfloat[]
	{\includegraphics[height=2.6cm,angle=0]{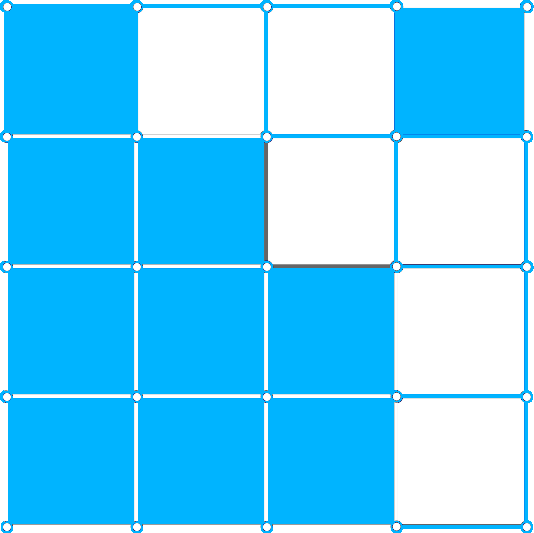}} 
	\hspace*{0.01\linewidth}
	\subfloat[]
	{\includegraphics[height=2.6cm,angle=0]{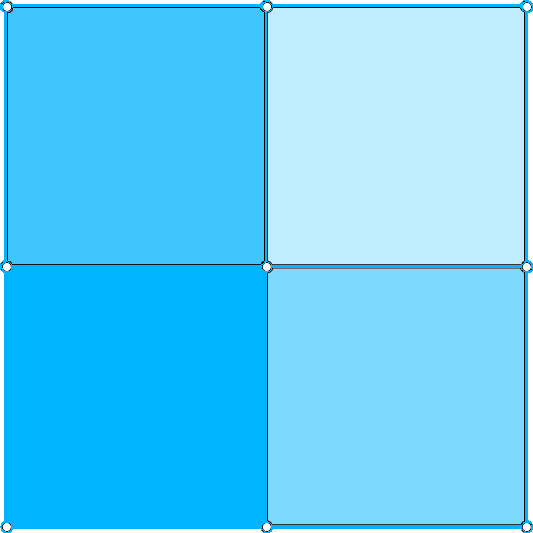}}
	\hspace*{0.01\linewidth}
	\subfloat[]
	{\includegraphics[height=2.6cm,angle=0]{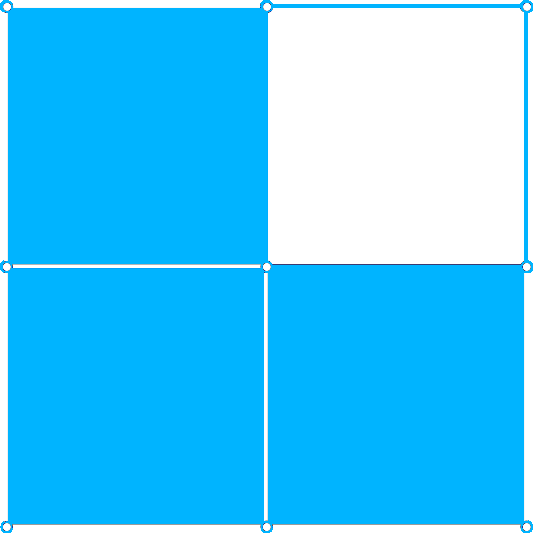}}
	\hspace*{0.04\linewidth}
	\subfloat[ ]
	{\includegraphics[height=3.4cm,angle=0]{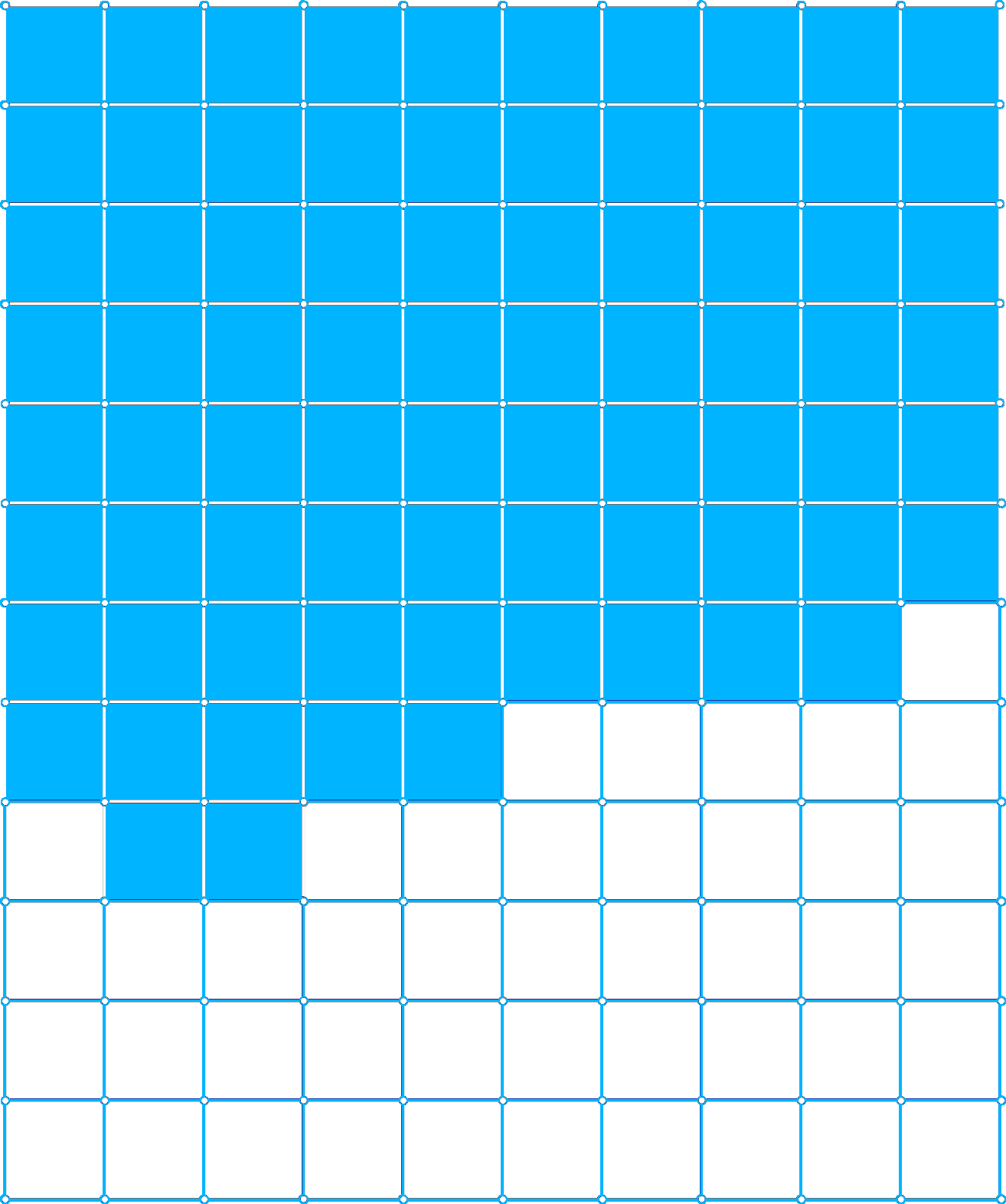}} \hspace*{0.02\linewidth}
	\subfloat[ ]
	{\includegraphics[height=3.4cm,angle=0]{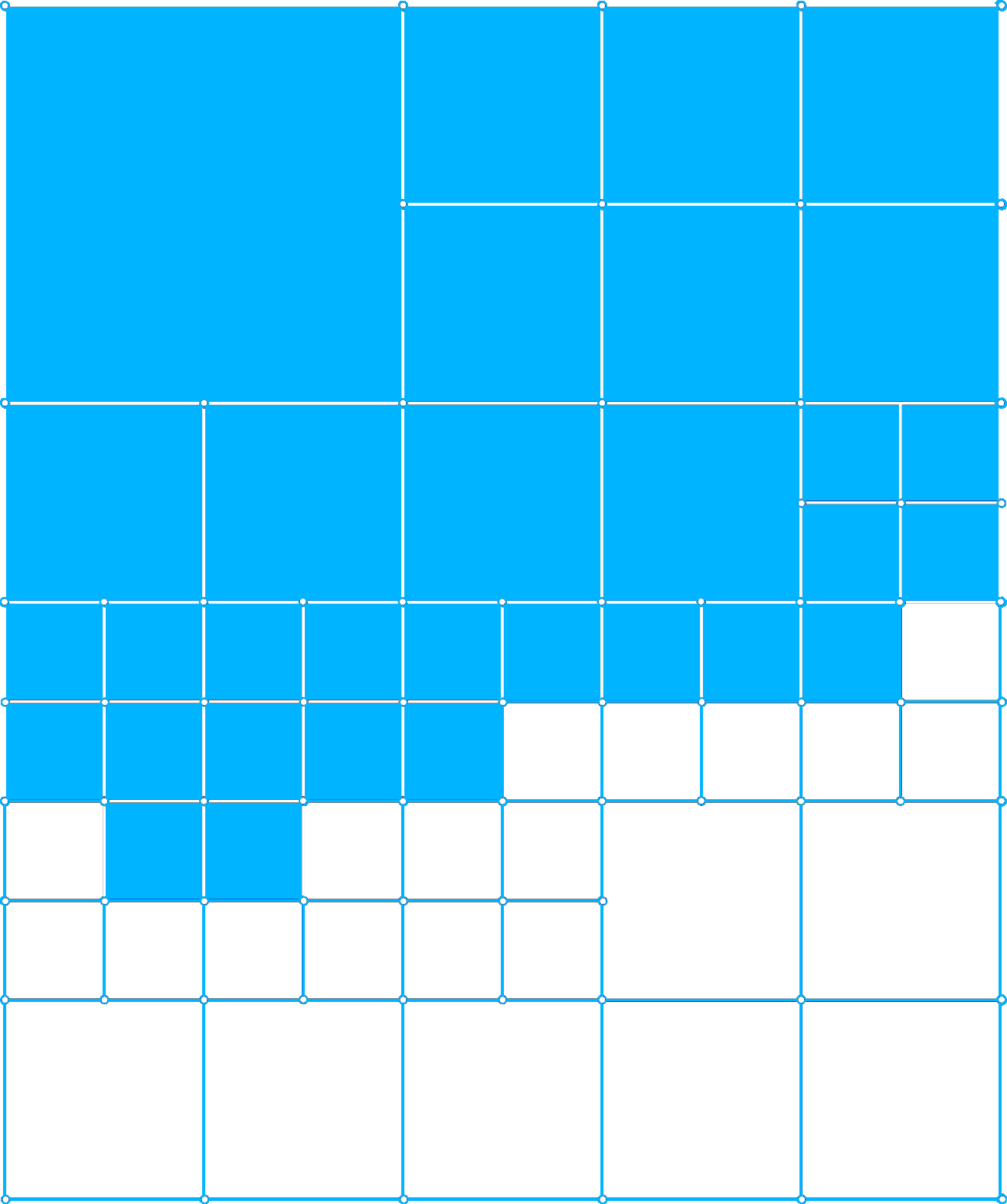}} 
	\caption{\textbf{Coarsening methods}. Resolution coarsening of a two-phase patch from (a) initial, high resolution into a coarsened grid (b) with new phases (mixture rule) and (c) with preserving the number of phases (''the-majority-wins'' rule). Adaptive mesh coarsening of (d) a uniform mesh into (b) a quadtree-type mesh.}
	\label{fig:coarsening-total}
\end{Figure}

The second variant overcomes the drawbacks of a coarsening that introduces new phases. For coarsened voxels at interfaces the quantity and properties of the initial phases are preserved following the rule, ''the-majority-wins''. In case of an equal count the newly created coarser voxel is endowed with those phase properties that shift the global phase ratio closer to the original one, compare the turn from (a) to (c) in Fig.~\ref{fig:coarsening-total}. This version corresponds to image segmentation which restores the number of real phases along with their sharp boundaries. Both variants are compared in \cite{Eidel.2020}.

\subsection{Adaptive mesh coarsening}   
\label{subsec:AdaptiveMeshCoarsening}
Adaptive mesh coarsening of the initially high resolution serves the purpose to reduce computational costs of discretizations following the highly resolved pixel/voxel grid. Therein, fine discretizations are preserved at interfaces for accuracy, and mesh coarsening is performed in the phase interiors for efficiency as visualized in Fig. \ref{fig:coarsening-total}, from (d) to (e) for the 2d case. In the transition from fine to coarse the gradient of element size should not be too steep in order to avoid a stiffening of the transition region due to the multitude of kinematical constraints applied to the hanging nodes on the transition elements \cite{Fischer.2020}. The attribute adaptive applies since the mesh coarsening in a preprocessing step anticipates the qualitative outcome of the simulation instead of following an a posteriori error distribution by mesh adaption. A thorough error analysis of quadtree-based mesh coarsening can be found in \cite{Fischer.2020}, \cite{Eidel.2020} in the context of homogenization. 
Quadtree-/octree-based finite element mesh generation was introduced in \cite{Yerry.1983} by Yerry and Shephard and an early overview can be found in Samet \cite{Samet.1990}. Methods like the Scaled Boundary FEM (SBFEM) and the Finite Cell Method (FCM) heavily rely on quadtree- and octree-meshing methods.

\subsection{Error estimation and error computation} 
\label{subsec:ErrorEstimationComputation} Error computation based on reference solutions for very fine discretizations controls the discretization error, but can become prohibitive. For this reason error estimation is required, which must be validated by comparison with actual errors. In this work a modified version of the reconstruction-based error indicator of Zienkiewicz and Zhu \cite{Zienkiewicz.1987, Zienkiewicz.1992, Zienkiewicz.1992b} is employed and measured by the energy norm. The energy norm of the solution $\bm u^h$ for a micro domain $\mathcal{B}_{\epsilon}$ is given by \eqref{eq:Enorm-solution}.

\begin{eqnarray}
||\,\bm u^h\,||^2_{A(\mathcal{B}_{\epsilon})} &=& \int_{\mathcal{B}_{\epsilon}} \mathbb{C}^\epsilon : \bm \varepsilon(\bm u^h):\bm \varepsilon(\bm u^h) \, \text{d}V \label{eq:Enorm-solution} 
\\
 e_{\text{mic}}^{2}:= || \bm e ||^2_{A(\mathcal{B}_{\epsilon})} 
%= \Vert \bm u^{0} - \bm u^h \Vert _{A(\Omega)} 
&=&  \int_{\mathcal{B}_{\epsilon}} \left( \boldsymbol{\sigma}^{\text{ref}} - \boldsymbol{\sigma}^h \right) : \left( \boldsymbol{\varepsilon}^{\text{ref}} - \boldsymbol{\varepsilon}^h \right) \, \text{d}V   
\label{eq:error_estimator_vs_reference_1}
\\
\bar e_{\text{mic}}^{2}:=|| \bar{\bm e} ||^2_{A({\color{black}\mathcal{B}_{\epsilon}})} 
&=&  \int_{{\color{black}\mathcal{B}_{\epsilon}}} \left( \boldsymbol{\sigma}^\star - \boldsymbol{\sigma}^h \right) : \left( \boldsymbol{\varepsilon}^\star - \boldsymbol{\varepsilon}^h \right) \, \text{d}V  
\label{eq:error_estimator} 
\\[2mm]
\Theta &=& || \bar{\bm e} ||_{A({\color{black}\mathcal{B}_{\epsilon}})} /
         || \bm e ||_{A(\mathcal{B}_{\epsilon})} 
\label{eq:Theta}         
\end{eqnarray}

The actual discretization error $|| \bm e ||_{A(\mathcal{B}_{\epsilon})}$ in terms of actual stresses $\boldsymbol{\sigma}^{\text{ref}}$ and strains $\boldsymbol{\varepsilon}^{\text{ref}}$ according to \eqref{eq:error_estimator_vs_reference_1} is obtained for a reference discretization, nominally with $h\rightarrow 0$, in practice for considerably finer discretizations than the mesh in use.   

Since the computation of the true error can become very expensive for large 3d problems, the estimation of errors obtained for the discretization in use is the method of choice. In the reconstruction-type error estimation of Zienkiewicz and Zhu according to \eqref{eq:error_estimator} the accuracy critically depends on the improved nodal stresses $\boldsymbol{\sigma}^\star$ and strains $\boldsymbol{\varepsilon}^\star$. The method for their computation is described in Sec. \ref{subsec:ImprovedStressCalculation} with a particular account of interfaces. 

The effectivity index $\Theta$ in \eqref{eq:Theta} as the ratio of the estimated error to the actual error is a measure for the accuracy of error estimation. For $h\rightarrow 0$ it must hold $\Theta \rightarrow 1$ for consistency. 

\subsection{Calculation of accurate/improved stresses}  
\label{subsec:ImprovedStressCalculation} 
Stress computation is of cardinal importance in the context of multiphase materials, in particular for error estimation. 
In the reconstruction-based error estimation, improved stresses ${\bm \sigma}^\star$ have direct impact on the quality of the error estimate as described in Sec. \ref{subsec:ErrorEstimationComputation}. For nodes in the phase interior they are calculated simply by averaging the values extrapolated from quadrature points of adjacent elements to that node. More sophisticated techniques exploiting superconvergence as proposed in Zienkiewicz and Zhu \cite{Zienkiewicz.1992, Zienkiewicz.1992b} have not shown a superior error estimate such that their computational overhead does not pay off in the present context. 
Stress averaging for a node at the discrete interface of different phases would ignore the stiffness mismatch and the corresponding stress jumps. For this reason, a phase distinction assigning multiple sets of stresses to a node avoids the falsifying effects of averaging. Moreover, the phase distinction in stress computation results in more accurate error estimates compared to a standard stress averaging that ignores interfaces \cite{Fischer.2020, Eidel.2020}.

%-------------------------------------------------------------------------------- 
%--------------------------------------------------------------------
\section{Preliminary Analysis for 2D}   
\label{sec:Concrete_2D} 
%--------------------------------------------------------------------
The overall aim of this section is to explore, to which extent the 2d simulation results are transferable to the 3d case with respect to size, resolution, and discretization, but also with regard to the quality of error estimation.   
 
The corresponding macro problem is chosen to be a square plate of edge length l=1000 mm, with a discretization of $20 \times 20$ quadrilateral elements with linear shape functions $p=1$. Dirichlet boundary conditions are applied to the left side of the cantilever plate, Neumann boundary conditions with a load $\bm g = [0, -100]^T$ N/mm$^2$ are applied to the opposite edge. The square micro domain is a slice of the 3d specimen with side length S (in 0.1mm) and subjected to plane strain conditions in the simulations and, if not otherwise stated, to PBC. The microdomain is located at $[960.57, 960,57]$ (mm) in the upper right corner of the plate. Figure \ref{fig:Concrete_structure}(b) shows the 2d slice taken from the original 3d concrete specimen.

\subsection{Quadtree-based adaptive mesh coarsening}
Adaptive mesh coarsening is applied to the initially uniform discretization. It is based on a quadtree-type discretization as described in \cite{Fischer.2020}; the rule is that at interfaces the fine resolution is maintained for accuracy, while in the interior of phases the mesh-coarsening is carried out for efficiency. 

\begin{figure}[htbp]
	\centering
	\subfloat[SRD320adap1]
	{\includegraphics[height=3.8cm, angle=0]{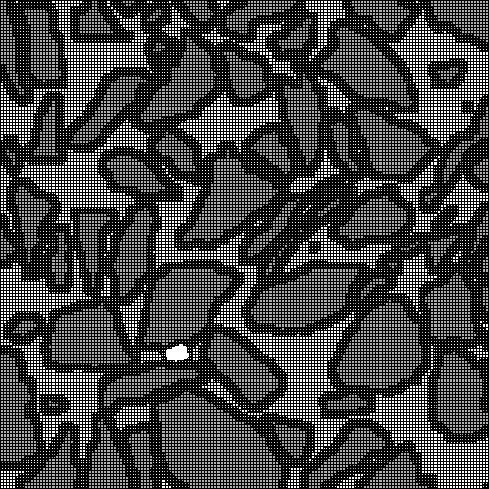}} \hspace*{0.04\linewidth}
	\subfloat[SRD320adap2]
	{\includegraphics[height=3.8cm, angle=0]{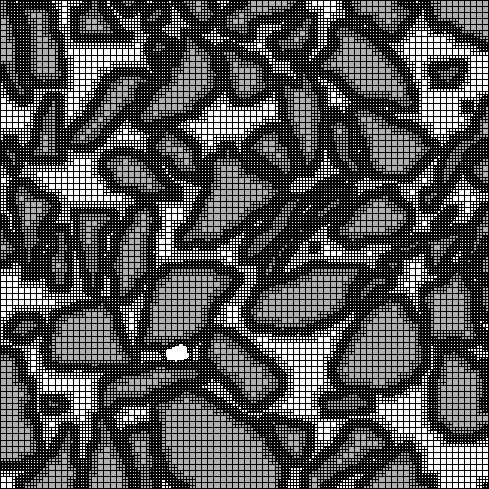}}
    \hspace*{0.04\linewidth}
	\subfloat[SRD320adap3]
	{\includegraphics[height=3.8cm, angle=0]{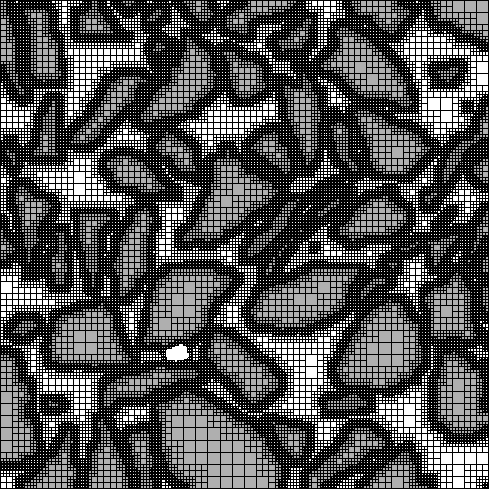}}
	\caption{\textbf{Specimen slice SRD320}. Adaptively coarsened meshes.}
	\label{fig:Concrete_2d_meshes}
\end{figure}
\begin{Table}[htbp]

	\begin{minipage}{16.5cm}  
		\centering
		\renewcommand{\arraystretch}{1.2} 
		\begin{tabular}{r c c c c}			
        \hline
               & \multicolumn{4}{c}{Adaptive mesh coarsening} \\ 
		SRD320 & 0  & 1  & 2  & 3 \\
		\hline
		ndof  & $205\,852$ & $108\,228$ & $95\,628$ & $94\,938$ \\
	    Factor & $1.0000$ & $0.5258$ & $0.4645$ & $0.4612$ \\
		Deactivated ndof  & $0$ & $6\,581$ & $8\,208$ & $8\,367$ \\
		\hline
		\end{tabular} 
	\end{minipage}
	\caption{\textbf{Specimen slice SRD320}. Number of degrees of freedom (ndof) of the original, uniform mesh (0) and the adaptively coarsened meshes (1--3), the corresponding reduction factor and the deactivated ndof for hanging nodes.}
	\label{tab:Concrete_coarsening_steps} 
\end{Table}

Table~\ref{tab:Concrete_coarsening_steps} provides the number of unknowns of the original mesh and the meshes resulting from the coarsening steps. After three coarsening steps the ndof is reduced to 46\% compared with the initial mesh.  

Here and in the following the ndof in the tables equals the ndof of the micro mesh minus the deactivated ndof of the hanging nodes.

\subsection{Validation of error estimation by comparison with actual errors} 
 
\begin{Table}[htbp]
	\begin{minipage}{16.5cm}  
		\centering
		\renewcommand{\arraystretch}{1.2} 
		 \resizebox{0.5\columnwidth}{!}{%
		\begin{tabular}{r r c c c c}
			\hline 		    
		     \multicolumn{2}{c}{SRD320} & \multicolumn{4}{c}{$e_{\text{mic}}$} \\
%		     \multicolumn{2}{c}{} & \multicolumn{4}{c}{adaptive coarsening} \\
		     \multicolumn{2}{c}{Reference D} & $0$ & $1$ & $2$ & $3$  \\
	        \hline
		   	 &  D640  &  $6.2297$  & $6.5747$ & $6.8850$ & $7.0670$ \\
		   	 &  & $1.000$ & $1.0554$ & $1.1052$ & $1.1344$ \\
		   	\cmidrule{3-6}
			 &  D1280   & $7.2830$ & $7.5803$ & $7.8509$ & $8.0109$ \\
		     & & $1.0000$ & $1.0408$ & $1.0780$ & $1.0999$ \\
		   	\cmidrule{3-6}
		     &  D1920  &  $7.5137$ & $7.8021$ & $8.0652$ & $8.2211$ \\
		   	 & & $1.000$ & $1.0384$ & $1.0734$ & $1.0942$ \\
		   	\cmidrule{3-6}
		     &  D2560   &  $7.6051$ & $7.8902$ & $8.1505$ &  $8.3048$ \\
		   	 & & $1.000$ & $1.0374$ & $1.0717$ & $1.0920$ \\
			\cmidrule{1-6}
			\multicolumn{2}{c}{} & \multicolumn{4}{c}{$\bar{e}_{\text{mic}}$} \\
			&  D320   & $5.8034$ & $6.1197$ & $6.3782$ & $6.4913$ \\
			& & $1.0000$ & $1.0545$ & $1.0991$ & $1.1185$ \\
			\hline
			& $\theta$(D2560) & $0.7631$ & $0.7756$ & $0.7826$ & $0.7816$ \\
			  \hline
		\end{tabular} 
		}
	\end{minipage}
	\caption{\textbf{Specimen slice SRD320}. Actual errors $e_{\text{mic}}$ and estimated errors $\bar{e}_{\text{mic}}$ for the uniform mesh (0) and the adaptively coarsened discretizations (1--3). Actual errors for various reference discretizations, SR320 D $\in$ \{640, 1280, 1920, 2560\}. Effectivity index $\Theta$ for D2560. Errors in $10^{-4}$ (Nmm).}
	\label{tab:Concrete_error_reference_microlevel} 
\end{Table} 
 
Table~\ref{tab:Concrete_error_reference_microlevel} shows the errors in the energy norm for the original, uniform mesh and the meshes from the three coarsening steps based on reference solutions for different discretizations. Even though the ndof is decreased by up to 54\%, the increase of the actual error remains below 10\%. The estimated errors are smaller than the calculated errors. The values of the effectivity index are close to 0.8 and indicate a good agreement of the estimated and the true errors for both uniform and adaptively coarsened meshes. A reference solution obtained by $h^{\text{ref}}=h/4$ (here at D1280) with $h$ being the current element size is enough to obtain a sufficiently accurate solution for the computation of the actual error. 
   
\begin{figure}[htbp]
	\centering
	\subfloat[$e_{\text{mic}}$ (\%) for D320uni]
	{\includegraphics[height=3.8cm, angle=0]{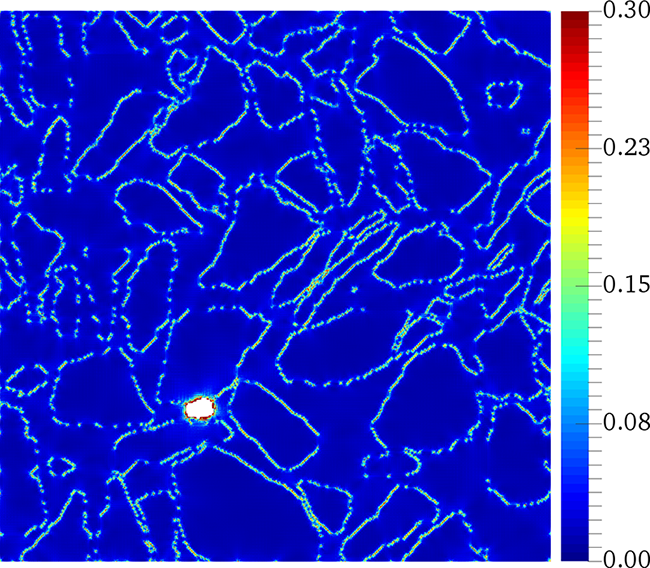}} \hspace*{0.02\linewidth}
	\subfloat[$\bar{e}_{\text{mic}}$ (\%) for D320uni]
	{\includegraphics[height=3.8cm, angle=0]{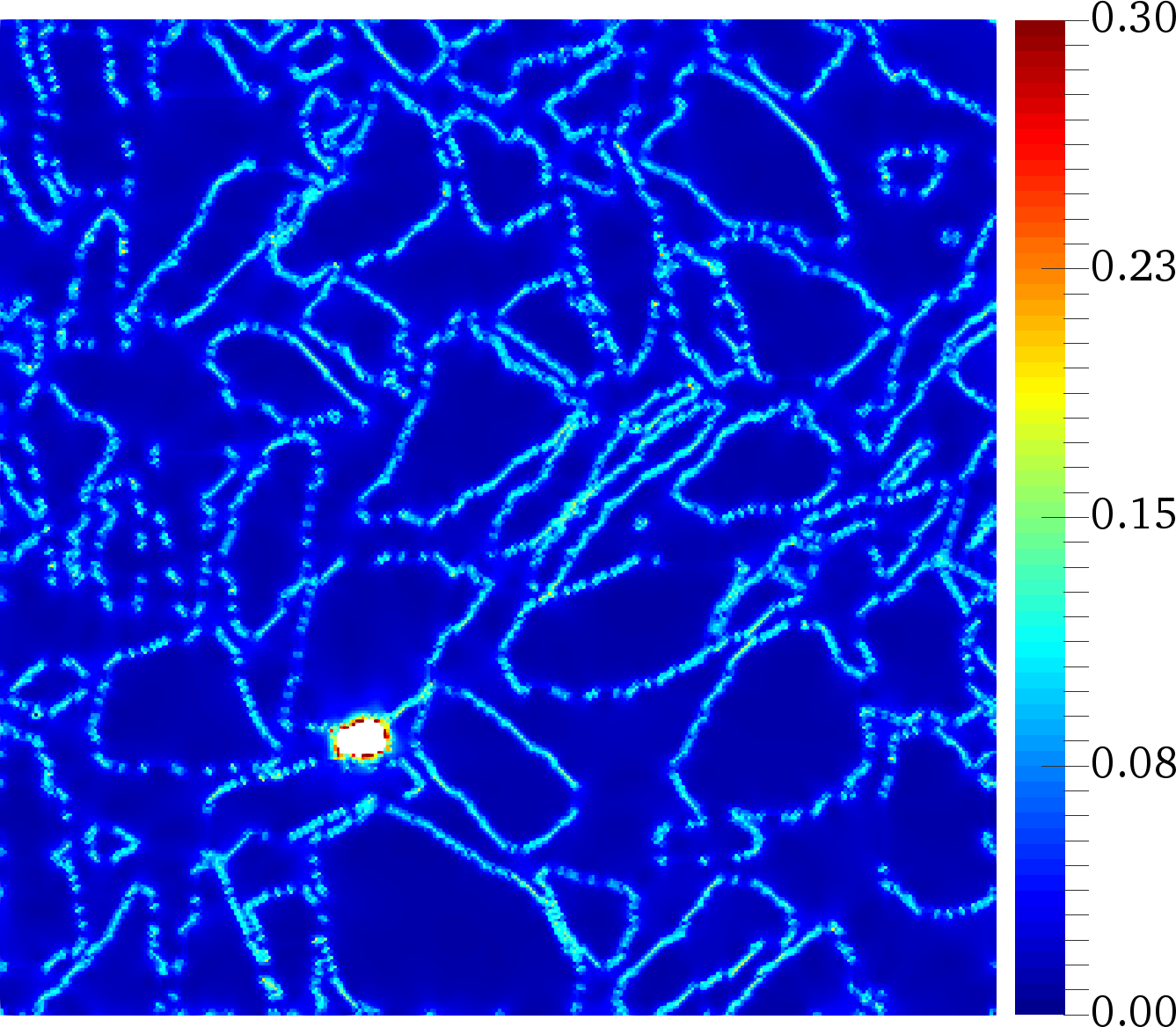}}
    \hspace*{0.04\linewidth}	
	\subfloat[$\varepsilon_{xx}$ for D320uni]
	{\includegraphics[height=3.8cm, angle=0]{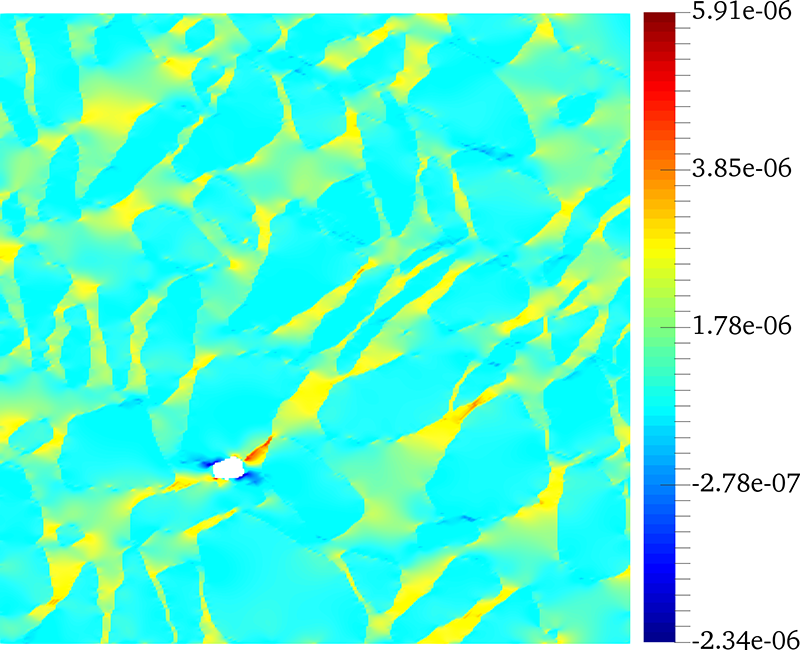}}
	\\
    \subfloat[$e_{\text{mic}}$ (\%) for D320adap3]
	{\includegraphics[height=3.8cm, angle=0]{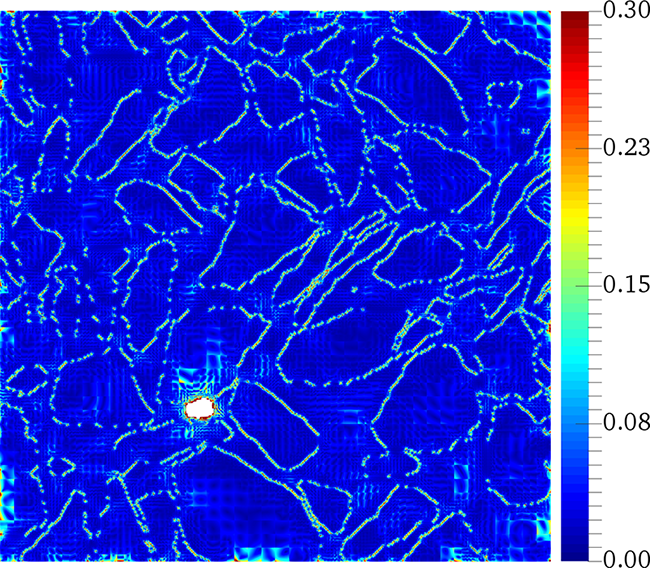}}  
	\hspace*{0.02\linewidth}
	\subfloat[$\bar{e}_{\text{mic}}$ (\%) for D320adap3]
	{\includegraphics[height=3.8cm, angle=0]{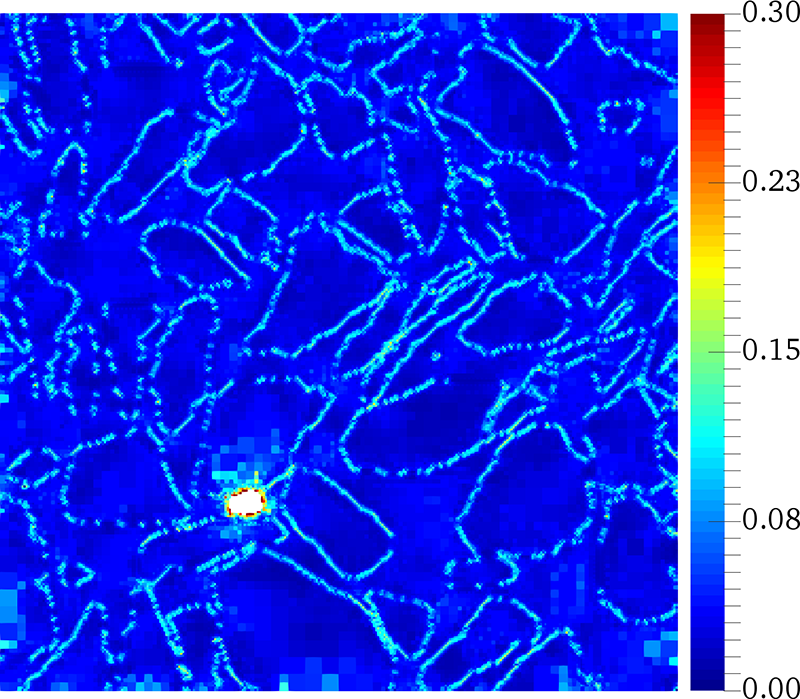}}
    \hspace*{0.04\linewidth}
	\subfloat[$\varepsilon_{xx}$ for D320adap3]
	{\includegraphics[height=3.8cm, angle=0]{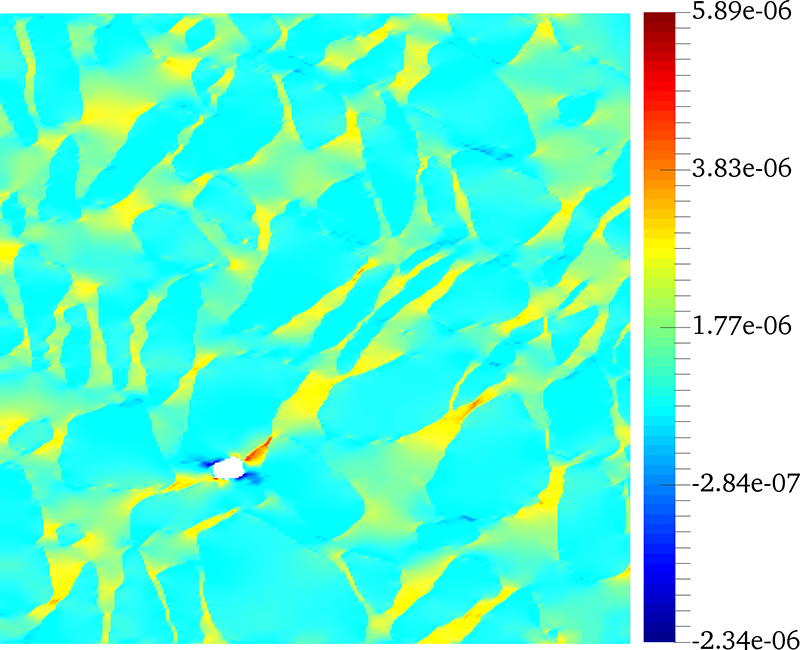}}
	\caption{\textbf{Specimen slice SR320}. For uniform and adaptively coarsened meshes, the distributions of the actual relative errors $|| {\bm e} ||_{A(\mathcal{B}_{e})} / || {\bm u} ||_{A(\mathcal{B}_{e})}$ in (a) and (d), and estimated relative errors $|| \bar{\bm e} ||_{A(\mathcal{B}_{e})} / || {\bm u} ||_{A(\mathcal{B}_{e})}$ in (b) and (e), and of the strain component $\varepsilon_{xx}$ in (c) and (f).} 
	\label{fig:Concrete_error_distribution}
\end{figure}

\begin{Figure}[htbp]
	\label{fig:Concrete_strain_distribution}
\end{Figure}
 
For the investigation of the error distributions on the micro domain the relative elementwise micro discretization error is analyzed. This relative error is the ratio of the computed or estimated error of an element to the element's energy norm. By doing so the influence of the element size is eliminated which is especially important for the non-uniform meshes.

The results in Fig.~\ref{fig:Concrete_error_distribution} indicate that (i) the error is largest at the aggregate-mortar interfaces, (ii) it is comparatively small in the interior of the two solid phases, which (iii) justifies the adaptive mesh-coarsening preserving fine resolution at interfaces and coarsening in the phase interior. Moreover it can be seen that (iv) the estimated errors are qualitatively and quantitatively in good agreement with the true errors, see the distributions and their maxima 
in Fig.~\ref{fig:Concrete_error_distribution}.

\subsection{Dependence of the 2d homogenized elastic constants on various parameters}

\subsubsection{Dependence on slice size} 

Figure~\ref{fig:component-elast-tensor_RAE_size} displays the coefficients of the homogenized elasticity tensor for different slice sizes, each with PBC. In the right of Fig.~\ref{fig:component-elast-tensor_RAE_size} some of them are shown.  %Fig.~\ref{fig:Concrete_RAE_size} shows three of them. 
As a result, the magnitude of the coefficients directly depends on the phase fraction of the aggregate as the stiffest phase having discrete peaks at S64 and S240. Since the finite element size coincides with the pixel size, the results for smaller samples exhibit larger discretization errors.

\begin{Figure}[htbp]
	\centering
    {\resizebox{0.46\columnwidth}{!}{% GNUPLOT: LaTeX picture with Postscript
\begingroup
  % Encoding inside the plot.  In the header of your document, this encoding
  % should to defined, e.g., by using
  % \usepackage[cp1252,<other encodings>]{inputenc}
  \inputencoding{cp1252}%
  \makeatletter
  \providecommand\color[2][]{%
    \GenericError{(gnuplot) \space\space\space\@spaces}{%
      Package color not loaded in conjunction with
      terminal option `colourtext'%
    }{See the gnuplot documentation for explanation.%
    }{Either use 'blacktext' in gnuplot or load the package
      color.sty in LaTeX.}%
    \renewcommand\color[2][]{}%
  }%
  \providecommand\includegraphics[2][]{%
    \GenericError{(gnuplot) \space\space\space\@spaces}{%
      Package graphicx or graphics not loaded%
    }{See the gnuplot documentation for explanation.%
    }{The gnuplot epslatex terminal needs graphicx.sty or graphics.sty.}%
    \renewcommand\includegraphics[2][]{}%
  }%
  \providecommand\rotatebox[2]{#2}%
  \@ifundefined{ifGPcolor}{%
    \newif\ifGPcolor
    \GPcolortrue
  }{}%
  \@ifundefined{ifGPblacktext}{%
    \newif\ifGPblacktext
    \GPblacktextfalse
  }{}%
  % define a \g@addto@macro without @ in the name:
  \let\gplgaddtomacro\g@addto@macro
  % define empty templates for all commands taking text:
  \gdef\gplbacktext{}%
  \gdef\gplfronttext{}%
  \makeatother
  \ifGPblacktext
    % no textcolor at all
    \def\colorrgb#1{}%
    \def\colorgray#1{}%
  \else
    % gray or color?
    \ifGPcolor
      \def\colorrgb#1{\color[rgb]{#1}}%
      \def\colorgray#1{\color[gray]{#1}}%
      \expandafter\def\csname LTw\endcsname{\color{white}}%
      \expandafter\def\csname LTb\endcsname{\color{black}}%
      \expandafter\def\csname LTa\endcsname{\color{black}}%
      \expandafter\def\csname LT0\endcsname{\color[rgb]{1,0,0}}%
      \expandafter\def\csname LT1\endcsname{\color[rgb]{0,1,0}}%
      \expandafter\def\csname LT2\endcsname{\color[rgb]{0,0,1}}%
      \expandafter\def\csname LT3\endcsname{\color[rgb]{1,0,1}}%
      \expandafter\def\csname LT4\endcsname{\color[rgb]{0,1,1}}%
      \expandafter\def\csname LT5\endcsname{\color[rgb]{1,1,0}}%
      \expandafter\def\csname LT6\endcsname{\color[rgb]{0,0,0}}%
      \expandafter\def\csname LT7\endcsname{\color[rgb]{1,0.3,0}}%
      \expandafter\def\csname LT8\endcsname{\color[rgb]{0.5,0.5,0.5}}%
    \else
      % gray
      \def\colorrgb#1{\color{black}}%
      \def\colorgray#1{\color[gray]{#1}}%
      \expandafter\def\csname LTw\endcsname{\color{white}}%
      \expandafter\def\csname LTb\endcsname{\color{black}}%
      \expandafter\def\csname LTa\endcsname{\color{black}}%
      \expandafter\def\csname LT0\endcsname{\color{black}}%
      \expandafter\def\csname LT1\endcsname{\color{black}}%
      \expandafter\def\csname LT2\endcsname{\color{black}}%
      \expandafter\def\csname LT3\endcsname{\color{black}}%
      \expandafter\def\csname LT4\endcsname{\color{black}}%
      \expandafter\def\csname LT5\endcsname{\color{black}}%
      \expandafter\def\csname LT6\endcsname{\color{black}}%
      \expandafter\def\csname LT7\endcsname{\color{black}}%
      \expandafter\def\csname LT8\endcsname{\color{black}}%
    \fi
  \fi
    \setlength{\unitlength}{0.0500bp}%
    \ifx\gptboxheight\undefined%
      \newlength{\gptboxheight}%
      \newlength{\gptboxwidth}%
      \newsavebox{\gptboxtext}%
    \fi%
    \setlength{\fboxrule}{0.5pt}%
    \setlength{\fboxsep}{1pt}%
\begin{picture}(6802.00,5102.00)%
    \gplgaddtomacro\gplbacktext{%
      \csname LTb\endcsname%%
      \put(548,4012){\makebox(0,0)[r]{\strut{}$33$}}%
      \csname LTb\endcsname%%
      \put(548,4345){\makebox(0,0)[r]{\strut{}$37$}}%
      \csname LTb\endcsname%%
      \put(548,4678){\makebox(0,0)[r]{\strut{}$41$}}%
      \csname LTb\endcsname%%
      \put(1133,3542){\makebox(0,0){\strut{} }}%
      \csname LTb\endcsname%%
      \put(1587,3542){\makebox(0,0){\strut{} }}%
      \csname LTb\endcsname%%
      \put(2040,3542){\makebox(0,0){\strut{} }}%
      \csname LTb\endcsname%%
      \put(2493,3542){\makebox(0,0){\strut{} }}%
      \csname LTb\endcsname%%
      \put(2947,3542){\makebox(0,0){\strut{} }}%
      \csname LTb\endcsname%%
      \put(3400,3542){\makebox(0,0){\strut{} }}%
      \csname LTb\endcsname%%
      \put(3853,3542){\makebox(0,0){\strut{} }}%
      \csname LTb\endcsname%%
      \put(4307,3542){\makebox(0,0){\strut{} }}%
      \csname LTb\endcsname%%
      \put(4760,3542){\makebox(0,0){\strut{} }}%
      \csname LTb\endcsname%%
      \put(5213,3542){\makebox(0,0){\strut{} }}%
      \csname LTb\endcsname%%
      \put(5667,3542){\makebox(0,0){\strut{} }}%
      \csname LTb\endcsname%%
      \put(6120,3542){\makebox(0,0){\strut{} }}%
    }%
    \gplgaddtomacro\gplfronttext{%
      \csname LTb\endcsname%%
      \put(5529,4672){\makebox(0,0)[r]{\strut{}$\mathbb{C}_{11}$}}%
      \csname LTb\endcsname%%
      \put(5529,4452){\makebox(0,0)[r]{\strut{}$\mathbb{C}_{22}$}}%
    }%
    \gplgaddtomacro\gplbacktext{%
      \csname LTb\endcsname%%
      \put(548,2895){\makebox(0,0)[r]{\strut{}$12$}}%
      \csname LTb\endcsname%%
      \put(548,3328){\makebox(0,0)[r]{\strut{}$14$}}%
      \csname LTb\endcsname%%
      \put(1133,2458){\makebox(0,0){\strut{} }}%
      \csname LTb\endcsname%%
      \put(1587,2458){\makebox(0,0){\strut{} }}%
      \csname LTb\endcsname%%
      \put(2040,2458){\makebox(0,0){\strut{} }}%
      \csname LTb\endcsname%%
      \put(2493,2458){\makebox(0,0){\strut{} }}%
      \csname LTb\endcsname%%
      \put(2947,2458){\makebox(0,0){\strut{} }}%
      \csname LTb\endcsname%%
      \put(3400,2458){\makebox(0,0){\strut{} }}%
      \csname LTb\endcsname%%
      \put(3853,2458){\makebox(0,0){\strut{} }}%
      \csname LTb\endcsname%%
      \put(4307,2458){\makebox(0,0){\strut{} }}%
      \csname LTb\endcsname%%
      \put(4760,2458){\makebox(0,0){\strut{} }}%
      \csname LTb\endcsname%%
      \put(5213,2458){\makebox(0,0){\strut{} }}%
      \csname LTb\endcsname%%
      \put(5667,2458){\makebox(0,0){\strut{} }}%
      \csname LTb\endcsname%%
      \put(6120,2458){\makebox(0,0){\strut{} }}%
    }%
    \gplgaddtomacro\gplfronttext{%
      \csname LTb\endcsname%%
      \put(130,3219){\rotatebox{-270}{\makebox(0,0){\strut{}$\mathbb{C}_{ij}$ (in GPa)}}}%
      \csname LTb\endcsname%%
      \put(5529,3588){\makebox(0,0)[r]{\strut{}$\mathbb{C}_{33}$}}%
    }%
    \gplgaddtomacro\gplbacktext{%
      \csname LTb\endcsname%%
      \put(548,1865){\makebox(0,0)[r]{\strut{}$8$}}%
      \csname LTb\endcsname%%
      \put(548,2407){\makebox(0,0)[r]{\strut{}$10$}}%
      \csname LTb\endcsname%%
      \put(1133,1374){\makebox(0,0){\strut{} }}%
      \csname LTb\endcsname%%
      \put(1587,1374){\makebox(0,0){\strut{} }}%
      \csname LTb\endcsname%%
      \put(2040,1374){\makebox(0,0){\strut{} }}%
      \csname LTb\endcsname%%
      \put(2493,1374){\makebox(0,0){\strut{} }}%
      \csname LTb\endcsname%%
      \put(2947,1374){\makebox(0,0){\strut{} }}%
      \csname LTb\endcsname%%
      \put(3400,1374){\makebox(0,0){\strut{} }}%
      \csname LTb\endcsname%%
      \put(3853,1374){\makebox(0,0){\strut{} }}%
      \csname LTb\endcsname%%
      \put(4307,1374){\makebox(0,0){\strut{} }}%
      \csname LTb\endcsname%%
      \put(4760,1374){\makebox(0,0){\strut{} }}%
      \csname LTb\endcsname%%
      \put(5213,1374){\makebox(0,0){\strut{} }}%
      \csname LTb\endcsname%%
      \put(5667,1374){\makebox(0,0){\strut{} }}%
      \csname LTb\endcsname%%
      \put(6120,1374){\makebox(0,0){\strut{} }}%
    }%
    \gplgaddtomacro\gplfronttext{%
      \csname LTb\endcsname%%
      \put(5529,2505){\makebox(0,0)[r]{\strut{}$\mathbb{C}_{12}$}}%
    }%
    \gplgaddtomacro\gplbacktext{%
      \colorrgb{0.00,0.00,1.00}%%
      \put(548,713){\makebox(0,0)[r]{\strut{}$0.55$}}%
      \colorrgb{0.00,0.00,1.00}%%
      \put(548,1052){\makebox(0,0)[r]{\strut{}$0.6$}}%
      \colorrgb{0.00,0.00,1.00}%%
      \put(548,1391){\makebox(0,0)[r]{\strut{}$0.65$}}%
      \csname LTb\endcsname%%
      \put(1133,290){\makebox(0,0){\strut{} 16}}%
      \csname LTb\endcsname%%
      \put(1587,290){\makebox(0,0){\strut{}32 }}%
      \csname LTb\endcsname%%
      \put(2040,290){\makebox(0,0){\strut{} 64}}%
      \csname LTb\endcsname%%
      \put(2493,290){\makebox(0,0){\strut{}128 }}%
      \csname LTb\endcsname%%
      \put(2947,290){\makebox(0,0){\strut{}160 }}%
      \csname LTb\endcsname%%
      \put(3400,290){\makebox(0,0){\strut{}200 }}%
      \csname LTb\endcsname%%
      \put(3853,290){\makebox(0,0){\strut{} 240}}%
      \csname LTb\endcsname%%
      \put(4307,290){\makebox(0,0){\strut{}256 }}%
      \csname LTb\endcsname%%
      \put(4760,290){\makebox(0,0){\strut{}280 }}%
      \csname LTb\endcsname%%
      \put(5213,290){\makebox(0,0){\strut{} 320}}%
      \csname LTb\endcsname%%
      \put(5667,290){\makebox(0,0){\strut{} 371}}%
    }%
    \gplgaddtomacro\gplfronttext{%
      \colorrgb{0.00,0.00,1.00}%%
      \put(-134,1052){\rotatebox{-270}{\makebox(0,0){\strut{}$\text{PF}$}}}%
      \csname LTb\endcsname%%
      \put(3400,-40){\makebox(0,0){\strut{}Size S}}%
      \csname LTb\endcsname%%
      \put(5529,1421){\makebox(0,0)[r]{\strut{}$\text{Aggregate phase fraction}$}}%
    }%
    \gplbacktext
    \put(0,0){\includegraphics{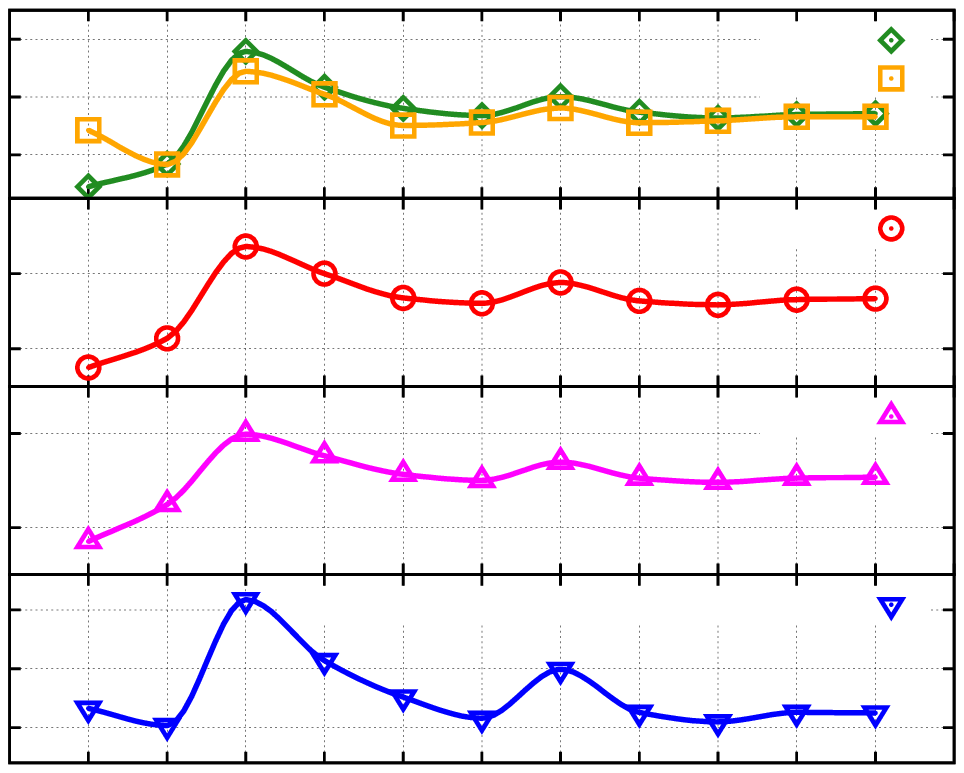}}%
    \gplfronttext
  \end{picture}%
\endgroup
}}
    {\includegraphics[height=5.2cm, angle=0]{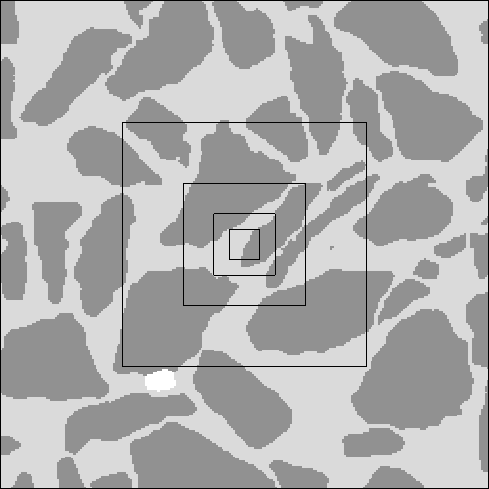}}
	\caption{\textbf{Homogenized 2d elasticity tensor for PBC}. $\mathbb{C}_{ij}$ for different slice sizes. Explicit numbers are given in the Supplement. Selected specimen sizes are S2$^n, n=4,5,6,7,8$.} 
	\label{fig:component-elast-tensor_RAE_size}
\end{Figure}

\begin{Figure}[htbp]
	\centering
     \resizebox{1.1\columnwidth}{!}{% GNUPLOT: LaTeX picture with Postscript
\begingroup
  % Encoding inside the plot.  In the header of your document, this encoding
  % should to defined, e.g., by using
  % \usepackage[cp1252,<other encodings>]{inputenc}
  \inputencoding{cp1252}%
  \makeatletter
  \providecommand\color[2][]{%
    \GenericError{(gnuplot) \space\space\space\@spaces}{%
      Package color not loaded in conjunction with
      terminal option `colourtext'%
    }{See the gnuplot documentation for explanation.%
    }{Either use 'blacktext' in gnuplot or load the package
      color.sty in LaTeX.}%
    \renewcommand\color[2][]{}%
  }%
  \providecommand\includegraphics[2][]{%
    \GenericError{(gnuplot) \space\space\space\@spaces}{%
      Package graphicx or graphics not loaded%
    }{See the gnuplot documentation for explanation.%
    }{The gnuplot epslatex terminal needs graphicx.sty or graphics.sty.}%
    \renewcommand\includegraphics[2][]{}%
  }%
  \providecommand\rotatebox[2]{#2}%
  \@ifundefined{ifGPcolor}{%
    \newif\ifGPcolor
    \GPcolortrue
  }{}%
  \@ifundefined{ifGPblacktext}{%
    \newif\ifGPblacktext
    \GPblacktextfalse
  }{}%
  % define a \g@addto@macro without @ in the name:
  \let\gplgaddtomacro\g@addto@macro
  % define empty templates for all commands taking text:
  \gdef\gplbacktext{}%
  \gdef\gplfronttext{}%
  \makeatother
  \ifGPblacktext
    % no textcolor at all
    \def\colorrgb#1{}%
    \def\colorgray#1{}%
  \else
    % gray or color?
    \ifGPcolor
      \def\colorrgb#1{\color[rgb]{#1}}%
      \def\colorgray#1{\color[gray]{#1}}%
      \expandafter\def\csname LTw\endcsname{\color{white}}%
      \expandafter\def\csname LTb\endcsname{\color{black}}%
      \expandafter\def\csname LTa\endcsname{\color{black}}%
      \expandafter\def\csname LT0\endcsname{\color[rgb]{1,0,0}}%
      \expandafter\def\csname LT1\endcsname{\color[rgb]{0,1,0}}%
      \expandafter\def\csname LT2\endcsname{\color[rgb]{0,0,1}}%
      \expandafter\def\csname LT3\endcsname{\color[rgb]{1,0,1}}%
      \expandafter\def\csname LT4\endcsname{\color[rgb]{0,1,1}}%
      \expandafter\def\csname LT5\endcsname{\color[rgb]{1,1,0}}%
      \expandafter\def\csname LT6\endcsname{\color[rgb]{0,0,0}}%
      \expandafter\def\csname LT7\endcsname{\color[rgb]{1,0.3,0}}%
      \expandafter\def\csname LT8\endcsname{\color[rgb]{0.5,0.5,0.5}}%
    \else
      % gray
      \def\colorrgb#1{\color{black}}%
      \def\colorgray#1{\color[gray]{#1}}%
      \expandafter\def\csname LTw\endcsname{\color{white}}%
      \expandafter\def\csname LTb\endcsname{\color{black}}%
      \expandafter\def\csname LTa\endcsname{\color{black}}%
      \expandafter\def\csname LT0\endcsname{\color{black}}%
      \expandafter\def\csname LT1\endcsname{\color{black}}%
      \expandafter\def\csname LT2\endcsname{\color{black}}%
      \expandafter\def\csname LT3\endcsname{\color{black}}%
      \expandafter\def\csname LT4\endcsname{\color{black}}%
      \expandafter\def\csname LT5\endcsname{\color{black}}%
      \expandafter\def\csname LT6\endcsname{\color{black}}%
      \expandafter\def\csname LT7\endcsname{\color{black}}%
      \expandafter\def\csname LT8\endcsname{\color{black}}%
    \fi
  \fi
    \setlength{\unitlength}{0.0500bp}%
    \ifx\gptboxheight\undefined%
      \newlength{\gptboxheight}%
      \newlength{\gptboxwidth}%
      \newsavebox{\gptboxtext}%
    \fi%
    \setlength{\fboxrule}{0.5pt}%
    \setlength{\fboxsep}{1pt}%
\begin{picture}(15874.00,3684.00)%
    \gplgaddtomacro\gplbacktext{%
      \csname LTb\endcsname%%
      \put(1455,368){\makebox(0,0)[r]{\strut{}$-20$}}%
      \csname LTb\endcsname%%
      \put(1455,759){\makebox(0,0)[r]{\strut{}$-15$}}%
      \csname LTb\endcsname%%
      \put(1455,1151){\makebox(0,0)[r]{\strut{}$-10$}}%
      \csname LTb\endcsname%%
      \put(1455,1542){\makebox(0,0)[r]{\strut{}$-5$}}%
      \csname LTb\endcsname%%
      \put(1455,1933){\makebox(0,0)[r]{\strut{}$0$}}%
      \csname LTb\endcsname%%
      \put(1455,2324){\makebox(0,0)[r]{\strut{}$5$}}%
      \csname LTb\endcsname%%
      \put(1455,2716){\makebox(0,0)[r]{\strut{}$10$}}%
      \csname LTb\endcsname%%
      \put(1455,3107){\makebox(0,0)[r]{\strut{}$15$}}%
      \csname LTb\endcsname%%
      \put(1455,3498){\makebox(0,0)[r]{\strut{}$20$}}%
      \csname LTb\endcsname%%
      \put(1587,148){\makebox(0,0){\strut{}$0$}}%
      \csname LTb\endcsname%%
      \put(1969,148){\makebox(0,0){\strut{}$50$}}%
      \csname LTb\endcsname%%
      \put(2351,148){\makebox(0,0){\strut{}$100$}}%
      \csname LTb\endcsname%%
      \put(2733,148){\makebox(0,0){\strut{}$150$}}%
      \csname LTb\endcsname%%
      \put(3115,148){\makebox(0,0){\strut{}$200$}}%
      \csname LTb\endcsname%%
      \put(3496,148){\makebox(0,0){\strut{}$250$}}%
      \csname LTb\endcsname%%
      \put(3878,148){\makebox(0,0){\strut{}$300$}}%
      \csname LTb\endcsname%%
      \put(4260,148){\makebox(0,0){\strut{}$350$}}%
      \put(4092,681){\makebox(0,0)[l]{\strut{}$\mathbb{C}_{11}$}}%
    }%
    \gplgaddtomacro\gplfronttext{%
      \csname LTb\endcsname%%
      \put(905,1933){\rotatebox{-270}{\makebox(0,0){\strut{}Deviation (in $\%$)}}}%
      \put(3114,-182){\makebox(0,0){\strut{}Size S}}%
      \csname LTb\endcsname%%
      \put(4051,3325){\makebox(0,0)[r]{\strut{}$\text{KUBC}$}}%
      \csname LTb\endcsname%%
      \put(4051,3105){\makebox(0,0)[r]{\strut{}$\text{PBC}$}}%
      \csname LTb\endcsname%%
      \put(4051,2885){\makebox(0,0)[r]{\strut{}$\text{SUBC}$}}%
    }%
    \gplgaddtomacro\gplbacktext{%
      \csname LTb\endcsname%%
      \put(4669,368){\makebox(0,0)[r]{\strut{} }}%
      \csname LTb\endcsname%%
      \put(4669,759){\makebox(0,0)[r]{\strut{} }}%
      \csname LTb\endcsname%%
      \put(4669,1151){\makebox(0,0)[r]{\strut{} }}%
      \csname LTb\endcsname%%
      \put(4669,1542){\makebox(0,0)[r]{\strut{} }}%
      \csname LTb\endcsname%%
      \put(4669,1933){\makebox(0,0)[r]{\strut{} }}%
      \csname LTb\endcsname%%
      \put(4669,2324){\makebox(0,0)[r]{\strut{} }}%
      \csname LTb\endcsname%%
      \put(4669,2716){\makebox(0,0)[r]{\strut{} }}%
      \csname LTb\endcsname%%
      \put(4669,3107){\makebox(0,0)[r]{\strut{} }}%
      \csname LTb\endcsname%%
      \put(5183,148){\makebox(0,0){\strut{}$50$}}%
      \csname LTb\endcsname%%
      \put(5565,148){\makebox(0,0){\strut{}$100$}}%
      \csname LTb\endcsname%%
      \put(5947,148){\makebox(0,0){\strut{}$150$}}%
      \csname LTb\endcsname%%
      \put(6329,148){\makebox(0,0){\strut{}$200$}}%
      \csname LTb\endcsname%%
      \put(6711,148){\makebox(0,0){\strut{}$250$}}%
      \csname LTb\endcsname%%
      \put(7093,148){\makebox(0,0){\strut{}$300$}}%
      \csname LTb\endcsname%%
      \put(7475,148){\makebox(0,0){\strut{}$350$}}%
      \put(7307,681){\makebox(0,0)[l]{\strut{}$\mathbb{C}_{33}$}}%
    }%
    \gplgaddtomacro\gplfronttext{%
      \csname LTb\endcsname%%
      \put(6329,-182){\makebox(0,0){\strut{}Size S}}%
      \csname LTb\endcsname%%
      \put(7266,3325){\makebox(0,0)[r]{\strut{}$\text{KUBC}$}}%
      \csname LTb\endcsname%%
      \put(7266,3105){\makebox(0,0)[r]{\strut{}$\text{PBC}$}}%
      \csname LTb\endcsname%%
      \put(7266,2885){\makebox(0,0)[r]{\strut{}$\text{SUBC}$}}%
    }%
    \gplgaddtomacro\gplbacktext{%
      \csname LTb\endcsname%%
      \put(7884,368){\makebox(0,0)[r]{\strut{} }}%
      \csname LTb\endcsname%%
      \put(7884,759){\makebox(0,0)[r]{\strut{} }}%
      \csname LTb\endcsname%%
      \put(7884,1151){\makebox(0,0)[r]{\strut{} }}%
      \csname LTb\endcsname%%
      \put(7884,1542){\makebox(0,0)[r]{\strut{} }}%
      \csname LTb\endcsname%%
      \put(7884,1933){\makebox(0,0)[r]{\strut{} }}%
      \csname LTb\endcsname%%
      \put(7884,2324){\makebox(0,0)[r]{\strut{} }}%
      \csname LTb\endcsname%%
      \put(7884,2716){\makebox(0,0)[r]{\strut{} }}%
      \csname LTb\endcsname%%
      \put(7884,3107){\makebox(0,0)[r]{\strut{} }}%
      \csname LTb\endcsname%%
      \put(8398,148){\makebox(0,0){\strut{}$50$}}%
      \csname LTb\endcsname%%
      \put(8780,148){\makebox(0,0){\strut{}$100$}}%
      \csname LTb\endcsname%%
      \put(9162,148){\makebox(0,0){\strut{}$150$}}%
      \csname LTb\endcsname%%
      \put(9544,148){\makebox(0,0){\strut{}$200$}}%
      \csname LTb\endcsname%%
      \put(9925,148){\makebox(0,0){\strut{}$250$}}%
      \csname LTb\endcsname%%
      \put(10307,148){\makebox(0,0){\strut{}$300$}}%
      \csname LTb\endcsname%%
      \put(10689,148){\makebox(0,0){\strut{}$350$}}%
      \put(10521,681){\makebox(0,0)[l]{\strut{}$\mathbb{C}_{22}$}}%
    }%
    \gplgaddtomacro\gplfronttext{%
      \csname LTb\endcsname%%
      \put(9543,-182){\makebox(0,0){\strut{}Size S}}%
      \csname LTb\endcsname%%
      \put(10480,3325){\makebox(0,0)[r]{\strut{}$\text{KUBC}$}}%
      \csname LTb\endcsname%%
      \put(10480,3105){\makebox(0,0)[r]{\strut{}$\text{PBC}$}}%
      \csname LTb\endcsname%%
      \put(10480,2885){\makebox(0,0)[r]{\strut{}$\text{SUBC}$}}%
    }%
    \gplgaddtomacro\gplbacktext{%
      \csname LTb\endcsname%%
      \put(11098,368){\makebox(0,0)[r]{\strut{} }}%
      \csname LTb\endcsname%%
      \put(11098,759){\makebox(0,0)[r]{\strut{} }}%
      \csname LTb\endcsname%%
      \put(11098,1151){\makebox(0,0)[r]{\strut{} }}%
      \csname LTb\endcsname%%
      \put(11098,1542){\makebox(0,0)[r]{\strut{} }}%
      \csname LTb\endcsname%%
      \put(11098,1933){\makebox(0,0)[r]{\strut{} }}%
      \csname LTb\endcsname%%
      \put(11098,2324){\makebox(0,0)[r]{\strut{} }}%
      \csname LTb\endcsname%%
      \put(11098,2716){\makebox(0,0)[r]{\strut{} }}%
      \csname LTb\endcsname%%
      \put(11098,3107){\makebox(0,0)[r]{\strut{} }}%
      \csname LTb\endcsname%%
      \put(11612,148){\makebox(0,0){\strut{}$50$}}%
      \csname LTb\endcsname%%
      \put(11994,148){\makebox(0,0){\strut{}$100$}}%
      \csname LTb\endcsname%%
      \put(12376,148){\makebox(0,0){\strut{}$150$}}%
      \csname LTb\endcsname%%
      \put(12758,148){\makebox(0,0){\strut{}$200$}}%
      \csname LTb\endcsname%%
      \put(13139,148){\makebox(0,0){\strut{}$250$}}%
      \csname LTb\endcsname%%
      \put(13521,148){\makebox(0,0){\strut{}$300$}}%
      \csname LTb\endcsname%%
      \put(13903,148){\makebox(0,0){\strut{}$350$}}%
      \put(13094,681){\makebox(0,0)[l]{\strut{}$\mathbb{C}_{12}=\mathbb{C}_{21}$}}%
    }%
    \gplgaddtomacro\gplfronttext{%
      \csname LTb\endcsname%%
      \put(12757,-182){\makebox(0,0){\strut{}Size S}}%
      \csname LTb\endcsname%%
      \put(13694,3325){\makebox(0,0)[r]{\strut{}$\text{KUBC}$}}%
      \csname LTb\endcsname%%
      \put(13694,3105){\makebox(0,0)[r]{\strut{}$\text{PBC}$}}%
      \csname LTb\endcsname%%
      \put(13694,2885){\makebox(0,0)[r]{\strut{}$\text{SUBC}$}}%
    }%
    \gplbacktext
    \put(0,0){\includegraphics{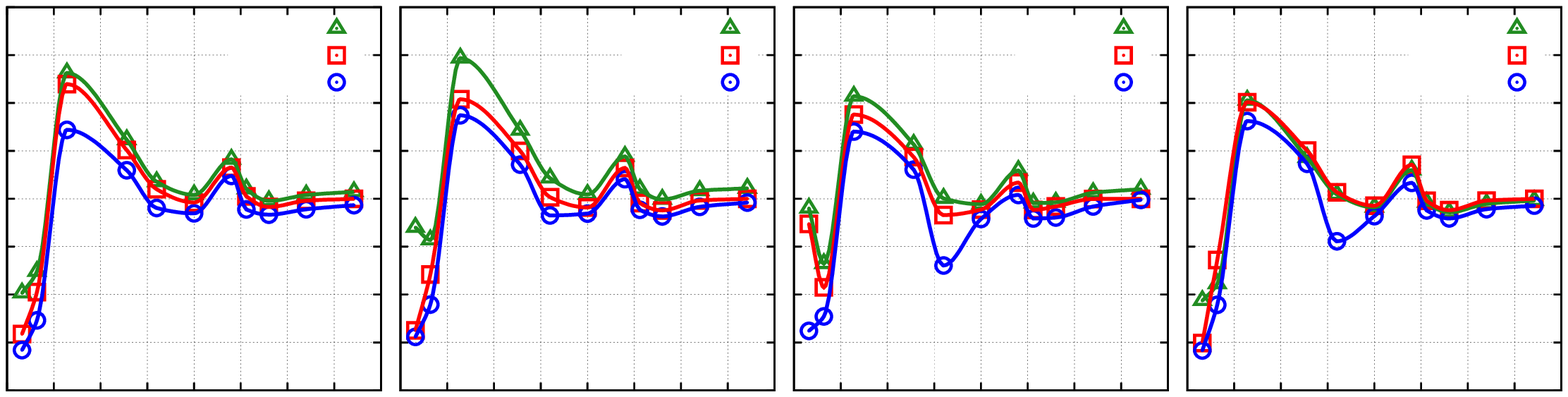}}%
    \gplfronttext
  \end{picture}%
\endgroup
}
     \\[2mm]
	\caption{\textbf{From apparent to effective properties in 2d}. Percentage deviations of  $\mathbb{C}_{ij}$ for KUBC, PBC and SUBC with different slice sizes from $\mathbb{C}_{ij}$(PBC, S371). 
	} 
	\label{fig:Apparent2Effective_RAE_size}
\end{Figure}

Figure~\ref{fig:Apparent2Effective_RAE_size} displays the percentage deviation in $\mathbb{C}_{ij}$ for KUBC, SUBC and PBC, where PBC(S371) is taken as the reference. The convergence characteristics, for SUBC from below, for KUBC from above, for PBC in between, indicate the transition from apparent to effective properties for sufficiently large S following the terminology used since Huet \cite{Huet.1990}. The well-known order\footnote{Inequalities between positive fourth-order tensors are understood in terms of quadratic forms, here in terms of energy densities, i.e. \, ${}^1\mathbb{C} \geq {}^2\mathbb{C} \, \Longleftrightarrow \, \bm \varepsilon :  {}^1\mathbb{C} \, \bm \varepsilon \geq \bm \varepsilon :  {}^2\mathbb{C} \, \bm \varepsilon \enskip \forall \, \bm \varepsilon \, .$} $\mathbb{C}$(KUBC) $\geq$ $\mathbb{C}$(PBC) $\geq$ $\mathbb{C}$(SUBC) is observed for all sizes in all components. The characteristics are not monotonic and the deviations do not decrease in a monotonic way, but show rapid convergence, with an outlier for S240; for size S200 and above the deviations stay below 2\%.

A setup, where a sample size automatically implies a size-dependent resolution and finite element discretization with corresponding errors, is not satisfactory and not necessary. In contrast, the finite element discretization $h$ can be chosen smaller than the pixel size $h_{\square}$ in either case, the resolution in image acquisition nowadays can be chosen almost at an arbitrary quality. For this reason Sec.~\ref{subsubsec:2d-resolution} separately analyzes the influence of resolution, and Sec.~\ref{subsubsec:2d-discretization} the influence of discretization. 
 
\subsubsection{Dependence on resolution}
\label{subsubsec:2d-resolution}

\begin{figure}[htbp]
	\centering
	\subfloat[S320-RD160 (0.2mm)]
	{\includegraphics[height=4.0cm, angle=0]{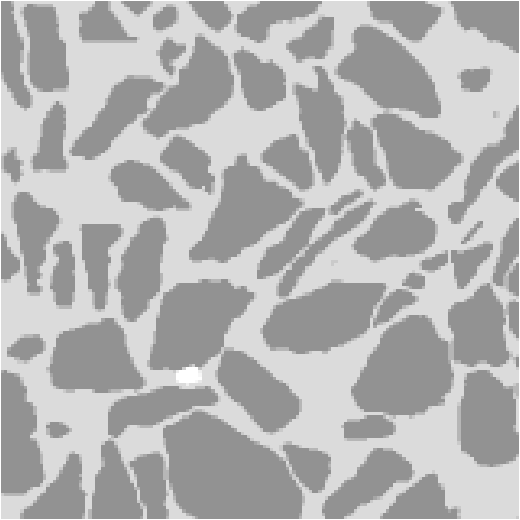}}
     \hspace*{0.03\linewidth}	
	\subfloat[S320-RD40 (0.8mm)]
	{\includegraphics[height=4.0cm, angle=0]{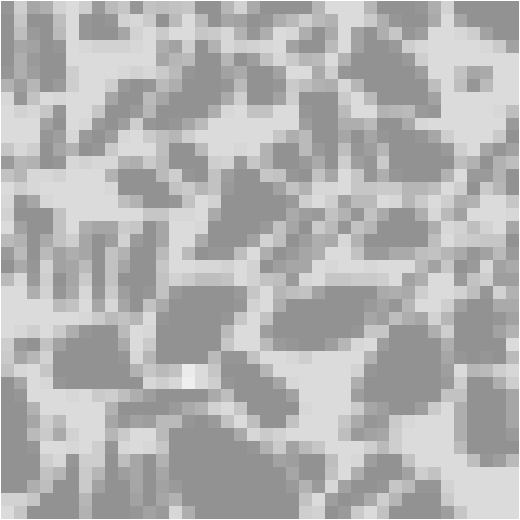}}
    \hspace*{0.03\linewidth}	
	\subfloat[S320-RD40 (0.8mm)]
	{\includegraphics[height=4.0cm, angle=0]{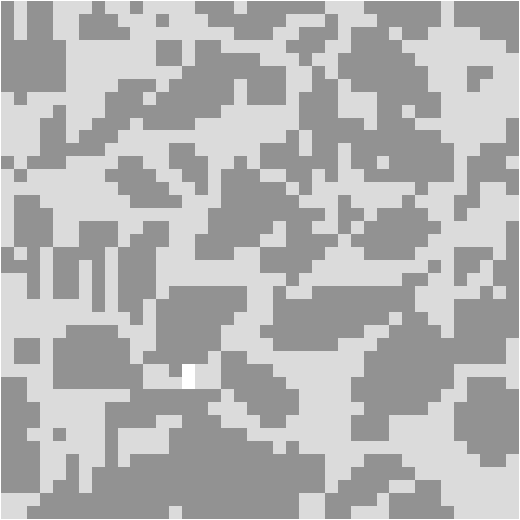}}
	\caption{\textbf{Slice of S320 in different resolutions}. For specimen SR320 three coarsening steps of resolution R (pixels per edge) are done, in (b) for phase averaging, in (c) for phase preserving coarsening. Pixel size in brackets.} 
	\label{fig:Concrete_resolution}
\end{figure}

Figure~\ref{fig:Concrete_resolution} shows the sample size S320 in different resolutions, where two types of resolution coarsening according to Sec.~\ref{subsec:ResolutionCoarsening} are carried out. The first type applies a mixture rule as shown in Fig.~\ref{fig:Concrete_resolution}(b). The second type follows the rule the-majority-wins, for the outcome see Fig.~\ref{fig:Concrete_resolution}(c). 

\begin{Figure}[htbp]
	\centering
	\subfloat[New phases]
     {\resizebox{0.44\columnwidth}{!}{% GNUPLOT: LaTeX picture with Postscript
\begingroup
  % Encoding inside the plot.  In the header of your document, this encoding
  % should to defined, e.g., by using
  % \usepackage[cp1252,<other encodings>]{inputenc}
  \inputencoding{cp1252}%
  \makeatletter
  \providecommand\color[2][]{%
    \GenericError{(gnuplot) \space\space\space\@spaces}{%
      Package color not loaded in conjunction with
      terminal option `colourtext'%
    }{See the gnuplot documentation for explanation.%
    }{Either use 'blacktext' in gnuplot or load the package
      color.sty in LaTeX.}%
    \renewcommand\color[2][]{}%
  }%
  \providecommand\includegraphics[2][]{%
    \GenericError{(gnuplot) \space\space\space\@spaces}{%
      Package graphicx or graphics not loaded%
    }{See the gnuplot documentation for explanation.%
    }{The gnuplot epslatex terminal needs graphicx.sty or graphics.sty.}%
    \renewcommand\includegraphics[2][]{}%
  }%
  \providecommand\rotatebox[2]{#2}%
  \@ifundefined{ifGPcolor}{%
    \newif\ifGPcolor
    \GPcolortrue
  }{}%
  \@ifundefined{ifGPblacktext}{%
    \newif\ifGPblacktext
    \GPblacktextfalse
  }{}%
  % define a \g@addto@macro without @ in the name:
  \let\gplgaddtomacro\g@addto@macro
  % define empty templates for all commands taking text:
  \gdef\gplbacktext{}%
  \gdef\gplfronttext{}%
  \makeatother
  \ifGPblacktext
    % no textcolor at all
    \def\colorrgb#1{}%
    \def\colorgray#1{}%
  \else
    % gray or color?
    \ifGPcolor
      \def\colorrgb#1{\color[rgb]{#1}}%
      \def\colorgray#1{\color[gray]{#1}}%
      \expandafter\def\csname LTw\endcsname{\color{white}}%
      \expandafter\def\csname LTb\endcsname{\color{black}}%
      \expandafter\def\csname LTa\endcsname{\color{black}}%
      \expandafter\def\csname LT0\endcsname{\color[rgb]{1,0,0}}%
      \expandafter\def\csname LT1\endcsname{\color[rgb]{0,1,0}}%
      \expandafter\def\csname LT2\endcsname{\color[rgb]{0,0,1}}%
      \expandafter\def\csname LT3\endcsname{\color[rgb]{1,0,1}}%
      \expandafter\def\csname LT4\endcsname{\color[rgb]{0,1,1}}%
      \expandafter\def\csname LT5\endcsname{\color[rgb]{1,1,0}}%
      \expandafter\def\csname LT6\endcsname{\color[rgb]{0,0,0}}%
      \expandafter\def\csname LT7\endcsname{\color[rgb]{1,0.3,0}}%
      \expandafter\def\csname LT8\endcsname{\color[rgb]{0.5,0.5,0.5}}%
    \else
      % gray
      \def\colorrgb#1{\color{black}}%
      \def\colorgray#1{\color[gray]{#1}}%
      \expandafter\def\csname LTw\endcsname{\color{white}}%
      \expandafter\def\csname LTb\endcsname{\color{black}}%
      \expandafter\def\csname LTa\endcsname{\color{black}}%
      \expandafter\def\csname LT0\endcsname{\color{black}}%
      \expandafter\def\csname LT1\endcsname{\color{black}}%
      \expandafter\def\csname LT2\endcsname{\color{black}}%
      \expandafter\def\csname LT3\endcsname{\color{black}}%
      \expandafter\def\csname LT4\endcsname{\color{black}}%
      \expandafter\def\csname LT5\endcsname{\color{black}}%
      \expandafter\def\csname LT6\endcsname{\color{black}}%
      \expandafter\def\csname LT7\endcsname{\color{black}}%
      \expandafter\def\csname LT8\endcsname{\color{black}}%
    \fi
  \fi
    \setlength{\unitlength}{0.0500bp}%
    \ifx\gptboxheight\undefined%
      \newlength{\gptboxheight}%
      \newlength{\gptboxwidth}%
      \newsavebox{\gptboxtext}%
    \fi%
    \setlength{\fboxrule}{0.5pt}%
    \setlength{\fboxsep}{1pt}%
\begin{picture}(6802.00,5102.00)%
    \gplgaddtomacro\gplbacktext{%
      \csname LTb\endcsname%%
      \put(548,3979){\makebox(0,0)[r]{\strut{}$35$}}%
      \csname LTb\endcsname%%
      \put(548,4195){\makebox(0,0)[r]{\strut{}$36$}}%
      \csname LTb\endcsname%%
      \put(548,4412){\makebox(0,0)[r]{\strut{}$37$}}%
      \csname LTb\endcsname%%
      \put(548,4628){\makebox(0,0)[r]{\strut{}$38$}}%
      \csname LTb\endcsname%%
      \put(548,4845){\makebox(0,0)[r]{\strut{}$39$}}%
      \csname LTb\endcsname%%
      \put(1768,3542){\makebox(0,0){\strut{} }}%
      \csname LTb\endcsname%%
      \put(2856,3542){\makebox(0,0){\strut{} }}%
      \csname LTb\endcsname%%
      \put(3944,3542){\makebox(0,0){\strut{} }}%
      \csname LTb\endcsname%%
      \put(5032,3542){\makebox(0,0){\strut{} }}%
    }%
    \gplgaddtomacro\gplfronttext{%
      \csname LTb\endcsname%%
      \put(5529,4672){\makebox(0,0)[r]{\strut{}$\mathbb{C}_{11}$}}%
      \csname LTb\endcsname%%
      \put(5529,4452){\makebox(0,0)[r]{\strut{}$\mathbb{C}_{22}$}}%
    }%
    \gplgaddtomacro\gplbacktext{%
      \csname LTb\endcsname%%
      \put(548,2895){\makebox(0,0)[r]{\strut{}$13.3$}}%
      \csname LTb\endcsname%%
      \put(548,3183){\makebox(0,0)[r]{\strut{}$13.7$}}%
      \csname LTb\endcsname%%
      \put(548,3472){\makebox(0,0)[r]{\strut{}$14.1$}}%
      \csname LTb\endcsname%%
      \put(1768,2458){\makebox(0,0){\strut{} }}%
      \csname LTb\endcsname%%
      \put(2856,2458){\makebox(0,0){\strut{} }}%
      \csname LTb\endcsname%%
      \put(3944,2458){\makebox(0,0){\strut{} }}%
      \csname LTb\endcsname%%
      \put(5032,2458){\makebox(0,0){\strut{} }}%
    }%
    \gplgaddtomacro\gplfronttext{%
      \csname LTb\endcsname%%
      \put(-134,3219){\rotatebox{-270}{\makebox(0,0){\strut{}$\mathbb{C}_{ij}$ (in GPa)}}}%
      \csname LTb\endcsname%%
      \put(5529,3588){\makebox(0,0)[r]{\strut{}$\mathbb{C}_{33}$}}%
    }%
    \gplgaddtomacro\gplbacktext{%
      \csname LTb\endcsname%%
      \put(548,1775){\makebox(0,0)[r]{\strut{}$9.1$}}%
      \csname LTb\endcsname%%
      \put(548,2136){\makebox(0,0)[r]{\strut{}$9.3$}}%
      \csname LTb\endcsname%%
      \put(548,2497){\makebox(0,0)[r]{\strut{}$9.5$}}%
      \csname LTb\endcsname%%
      \put(1768,1374){\makebox(0,0){\strut{} }}%
      \csname LTb\endcsname%%
      \put(2856,1374){\makebox(0,0){\strut{} }}%
      \csname LTb\endcsname%%
      \put(3944,1374){\makebox(0,0){\strut{} }}%
      \csname LTb\endcsname%%
      \put(5032,1374){\makebox(0,0){\strut{} }}%
    }%
    \gplgaddtomacro\gplfronttext{%
      \csname LTb\endcsname%%
      \put(5529,2505){\makebox(0,0)[r]{\strut{}$\mathbb{C}_{12}$}}%
    }%
    \gplgaddtomacro\gplbacktext{%
      \colorrgb{0.00,0.00,1.00}%%
      \put(548,781){\makebox(0,0)[r]{\strut{}$0.55$}}%
      \colorrgb{0.00,0.00,1.00}%%
      \put(548,1323){\makebox(0,0)[r]{\strut{}$0.57$}}%
      \csname LTb\endcsname%%
      \put(1768,290){\makebox(0,0){\strut{}320}}%
      \csname LTb\endcsname%%
      \put(2856,290){\makebox(0,0){\strut{}160}}%
      \csname LTb\endcsname%%
      \put(3944,290){\makebox(0,0){\strut{}80}}%
      \csname LTb\endcsname%%
      \put(5032,290){\makebox(0,0){\strut{}40}}%
    }%
    \gplgaddtomacro\gplfronttext{%
      \colorrgb{0.00,0.00,1.00}%%
      \put(-134,1052){\rotatebox{-270}{\makebox(0,0){\strut{}$\text{PF}$}}}%
      \csname LTb\endcsname%%
      \put(3400,-40){\makebox(0,0){\strut{}Resolution R}}%
      \csname LTb\endcsname%%
      \put(5529,1421){\makebox(0,0)[r]{\strut{}$\text{Aggregate phase fraction}$}}%
    }%
    \gplbacktext
    \put(0,0){\includegraphics{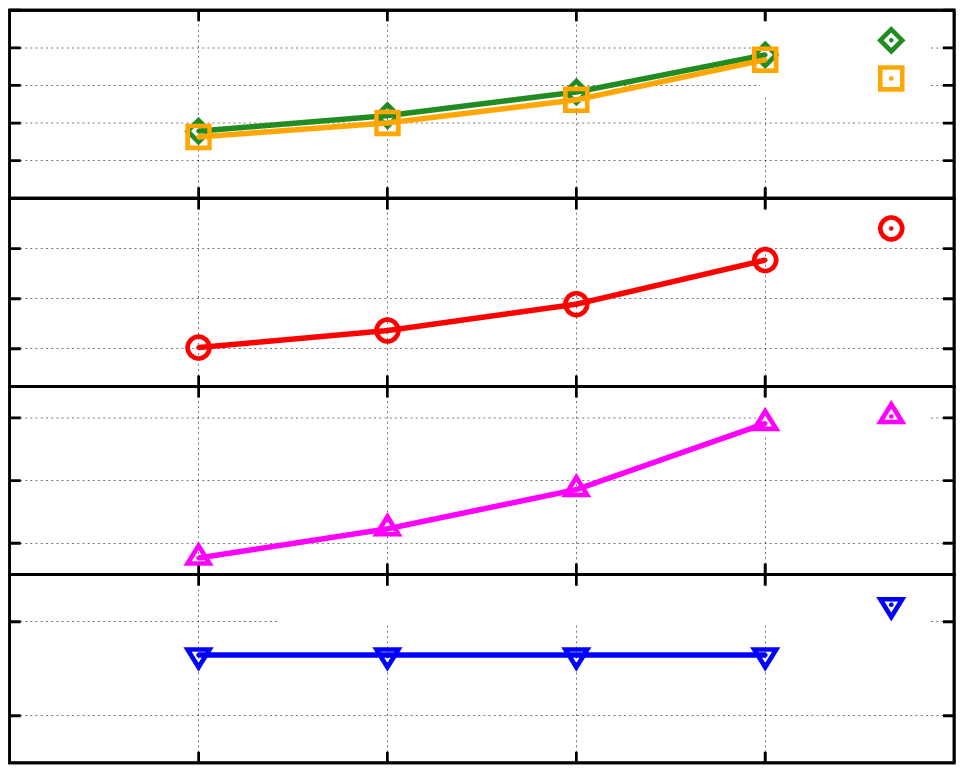}}%
    \gplfronttext
  \end{picture}%
\endgroup
}}
     \hspace*{4mm}
     \subfloat[Phase preserving]
     {\resizebox{0.44\columnwidth}{!}{% GNUPLOT: LaTeX picture with Postscript
\begingroup
  % Encoding inside the plot.  In the header of your document, this encoding
  % should to defined, e.g., by using
  % \usepackage[cp1252,<other encodings>]{inputenc}
  \inputencoding{cp1252}%
  \makeatletter
  \providecommand\color[2][]{%
    \GenericError{(gnuplot) \space\space\space\@spaces}{%
      Package color not loaded in conjunction with
      terminal option `colourtext'%
    }{See the gnuplot documentation for explanation.%
    }{Either use 'blacktext' in gnuplot or load the package
      color.sty in LaTeX.}%
    \renewcommand\color[2][]{}%
  }%
  \providecommand\includegraphics[2][]{%
    \GenericError{(gnuplot) \space\space\space\@spaces}{%
      Package graphicx or graphics not loaded%
    }{See the gnuplot documentation for explanation.%
    }{The gnuplot epslatex terminal needs graphicx.sty or graphics.sty.}%
    \renewcommand\includegraphics[2][]{}%
  }%
  \providecommand\rotatebox[2]{#2}%
  \@ifundefined{ifGPcolor}{%
    \newif\ifGPcolor
    \GPcolortrue
  }{}%
  \@ifundefined{ifGPblacktext}{%
    \newif\ifGPblacktext
    \GPblacktextfalse
  }{}%
  % define a \g@addto@macro without @ in the name:
  \let\gplgaddtomacro\g@addto@macro
  % define empty templates for all commands taking text:
  \gdef\gplbacktext{}%
  \gdef\gplfronttext{}%
  \makeatother
  \ifGPblacktext
    % no textcolor at all
    \def\colorrgb#1{}%
    \def\colorgray#1{}%
  \else
    % gray or color?
    \ifGPcolor
      \def\colorrgb#1{\color[rgb]{#1}}%
      \def\colorgray#1{\color[gray]{#1}}%
      \expandafter\def\csname LTw\endcsname{\color{white}}%
      \expandafter\def\csname LTb\endcsname{\color{black}}%
      \expandafter\def\csname LTa\endcsname{\color{black}}%
      \expandafter\def\csname LT0\endcsname{\color[rgb]{1,0,0}}%
      \expandafter\def\csname LT1\endcsname{\color[rgb]{0,1,0}}%
      \expandafter\def\csname LT2\endcsname{\color[rgb]{0,0,1}}%
      \expandafter\def\csname LT3\endcsname{\color[rgb]{1,0,1}}%
      \expandafter\def\csname LT4\endcsname{\color[rgb]{0,1,1}}%
      \expandafter\def\csname LT5\endcsname{\color[rgb]{1,1,0}}%
      \expandafter\def\csname LT6\endcsname{\color[rgb]{0,0,0}}%
      \expandafter\def\csname LT7\endcsname{\color[rgb]{1,0.3,0}}%
      \expandafter\def\csname LT8\endcsname{\color[rgb]{0.5,0.5,0.5}}%
    \else
      % gray
      \def\colorrgb#1{\color{black}}%
      \def\colorgray#1{\color[gray]{#1}}%
      \expandafter\def\csname LTw\endcsname{\color{white}}%
      \expandafter\def\csname LTb\endcsname{\color{black}}%
      \expandafter\def\csname LTa\endcsname{\color{black}}%
      \expandafter\def\csname LT0\endcsname{\color{black}}%
      \expandafter\def\csname LT1\endcsname{\color{black}}%
      \expandafter\def\csname LT2\endcsname{\color{black}}%
      \expandafter\def\csname LT3\endcsname{\color{black}}%
      \expandafter\def\csname LT4\endcsname{\color{black}}%
      \expandafter\def\csname LT5\endcsname{\color{black}}%
      \expandafter\def\csname LT6\endcsname{\color{black}}%
      \expandafter\def\csname LT7\endcsname{\color{black}}%
      \expandafter\def\csname LT8\endcsname{\color{black}}%
    \fi
  \fi
    \setlength{\unitlength}{0.0500bp}%
    \ifx\gptboxheight\undefined%
      \newlength{\gptboxheight}%
      \newlength{\gptboxwidth}%
      \newsavebox{\gptboxtext}%
    \fi%
    \setlength{\fboxrule}{0.5pt}%
    \setlength{\fboxsep}{1pt}%
\begin{picture}(6802.00,5102.00)%
    \gplgaddtomacro\gplbacktext{%
      \csname LTb\endcsname%%
      \put(548,4033){\makebox(0,0)[r]{\strut{}$35.5$}}%
      \csname LTb\endcsname%%
      \put(548,4304){\makebox(0,0)[r]{\strut{}$36$}}%
      \csname LTb\endcsname%%
      \put(548,4574){\makebox(0,0)[r]{\strut{}$36.5$}}%
      \csname LTb\endcsname%%
      \put(1768,3542){\makebox(0,0){\strut{} }}%
      \csname LTb\endcsname%%
      \put(2856,3542){\makebox(0,0){\strut{} }}%
      \csname LTb\endcsname%%
      \put(3944,3542){\makebox(0,0){\strut{} }}%
      \csname LTb\endcsname%%
      \put(5032,3542){\makebox(0,0){\strut{} }}%
    }%
    \gplgaddtomacro\gplfronttext{%
      \csname LTb\endcsname%%
      \put(5529,4672){\makebox(0,0)[r]{\strut{}$\mathbb{C}_{11}$}}%
      \csname LTb\endcsname%%
      \put(5529,4452){\makebox(0,0)[r]{\strut{}$\mathbb{C}_{22}$}}%
    }%
    \gplgaddtomacro\gplbacktext{%
      \csname LTb\endcsname%%
      \put(548,2949){\makebox(0,0)[r]{\strut{}$13.2$}}%
      \csname LTb\endcsname%%
      \put(548,3220){\makebox(0,0)[r]{\strut{}$13.4$}}%
      \csname LTb\endcsname%%
      \put(548,3490){\makebox(0,0)[r]{\strut{}$13.6$}}%
      \csname LTb\endcsname%%
      \put(1768,2458){\makebox(0,0){\strut{} }}%
      \csname LTb\endcsname%%
      \put(2856,2458){\makebox(0,0){\strut{} }}%
      \csname LTb\endcsname%%
      \put(3944,2458){\makebox(0,0){\strut{} }}%
      \csname LTb\endcsname%%
      \put(5032,2458){\makebox(0,0){\strut{} }}%
    }%
    \gplgaddtomacro\gplfronttext{%
      \csname LTb\endcsname%%
      \put(-134,3219){\rotatebox{-270}{\makebox(0,0){\strut{}$\mathbb{C}_{ij}$ (in GPa)}}}%
      \csname LTb\endcsname%%
      \put(5529,3588){\makebox(0,0)[r]{\strut{}$\mathbb{C}_{33}$}}%
    }%
    \gplgaddtomacro\gplbacktext{%
      \csname LTb\endcsname%%
      \put(548,1865){\makebox(0,0)[r]{\strut{}$8.9$}}%
      \csname LTb\endcsname%%
      \put(548,2136){\makebox(0,0)[r]{\strut{}$9$}}%
      \csname LTb\endcsname%%
      \put(548,2407){\makebox(0,0)[r]{\strut{}$9.1$}}%
      \csname LTb\endcsname%%
      \put(1768,1374){\makebox(0,0){\strut{} }}%
      \csname LTb\endcsname%%
      \put(2856,1374){\makebox(0,0){\strut{} }}%
      \csname LTb\endcsname%%
      \put(3944,1374){\makebox(0,0){\strut{} }}%
      \csname LTb\endcsname%%
      \put(5032,1374){\makebox(0,0){\strut{} }}%
    }%
    \gplgaddtomacro\gplfronttext{%
      \csname LTb\endcsname%%
      \put(5529,2505){\makebox(0,0)[r]{\strut{}$\mathbb{C}_{12}$}}%
    }%
    \gplgaddtomacro\gplbacktext{%
      \colorrgb{0.00,0.00,1.00}%%
      \put(548,781){\makebox(0,0)[r]{\strut{}$0.55$}}%
      \colorrgb{0.00,0.00,1.00}%%
      \put(548,1052){\makebox(0,0)[r]{\strut{}$0.56$}}%
      \colorrgb{0.00,0.00,1.00}%%
      \put(548,1323){\makebox(0,0)[r]{\strut{}$0.57$}}%
      \csname LTb\endcsname%%
      \put(1768,290){\makebox(0,0){\strut{}320}}%
      \csname LTb\endcsname%%
      \put(2856,290){\makebox(0,0){\strut{}160}}%
      \csname LTb\endcsname%%
      \put(3944,290){\makebox(0,0){\strut{}80}}%
      \csname LTb\endcsname%%
      \put(5032,290){\makebox(0,0){\strut{}40}}%
    }%
    \gplgaddtomacro\gplfronttext{%
      \colorrgb{0.00,0.00,1.00}%%
      \put(-134,1052){\rotatebox{-270}{\makebox(0,0){\strut{}$\text{PF}$}}}%
      \csname LTb\endcsname%%
      \put(3400,-40){\makebox(0,0){\strut{}Resolution R}}%
      \csname LTb\endcsname%%
      \put(5529,1421){\makebox(0,0)[r]{\strut{}$\text{Aggregate phase fraction}$}}%
    }%
    \gplbacktext
    \put(0,0){\includegraphics{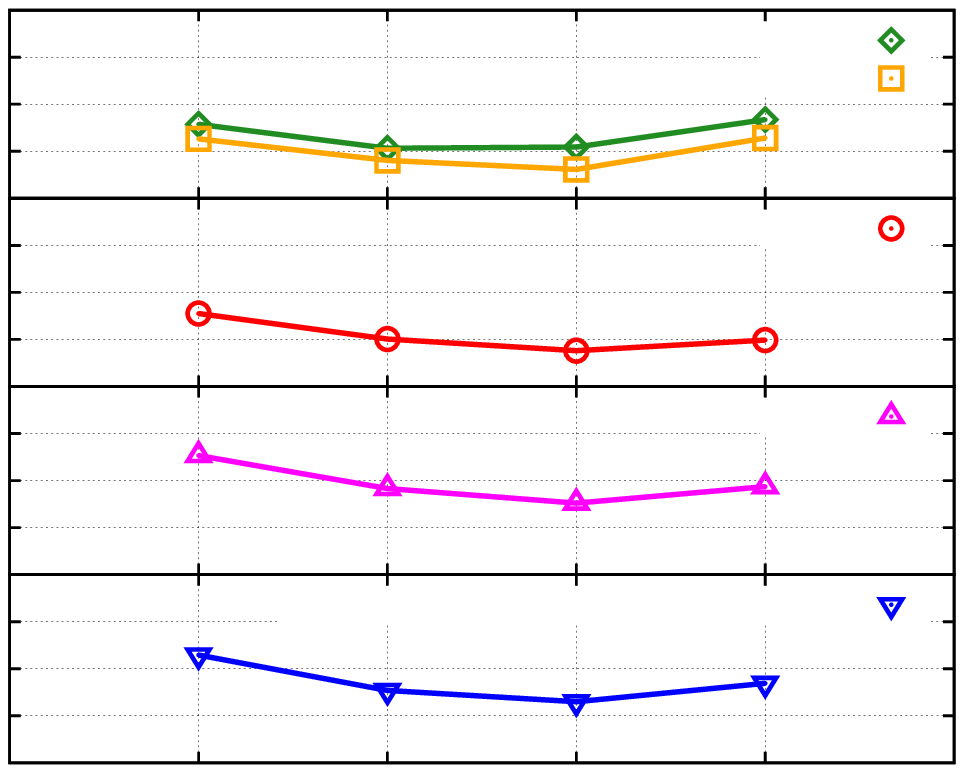}}%
    \gplfronttext
  \end{picture}%
\endgroup
}}
	\caption{\textbf{Resolution-dependent 2d homogenized elasticity tensor}. Components $\mathbb{C}_{ij}$ and phase fraction (PF) of the stiff aggregate phase for size S320 at different resolutions R but each with the same discretization D320, hence S320-R\,$i$-D320, (a) with new phases at interfaces, (b) for the case preserving the number of phases. Numbers are given in the Supplement.} 
	\label{fig:component-elast-tensor_resolution}
\end{Figure} 

Image coarsening introduces a resolution error and, if the finite element discretization follows the pixel grid, additionally an increase in the discretization error will be induced. For a novel concept that decomposes micro errors into a resolution part and a standard discretization part we refer to \cite{Eidel.2020}. Here, however, the discretization D320 is kept constant during the coarsening procedure which has the advantage of keeping the discretization error constant and thereby isolates the effect of resolution coarsening. 

Figure~\ref{fig:component-elast-tensor_resolution} displays the corresponding results for the homogenized elasticity tensor. In Fig.~\ref{fig:component-elast-tensor_resolution}(a) for the case of averaging phase properties it is obvious that although the volume average of the Young's modulus $<E>$ remains constant (here in terms of the aggregate phase fraction ''on average'') in coarsening, the homogenized elastic constants show larger changes compared to the phase-preserving case of Fig.~\ref{fig:component-elast-tensor_resolution}(b). In the latter case, the components of the homogenized elasticity tensor follow quantitatively the stiffest phase fraction. Helpful in the preservation of the phase fractions is that in case of an equal count of pixels of different properties being merged, the attributed property to the novel voxel shifts the global phase fraction closer to the original one.  
  
\subsubsection{Dependence on discretization}
\label{subsubsec:2d-discretization} 

Figure~\ref{fig:component-elast-tensor_discretization} displays for the same sample size the influence of discretization by subdividing each pixel into $2^2$, $4^2$, $6^2$ square sub-elements. Explicit numbers for that case are listed in the Supplement. The influence is rather minor, most notably in comparison with the influence of the image resolution shown in Fig.~\ref{fig:component-elast-tensor_resolution}.

\begin{Figure}[htbp]
	\centering
     \resizebox{0.45\columnwidth}{!}{% GNUPLOT: LaTeX picture with Postscript
\begingroup
  % Encoding inside the plot.  In the header of your document, this encoding
  % should to defined, e.g., by using
  % \usepackage[cp1252,<other encodings>]{inputenc}
  \inputencoding{cp1252}%
  \makeatletter
  \providecommand\color[2][]{%
    \GenericError{(gnuplot) \space\space\space\@spaces}{%
      Package color not loaded in conjunction with
      terminal option `colourtext'%
    }{See the gnuplot documentation for explanation.%
    }{Either use 'blacktext' in gnuplot or load the package
      color.sty in LaTeX.}%
    \renewcommand\color[2][]{}%
  }%
  \providecommand\includegraphics[2][]{%
    \GenericError{(gnuplot) \space\space\space\@spaces}{%
      Package graphicx or graphics not loaded%
    }{See the gnuplot documentation for explanation.%
    }{The gnuplot epslatex terminal needs graphicx.sty or graphics.sty.}%
    \renewcommand\includegraphics[2][]{}%
  }%
  \providecommand\rotatebox[2]{#2}%
  \@ifundefined{ifGPcolor}{%
    \newif\ifGPcolor
    \GPcolortrue
  }{}%
  \@ifundefined{ifGPblacktext}{%
    \newif\ifGPblacktext
    \GPblacktextfalse
  }{}%
  % define a \g@addto@macro without @ in the name:
  \let\gplgaddtomacro\g@addto@macro
  % define empty templates for all commands taking text:
  \gdef\gplbacktext{}%
  \gdef\gplfronttext{}%
  \makeatother
  \ifGPblacktext
    % no textcolor at all
    \def\colorrgb#1{}%
    \def\colorgray#1{}%
  \else
    % gray or color?
    \ifGPcolor
      \def\colorrgb#1{\color[rgb]{#1}}%
      \def\colorgray#1{\color[gray]{#1}}%
      \expandafter\def\csname LTw\endcsname{\color{white}}%
      \expandafter\def\csname LTb\endcsname{\color{black}}%
      \expandafter\def\csname LTa\endcsname{\color{black}}%
      \expandafter\def\csname LT0\endcsname{\color[rgb]{1,0,0}}%
      \expandafter\def\csname LT1\endcsname{\color[rgb]{0,1,0}}%
      \expandafter\def\csname LT2\endcsname{\color[rgb]{0,0,1}}%
      \expandafter\def\csname LT3\endcsname{\color[rgb]{1,0,1}}%
      \expandafter\def\csname LT4\endcsname{\color[rgb]{0,1,1}}%
      \expandafter\def\csname LT5\endcsname{\color[rgb]{1,1,0}}%
      \expandafter\def\csname LT6\endcsname{\color[rgb]{0,0,0}}%
      \expandafter\def\csname LT7\endcsname{\color[rgb]{1,0.3,0}}%
      \expandafter\def\csname LT8\endcsname{\color[rgb]{0.5,0.5,0.5}}%
    \else
      % gray
      \def\colorrgb#1{\color{black}}%
      \def\colorgray#1{\color[gray]{#1}}%
      \expandafter\def\csname LTw\endcsname{\color{white}}%
      \expandafter\def\csname LTb\endcsname{\color{black}}%
      \expandafter\def\csname LTa\endcsname{\color{black}}%
      \expandafter\def\csname LT0\endcsname{\color{black}}%
      \expandafter\def\csname LT1\endcsname{\color{black}}%
      \expandafter\def\csname LT2\endcsname{\color{black}}%
      \expandafter\def\csname LT3\endcsname{\color{black}}%
      \expandafter\def\csname LT4\endcsname{\color{black}}%
      \expandafter\def\csname LT5\endcsname{\color{black}}%
      \expandafter\def\csname LT6\endcsname{\color{black}}%
      \expandafter\def\csname LT7\endcsname{\color{black}}%
      \expandafter\def\csname LT8\endcsname{\color{black}}%
    \fi
  \fi
    \setlength{\unitlength}{0.0500bp}%
    \ifx\gptboxheight\undefined%
      \newlength{\gptboxheight}%
      \newlength{\gptboxwidth}%
      \newsavebox{\gptboxtext}%
    \fi%
    \setlength{\fboxrule}{0.5pt}%
    \setlength{\fboxsep}{1pt}%
\begin{picture}(6802.00,5102.00)%
    \gplgaddtomacro\gplbacktext{%
      \csname LTb\endcsname%%
      \put(548,3762){\makebox(0,0)[r]{\strut{}$35.6$}}%
      \csname LTb\endcsname%%
      \put(548,4123){\makebox(0,0)[r]{\strut{}$35.7$}}%
      \csname LTb\endcsname%%
      \put(548,4484){\makebox(0,0)[r]{\strut{}$35.8$}}%
      \csname LTb\endcsname%%
      \put(1768,3181){\makebox(0,0){\strut{} }}%
      \csname LTb\endcsname%%
      \put(2856,3181){\makebox(0,0){\strut{} }}%
      \csname LTb\endcsname%%
      \put(3944,3181){\makebox(0,0){\strut{} }}%
      \csname LTb\endcsname%%
      \put(5032,3181){\makebox(0,0){\strut{} }}%
    }%
    \gplgaddtomacro\gplfronttext{%
      \csname LTb\endcsname%%
      \put(5529,4672){\makebox(0,0)[r]{\strut{}$\mathbb{C}_{11}$}}%
      \csname LTb\endcsname%%
      \put(5529,4452){\makebox(0,0)[r]{\strut{}$\mathbb{C}_{22}$}}%
    }%
    \gplgaddtomacro\gplbacktext{%
      \csname LTb\endcsname%%
      \put(548,2316){\makebox(0,0)[r]{\strut{}$13.25$}}%
      \csname LTb\endcsname%%
      \put(548,2678){\makebox(0,0)[r]{\strut{}$13.3$}}%
      \csname LTb\endcsname%%
      \put(548,3039){\makebox(0,0)[r]{\strut{}$13.35$}}%
      \csname LTb\endcsname%%
      \put(1768,1735){\makebox(0,0){\strut{} }}%
      \csname LTb\endcsname%%
      \put(2856,1735){\makebox(0,0){\strut{} }}%
      \csname LTb\endcsname%%
      \put(3944,1735){\makebox(0,0){\strut{} }}%
      \csname LTb\endcsname%%
      \put(5032,1735){\makebox(0,0){\strut{} }}%
    }%
    \gplgaddtomacro\gplfronttext{%
      \csname LTb\endcsname%%
      \put(-266,2677){\rotatebox{-270}{\makebox(0,0){\strut{}$\mathbb{C}_{ij}$ (in GPa)}}}%
      \csname LTb\endcsname%%
      \put(5529,3227){\makebox(0,0)[r]{\strut{}$\mathbb{C}_{33}$}}%
    }%
    \gplgaddtomacro\gplbacktext{%
      \csname LTb\endcsname%%
      \put(548,992){\makebox(0,0)[r]{\strut{}$9.055$}}%
      \csname LTb\endcsname%%
      \put(548,1473){\makebox(0,0)[r]{\strut{}$9.06$}}%
      \csname LTb\endcsname%%
      \put(1768,290){\makebox(0,0){\strut{}320}}%
      \csname LTb\endcsname%%
      \put(2856,290){\makebox(0,0){\strut{}640}}%
      \csname LTb\endcsname%%
      \put(3944,290){\makebox(0,0){\strut{}1280}}%
      \csname LTb\endcsname%%
      \put(5032,290){\makebox(0,0){\strut{}1920}}%
    }%
    \gplgaddtomacro\gplfronttext{%
      \csname LTb\endcsname%%
      \put(3400,-40){\makebox(0,0){\strut{}Discretization D}}%
      \csname LTb\endcsname%%
      \put(5529,1782){\makebox(0,0)[r]{\strut{}$\mathbb{C}_{12}$}}%
    }%
    \gplbacktext
    \put(0,0){\includegraphics{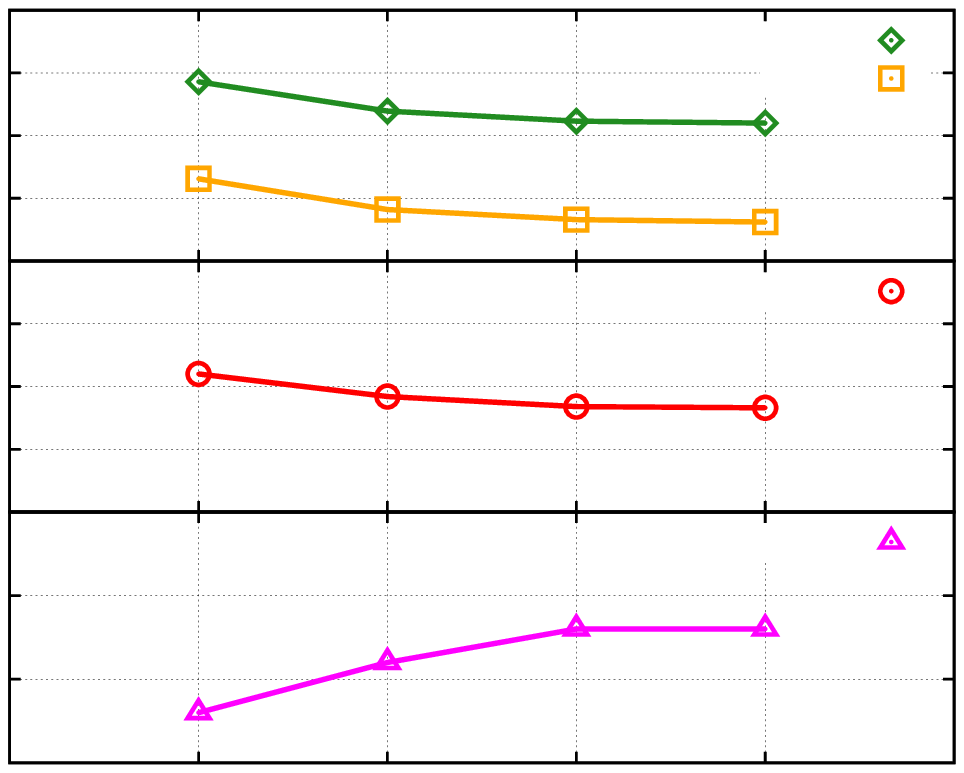}}%
    \gplfronttext
  \end{picture}%
\endgroup
}
	\caption{\textbf{Discretization-dependent 2d homogenized elasticity tensor}. Components $\mathbb{C}_{ij}$ of the size and resolution SR320 but each which different discretizations D.} 
	\label{fig:component-elast-tensor_discretization}
\end{Figure}

\subsection{Check of isotropy-2D}
\label{subsec:test-for-isotropy}

The specimen size S320 is selected to measure the deviation from isotropy. The isotropic elasticity law in terms of the compliance matrix for plane strain conditions reads 

\begin{equation}
\begin{bmatrix}
\varepsilon_{11}\\ 
\varepsilon_{22}\\ 
2\varepsilon_{12}
\end{bmatrix} = 
\underbrace{\begin{pmatrix}
	\frac{1-\nu^2}{E} & -\frac{\nu(1+\nu)}{E} & 0\\ 
	& \phantom{-}\frac{1-\nu^2}{E} & 0\\ 
	\text{sym.} &  & \frac{2(1+\nu)}{E}
	\end{pmatrix}}_{\substack{\text{Elastic compliance} \\ \text{matrix \,} \mathbf{\mathbb S}}}
\begin{bmatrix}
\sigma_{11}\\ 
\sigma_{22}\\ 
\sigma_{12}
\end{bmatrix} \, .
\label{2D isotropy elasticity law}
\end{equation} 

The parameters $E$ and $\nu$ --and therefore the shear modulus $G=E/(2(1+\nu))$-- are identified by the homogenized elastic coefficients $\mathbb S_{11}$ and $\mathbb S_{12}$. The remaining  coefficients are used to validate the hypothesis of isotropy. Equation \eqref{2D isotropy elasticity law} implies 
(i) $\mathbb S_{11}=\mathbb S_{22}$, (ii) $\mathbb S_{33}=1/G$, (iii) $\mathbb S_{13}=0$, (iv) $\mathbb S_{23}=0$\, or in terms of the stiffness matrix (i) $\mathbb C_{11}=\mathbb C_{22}$, (ii) $\mathbb C_{33}=G$, (iii) $\mathbb C_{13}=0$, (iv) $\mathbb C_{23}=0$. Conditions (i)-(iv) are used to check the homogenized elasticity tensor with respect to isotropy, \cite{Eidel.2019}.

\begin{Table}[htbp]
\centering
\renewcommand{\arraystretch}{1.2}
 \resizebox{0.62\columnwidth}{!}{%
\begin{tabular}{cc|cccc}
\hline 
 \multicolumn{2}{c}{Identification} & \multicolumn{4}{c}{Check of isotropy} \\
\hline 
\rule{0pt}{25pt} 
 $E$ (MPa)  & $\nu$ & $\displaystyle \frac{\mathbb C_{11}-\mathbb C_{22}}{\mathbb C_{11}}$ & $\displaystyle \frac{\vert \mathbb C_{33}-G \vert}{G}$ & $\displaystyle \frac{\vert \mathbb C_{13} \vert}{\mathbb C_{11}}$ & $\displaystyle \frac{\vert \mathbb C_{23} \vert}{\mathbb C_{11}}$    \\
\hline 
 32111.77 & 0.203 & 0.43 $\%$  & 0.30 $\%$ & 0.02 $\%$  & 0.11 $\%$   \\
\hline
\end{tabular} 
}
\caption{\textbf{Identification}. Elastic constants for S320 assuming isotropy and deviation [$\%$] of the remaining coefficients from isotropy.} 	
\label{tab:Deviation Isotropy}
\end{Table}
Since the maximal deviation of the elastic coefficients from the isotropic ones is less than 0.5\% and $\mathbb C_{13}$ and $\mathbb C_{23}$ virtually vanish, the elasticity law of the specimen S320 can be reasonably regarded as isotropic.

%--------------------------------------- 3 D ----------------------------------------------

\section{Analysis for 3D}
\label{sec:Concrete_3D_elastic}

For the 3d simulations with up to 371$^3$ voxels (more than 150 million unknowns) iterative solvers along with preconditioners of the PETSc library are used based on MPI-parallelization employing up to 200 compute nodes, each with 256 GB RAM and 64 cores.  

The macro problem is a 3d cantilever beam of length $L=5000$~mm, height $B=1000$~mm and width $D=100$~mm, see Fig.~\ref{fig:mic-mac-transition}. It is loaded at $x=L$ by a line-load of $q_{0y}=-40.0$~N/mm. The considered volume of the microstructure is centered at the point with $x=y=z=2.1$~mm, where the computation of stress, strain, and discretization errors are carried out. 

For the 3d concrete specimen various subvolumes of cubic shape are considered for different resolutions R and discretizations D. 
 
\subsection{Specimen size and adaptive mesh-coarsening}
\label{subsubsec:SpecimenSize-MeshCoarsening}
 
Table \ref{tab:Concrete3D_coarsening_steps} shows the outcome of the octree-based adaptive mesh-coarsening applied to the slightly reduced specimen size S368. The number of unknowns of initially more than 150 millions is considerably reduced in three steps to less than 55\% which is the end in the adaptive D-dimension of the SRD space. The remaining problem size however is still quite large and shall be lowered by reductions in size S in the following.  

\begin{Table}[htbp]
	\begin{minipage}{16.5cm}  
		\centering
		\renewcommand{\arraystretch}{1.2} 
		 \resizebox{0.62\columnwidth}{!}{%
		\begin{tabular}{r c c c c}			
        \hline
                            & \multicolumn{4}{c}{Adaptive mesh coarsening}    \\
		SRD368              & 0          & 1          & 2         & 3         \\
		\hline
		ndof                & $150\,730\,227$ & $84\,301\,767$ & $82\,651\,488$ & $82\,649\,319$ \\
		Factor              & $1.0000$   & $0.5593$   & $0.5484$  & $0.5483$  \\
		Deactivated ndof    & $0$        & $24\,357\,456$   & $26\,241\,495$  & $26\,250\,483$  \\
		\hline
		\end{tabular} 
		}
	\end{minipage}
	\caption{\textbf{Specimen SRD368}. Values of ndof of the uniform mesh and the coarsened meshes, the corresponding reduction factor and the deactivated ndof for hanging nodes.}
	\label{tab:Concrete3D_coarsening_steps} 
\end{Table}

\begin{Figure}[htbp]
\begin{minipage}[c]{8.5cm}
\centering
\resizebox{0.98\columnwidth}{!}{% GNUPLOT: LaTeX picture with Postscript
\begingroup
  % Encoding inside the plot.  In the header of your document, this encoding
  % should to defined, e.g., by using
  % \usepackage[cp1252,<other encodings>]{inputenc}
  \inputencoding{cp1252}%
  \makeatletter
  \providecommand\color[2][]{%
    \GenericError{(gnuplot) \space\space\space\@spaces}{%
      Package color not loaded in conjunction with
      terminal option `colourtext'%
    }{See the gnuplot documentation for explanation.%
    }{Either use 'blacktext' in gnuplot or load the package
      color.sty in LaTeX.}%
    \renewcommand\color[2][]{}%
  }%
  \providecommand\includegraphics[2][]{%
    \GenericError{(gnuplot) \space\space\space\@spaces}{%
      Package graphicx or graphics not loaded%
    }{See the gnuplot documentation for explanation.%
    }{The gnuplot epslatex terminal needs graphicx.sty or graphics.sty.}%
    \renewcommand\includegraphics[2][]{}%
  }%
  \providecommand\rotatebox[2]{#2}%
  \@ifundefined{ifGPcolor}{%
    \newif\ifGPcolor
    \GPcolortrue
  }{}%
  \@ifundefined{ifGPblacktext}{%
    \newif\ifGPblacktext
    \GPblacktextfalse
  }{}%
  % define a \g@addto@macro without @ in the name:
  \let\gplgaddtomacro\g@addto@macro
  % define empty templates for all commands taking text:
  \gdef\gplbacktext{}%
  \gdef\gplfronttext{}%
  \makeatother
  \ifGPblacktext
    % no textcolor at all
    \def\colorrgb#1{}%
    \def\colorgray#1{}%
  \else
    % gray or color?
    \ifGPcolor
      \def\colorrgb#1{\color[rgb]{#1}}%
      \def\colorgray#1{\color[gray]{#1}}%
      \expandafter\def\csname LTw\endcsname{\color{white}}%
      \expandafter\def\csname LTb\endcsname{\color{black}}%
      \expandafter\def\csname LTa\endcsname{\color{black}}%
      \expandafter\def\csname LT0\endcsname{\color[rgb]{1,0,0}}%
      \expandafter\def\csname LT1\endcsname{\color[rgb]{0,1,0}}%
      \expandafter\def\csname LT2\endcsname{\color[rgb]{0,0,1}}%
      \expandafter\def\csname LT3\endcsname{\color[rgb]{1,0,1}}%
      \expandafter\def\csname LT4\endcsname{\color[rgb]{0,1,1}}%
      \expandafter\def\csname LT5\endcsname{\color[rgb]{1,1,0}}%
      \expandafter\def\csname LT6\endcsname{\color[rgb]{0,0,0}}%
      \expandafter\def\csname LT7\endcsname{\color[rgb]{1,0.3,0}}%
      \expandafter\def\csname LT8\endcsname{\color[rgb]{0.5,0.5,0.5}}%
    \else
      % gray
      \def\colorrgb#1{\color{black}}%
      \def\colorgray#1{\color[gray]{#1}}%
      \expandafter\def\csname LTw\endcsname{\color{white}}%
      \expandafter\def\csname LTb\endcsname{\color{black}}%
      \expandafter\def\csname LTa\endcsname{\color{black}}%
      \expandafter\def\csname LT0\endcsname{\color{black}}%
      \expandafter\def\csname LT1\endcsname{\color{black}}%
      \expandafter\def\csname LT2\endcsname{\color{black}}%
      \expandafter\def\csname LT3\endcsname{\color{black}}%
      \expandafter\def\csname LT4\endcsname{\color{black}}%
      \expandafter\def\csname LT5\endcsname{\color{black}}%
      \expandafter\def\csname LT6\endcsname{\color{black}}%
      \expandafter\def\csname LT7\endcsname{\color{black}}%
      \expandafter\def\csname LT8\endcsname{\color{black}}%
    \fi
  \fi
    \setlength{\unitlength}{0.0500bp}%
    \ifx\gptboxheight\undefined%
      \newlength{\gptboxheight}%
      \newlength{\gptboxwidth}%
      \newsavebox{\gptboxtext}%
    \fi%
    \setlength{\fboxrule}{0.5pt}%
    \setlength{\fboxsep}{1pt}%
\begin{picture}(6802.00,6236.00)%
    \gplgaddtomacro\gplbacktext{%
      \csname LTb\endcsname%%
      \put(548,4365){\makebox(0,0)[r]{\strut{}$44$}}%
      \csname LTb\endcsname%%
      \put(548,4988){\makebox(0,0)[r]{\strut{}$48$}}%
      \csname LTb\endcsname%%
      \put(548,5611){\makebox(0,0)[r]{\strut{}$52$}}%
      \csname LTb\endcsname%%
      \put(1175,3833){\makebox(0,0){\strut{} }}%
      \csname LTb\endcsname%%
      \put(1669,3833){\makebox(0,0){\strut{} }}%
      \csname LTb\endcsname%%
      \put(2164,3833){\makebox(0,0){\strut{} }}%
      \csname LTb\endcsname%%
      \put(2658,3833){\makebox(0,0){\strut{} }}%
      \csname LTb\endcsname%%
      \put(3153,3833){\makebox(0,0){\strut{} }}%
      \csname LTb\endcsname%%
      \put(3647,3833){\makebox(0,0){\strut{} }}%
      \csname LTb\endcsname%%
      \put(4142,3833){\makebox(0,0){\strut{} }}%
      \csname LTb\endcsname%%
      \put(4636,3833){\makebox(0,0){\strut{} }}%
      \csname LTb\endcsname%%
      \put(5131,3833){\makebox(0,0){\strut{} }}%
      \csname LTb\endcsname%%
      \put(5625,3833){\makebox(0,0){\strut{} }}%
    }%
    \gplgaddtomacro\gplfronttext{%
      \csname LTb\endcsname%%
      \put(5529,5750){\makebox(0,0)[r]{\strut{}$\mathbb{C}_{11}$}}%
      \csname LTb\endcsname%%
      \put(5529,5530){\makebox(0,0)[r]{\strut{}$\mathbb{C}_{22}$}}%
      \csname LTb\endcsname%%
      \put(5529,5310){\makebox(0,0)[r]{\strut{}$\mathbb{C}_{33}$}}%
    }%
    \gplgaddtomacro\gplbacktext{%
      \csname LTb\endcsname%%
      \put(548,3274){\makebox(0,0)[r]{\strut{}$18$}}%
      \csname LTb\endcsname%%
      \put(548,3585){\makebox(0,0)[r]{\strut{}$20$}}%
      \csname LTb\endcsname%%
      \put(548,3896){\makebox(0,0)[r]{\strut{}$22$}}%
      \csname LTb\endcsname%%
      \put(1175,2898){\makebox(0,0){\strut{} }}%
      \csname LTb\endcsname%%
      \put(1669,2898){\makebox(0,0){\strut{} }}%
      \csname LTb\endcsname%%
      \put(2164,2898){\makebox(0,0){\strut{} }}%
      \csname LTb\endcsname%%
      \put(2658,2898){\makebox(0,0){\strut{} }}%
      \csname LTb\endcsname%%
      \put(3153,2898){\makebox(0,0){\strut{} }}%
      \csname LTb\endcsname%%
      \put(3647,2898){\makebox(0,0){\strut{} }}%
      \csname LTb\endcsname%%
      \put(4142,2898){\makebox(0,0){\strut{} }}%
      \csname LTb\endcsname%%
      \put(4636,2898){\makebox(0,0){\strut{} }}%
      \csname LTb\endcsname%%
      \put(5131,2898){\makebox(0,0){\strut{} }}%
      \csname LTb\endcsname%%
      \put(5625,2898){\makebox(0,0){\strut{} }}%
    }%
    \gplgaddtomacro\gplfronttext{%
      \csname LTb\endcsname%%
      \put(130,3585){\rotatebox{-270}{\makebox(0,0){\strut{}$\mathbb{C}_{ij}$ (in GPa)}}}%
      \csname LTb\endcsname%%
      \put(5529,3879){\makebox(0,0)[r]{\strut{}$\mathbb{C}_{12}$}}%
      \csname LTb\endcsname%%
      \put(5529,3659){\makebox(0,0)[r]{\strut{}$\mathbb{C}_{13}$}}%
      \csname LTb\endcsname%%
      \put(5529,3439){\makebox(0,0)[r]{\strut{}$\mathbb{C}_{23}$}}%
    }%
    \gplgaddtomacro\gplbacktext{%
      \csname LTb\endcsname%%
      \put(548,2416){\makebox(0,0)[r]{\strut{}$13$}}%
      \csname LTb\endcsname%%
      \put(548,2650){\makebox(0,0)[r]{\strut{}$14$}}%
      \csname LTb\endcsname%%
      \put(548,2883){\makebox(0,0)[r]{\strut{}$15$}}%
      \csname LTb\endcsname%%
      \put(1175,1962){\makebox(0,0){\strut{} }}%
      \csname LTb\endcsname%%
      \put(1669,1962){\makebox(0,0){\strut{} }}%
      \csname LTb\endcsname%%
      \put(2164,1962){\makebox(0,0){\strut{} }}%
      \csname LTb\endcsname%%
      \put(2658,1962){\makebox(0,0){\strut{} }}%
      \csname LTb\endcsname%%
      \put(3153,1962){\makebox(0,0){\strut{} }}%
      \csname LTb\endcsname%%
      \put(3647,1962){\makebox(0,0){\strut{} }}%
      \csname LTb\endcsname%%
      \put(4142,1962){\makebox(0,0){\strut{} }}%
      \csname LTb\endcsname%%
      \put(4636,1962){\makebox(0,0){\strut{} }}%
      \csname LTb\endcsname%%
      \put(5131,1962){\makebox(0,0){\strut{} }}%
      \csname LTb\endcsname%%
      \put(5625,1962){\makebox(0,0){\strut{} }}%
    }%
    \gplgaddtomacro\gplfronttext{%
      \csname LTb\endcsname%%
      \put(5529,2944){\makebox(0,0)[r]{\strut{}$\mathbb{C}_{44}$}}%
      \csname LTb\endcsname%%
      \put(5529,2724){\makebox(0,0)[r]{\strut{}$\mathbb{C}_{55}$}}%
      \csname LTb\endcsname%%
      \put(5529,2504){\makebox(0,0)[r]{\strut{}$\mathbb{C}_{66}$}}%
    }%
    \gplgaddtomacro\gplbacktext{%
      \colorrgb{0.00,0.00,1.00}%%
      \put(548,836){\makebox(0,0)[r]{\strut{}$0.55$}}%
      \colorrgb{0.00,0.00,1.00}%%
      \put(548,1403){\makebox(0,0)[r]{\strut{}$0.63$}}%
      \colorrgb{0.00,0.00,1.00}%%
      \put(548,1969){\makebox(0,0)[r]{\strut{}$0.71$}}%
      \csname LTb\endcsname%%
      \put(1175,403){\makebox(0,0){\strut{}16}}%
      \csname LTb\endcsname%%
      \put(1669,403){\makebox(0,0){\strut{}32}}%
      \csname LTb\endcsname%%
      \put(2164,403){\makebox(0,0){\strut{}64}}%
      \csname LTb\endcsname%%
      \put(2658,403){\makebox(0,0){\strut{}128}}%
      \csname LTb\endcsname%%
      \put(3153,403){\makebox(0,0){\strut{}150}}%
      \csname LTb\endcsname%%
      \put(3647,403){\makebox(0,0){\strut{}200}}%
      \csname LTb\endcsname%%
      \put(4142,403){\makebox(0,0){\strut{}256}}%
      \csname LTb\endcsname%%
      \put(4636,403){\makebox(0,0){\strut{}300}}%
      \csname LTb\endcsname%%
      \put(5131,403){\makebox(0,0){\strut{}320}}%
      \csname LTb\endcsname%%
      \put(5625,403){\makebox(0,0){\strut{}371}}%
    }%
    \gplgaddtomacro\gplfronttext{%
      \colorrgb{0.00,0.00,1.00}%%
      \put(-134,1402){\rotatebox{-270}{\makebox(0,0){\strut{}$\text{PF}$}}}%
      \csname LTb\endcsname%%
      \put(3400,73){\makebox(0,0){\strut{}Size S}}%
      \csname LTb\endcsname%%
      \put(5529,2009){\makebox(0,0)[r]{\strut{}$\text{Aggregate phase fraction}$}}%
    }%
    \gplbacktext
    \put(0,0){\includegraphics{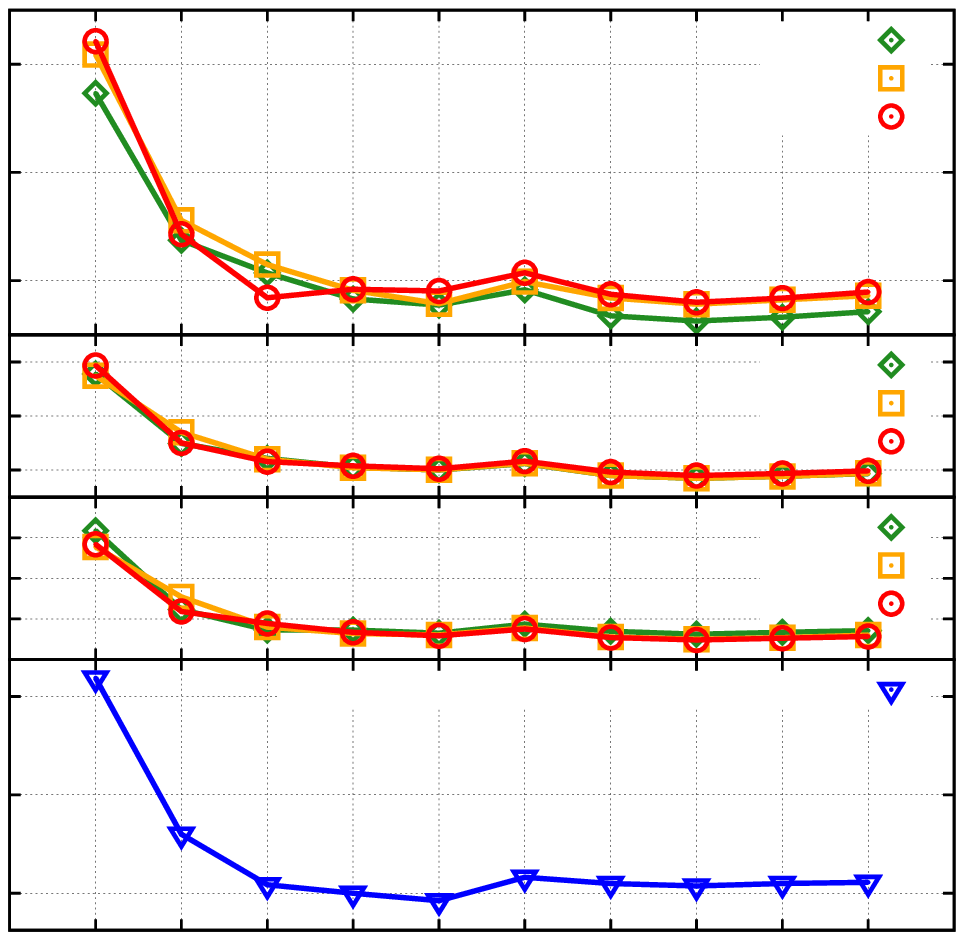}}%
    \gplfronttext
  \end{picture}%
\endgroup
}
\end{minipage}
\begin{minipage}[c]{7.5cm}
\centering
{\includegraphics[height=2.5cm, angle=0]{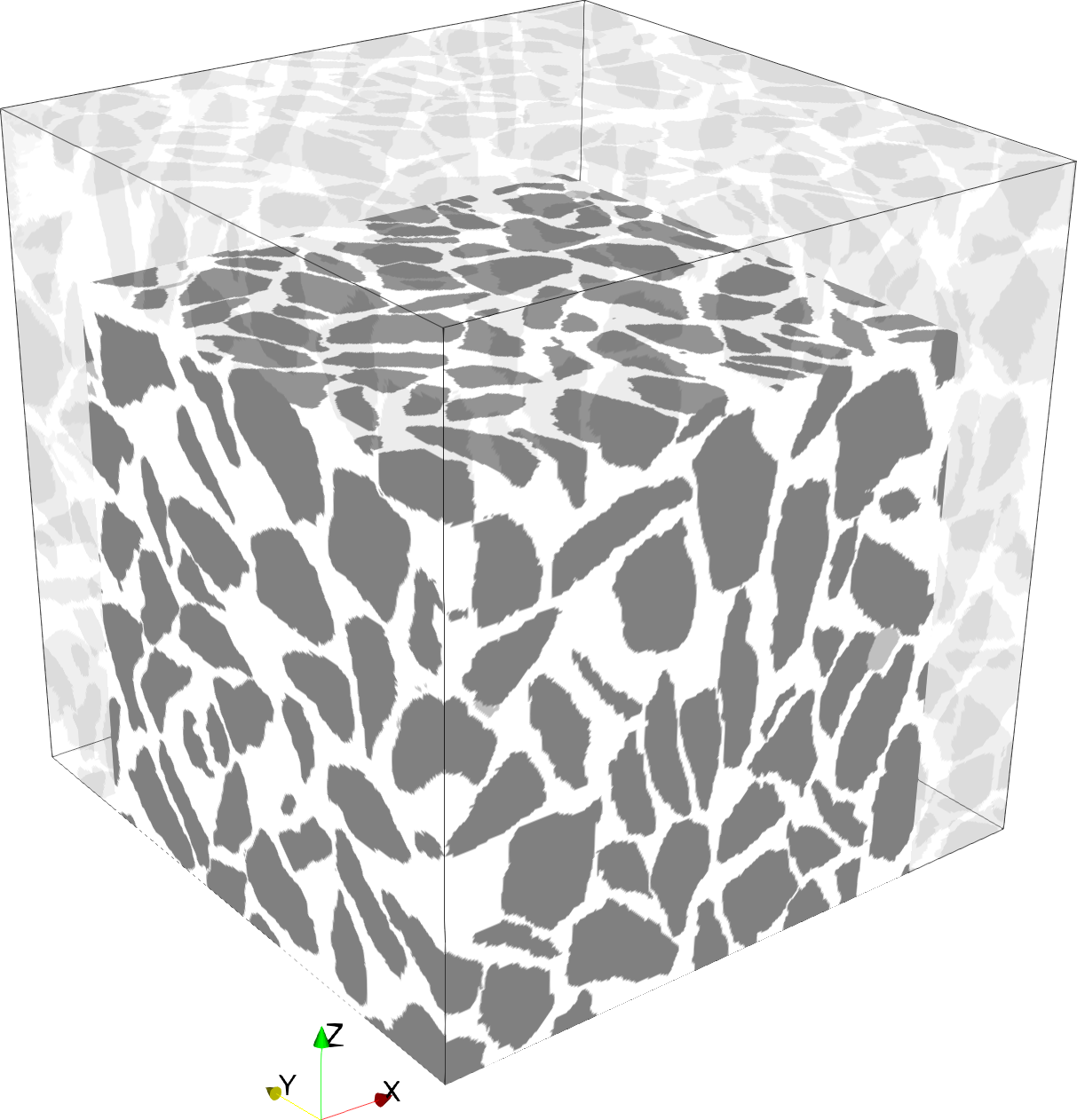}} 
{\includegraphics[height=2.5cm, angle=0]{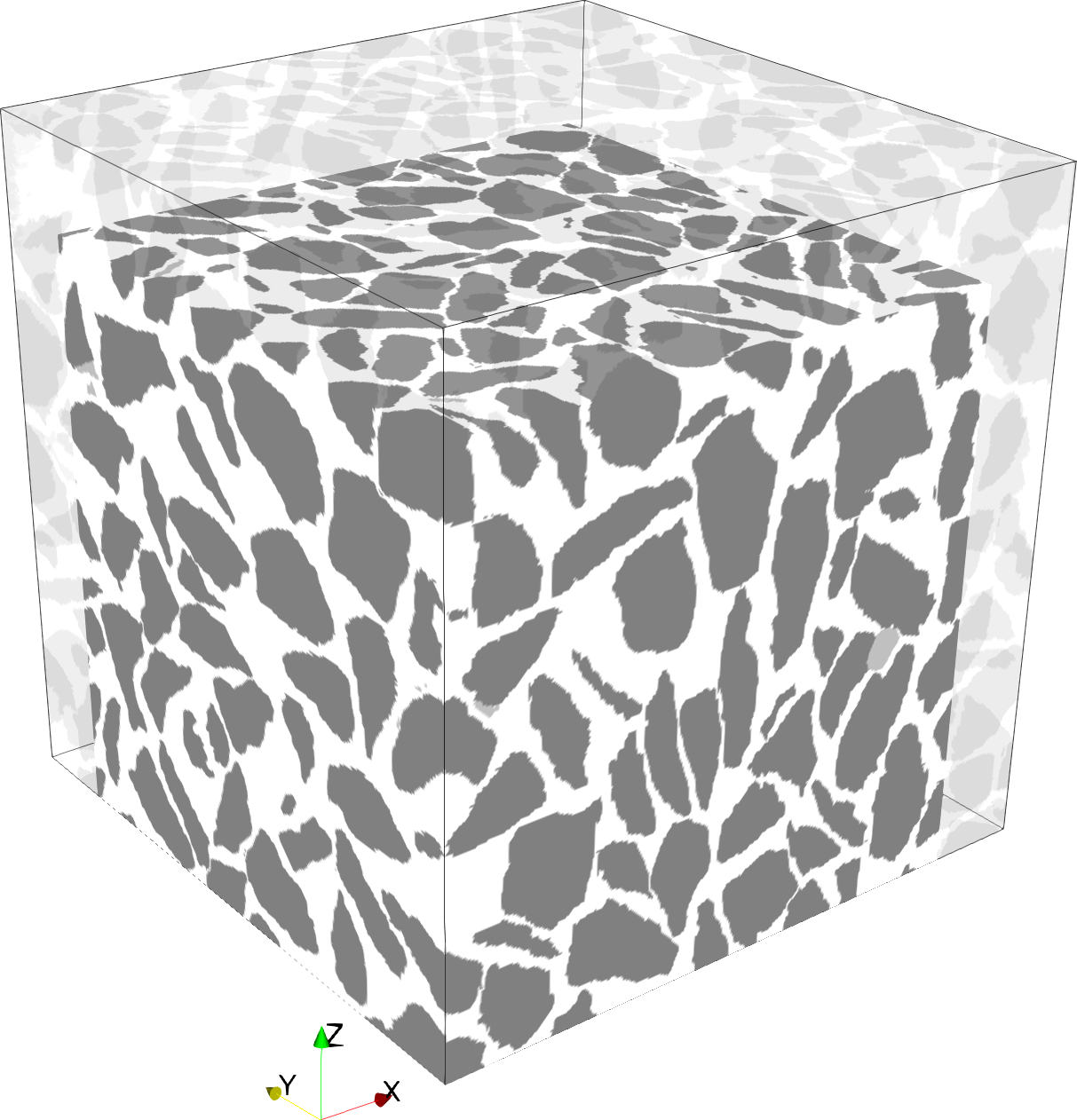}} 
%{\includegraphics[height=2.5cm, angle=0]{Concrete_wo_mesh.png}} \\
{\includegraphics[height=2.5cm, angle=0]{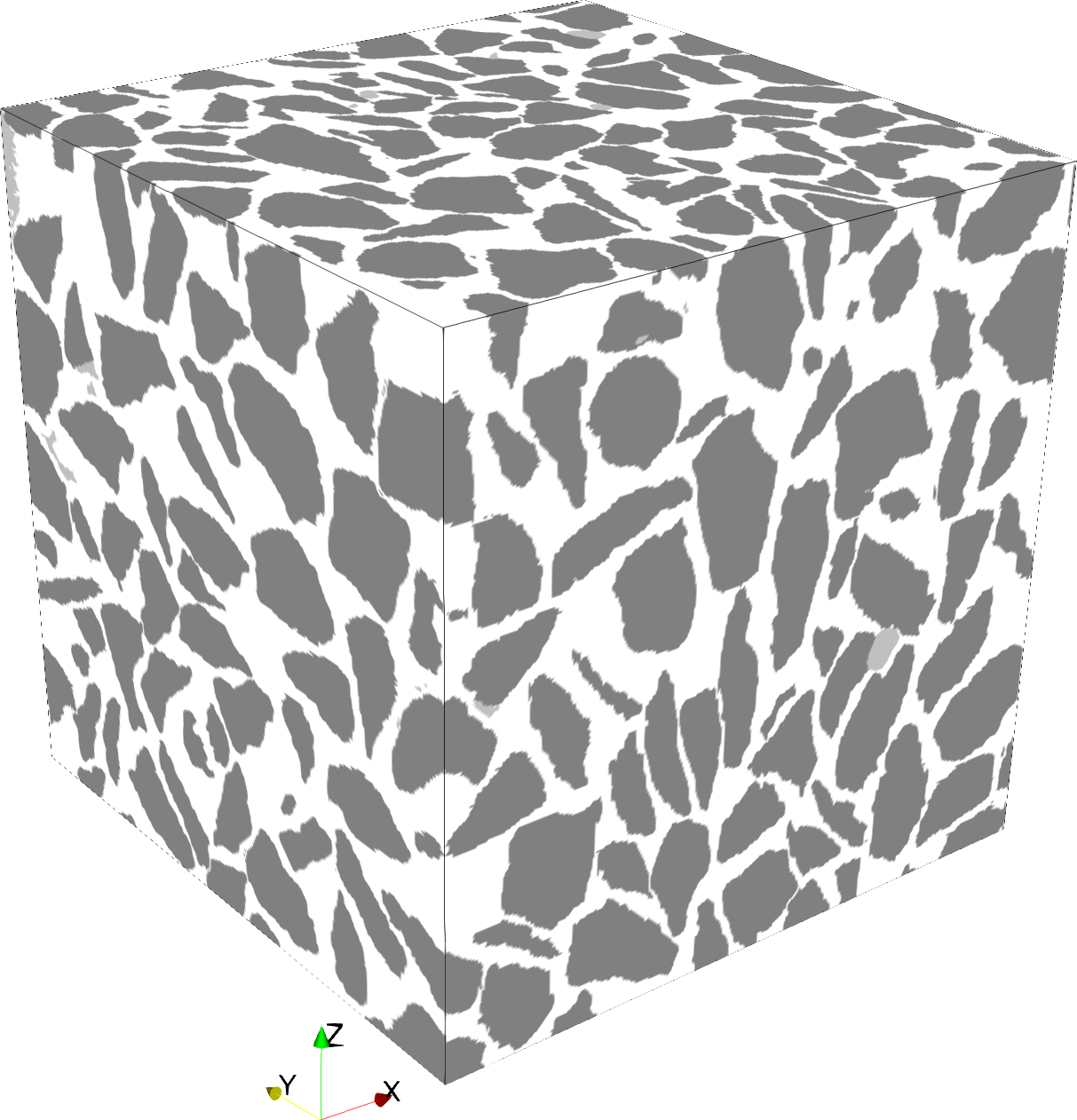}} \\
{\includegraphics[height=2.5cm, angle=0]{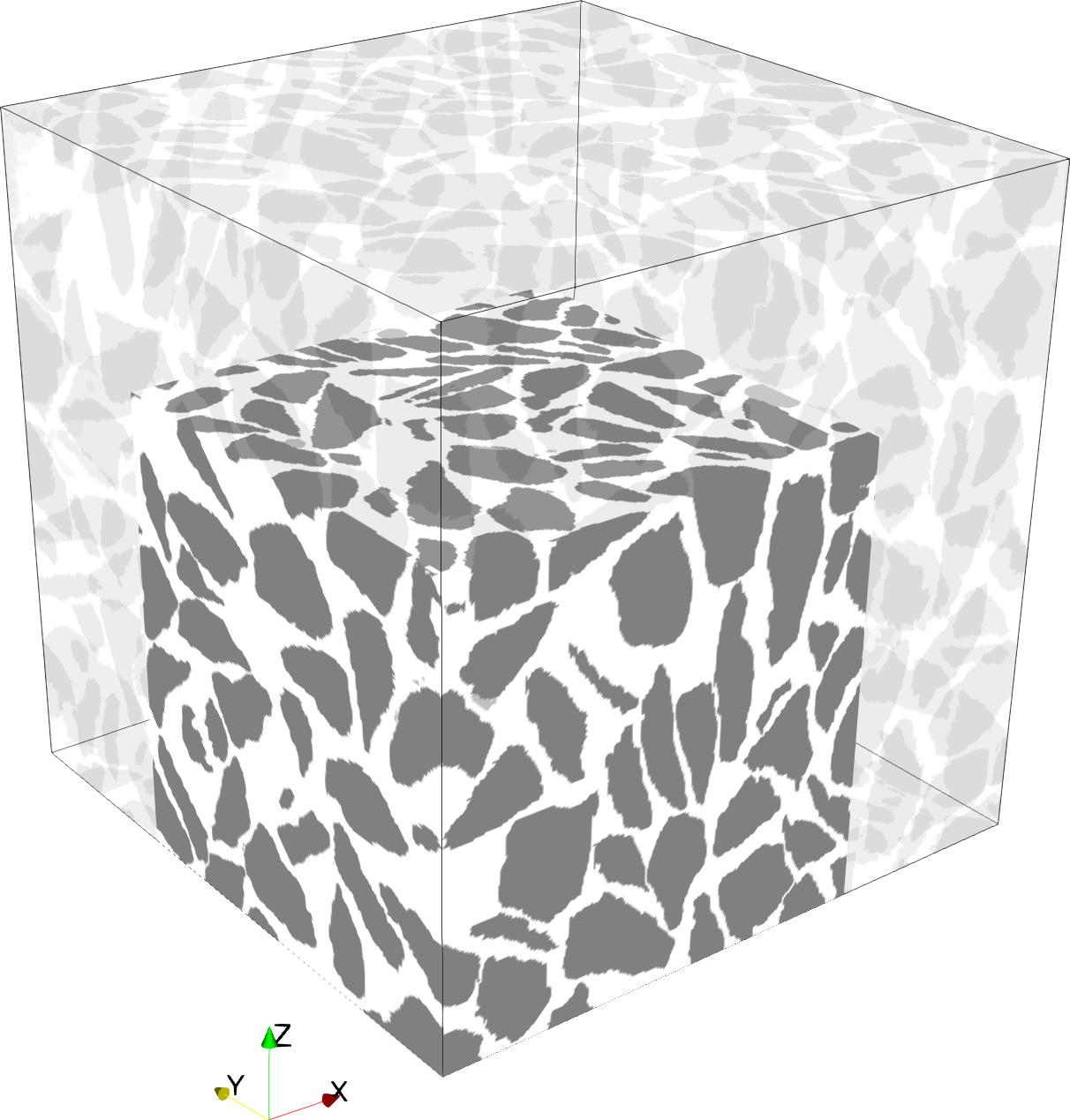}}  
{\includegraphics[height=2.5cm, angle=0]{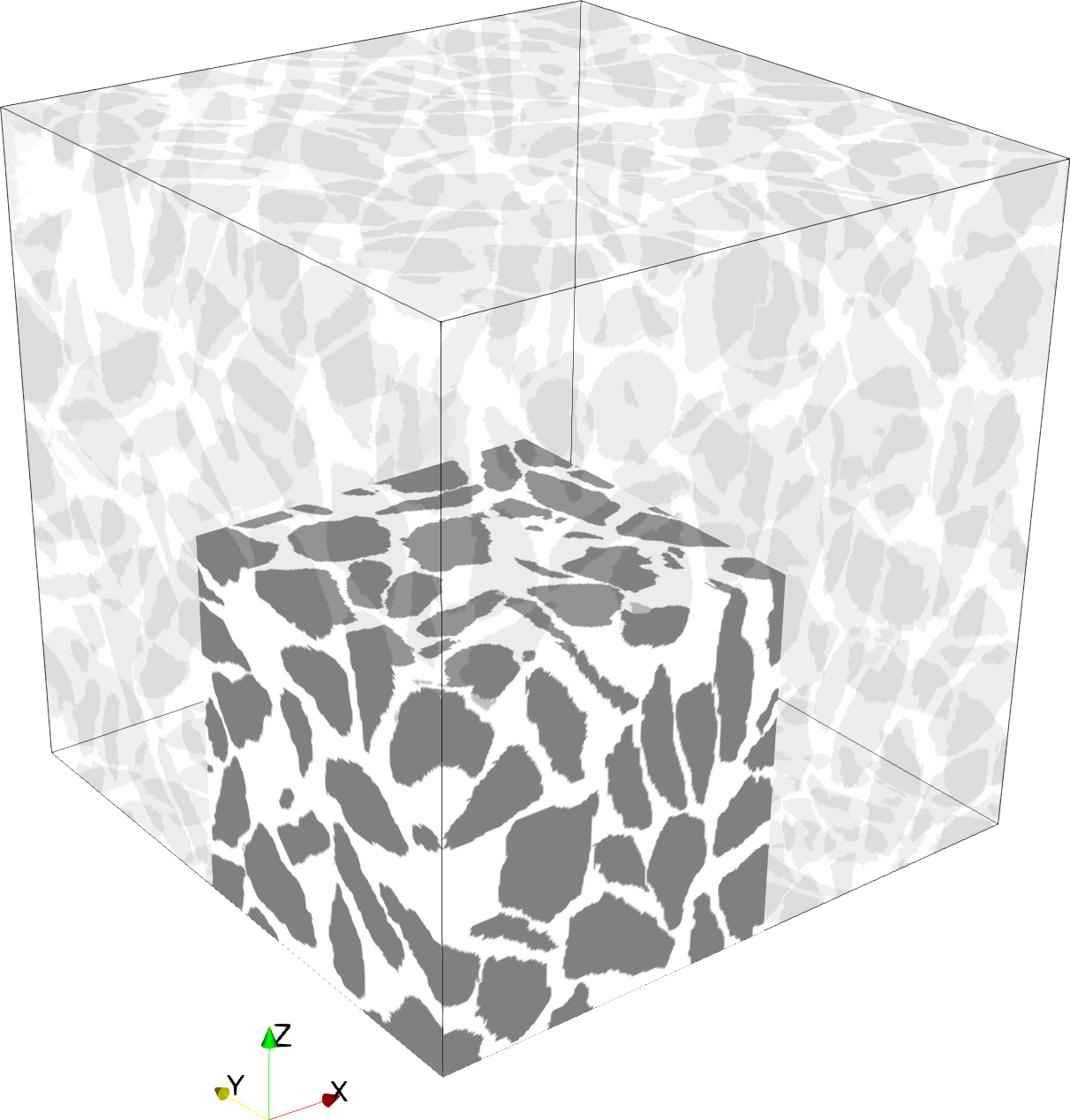}} 
{\includegraphics[height=2.5cm, angle=0]{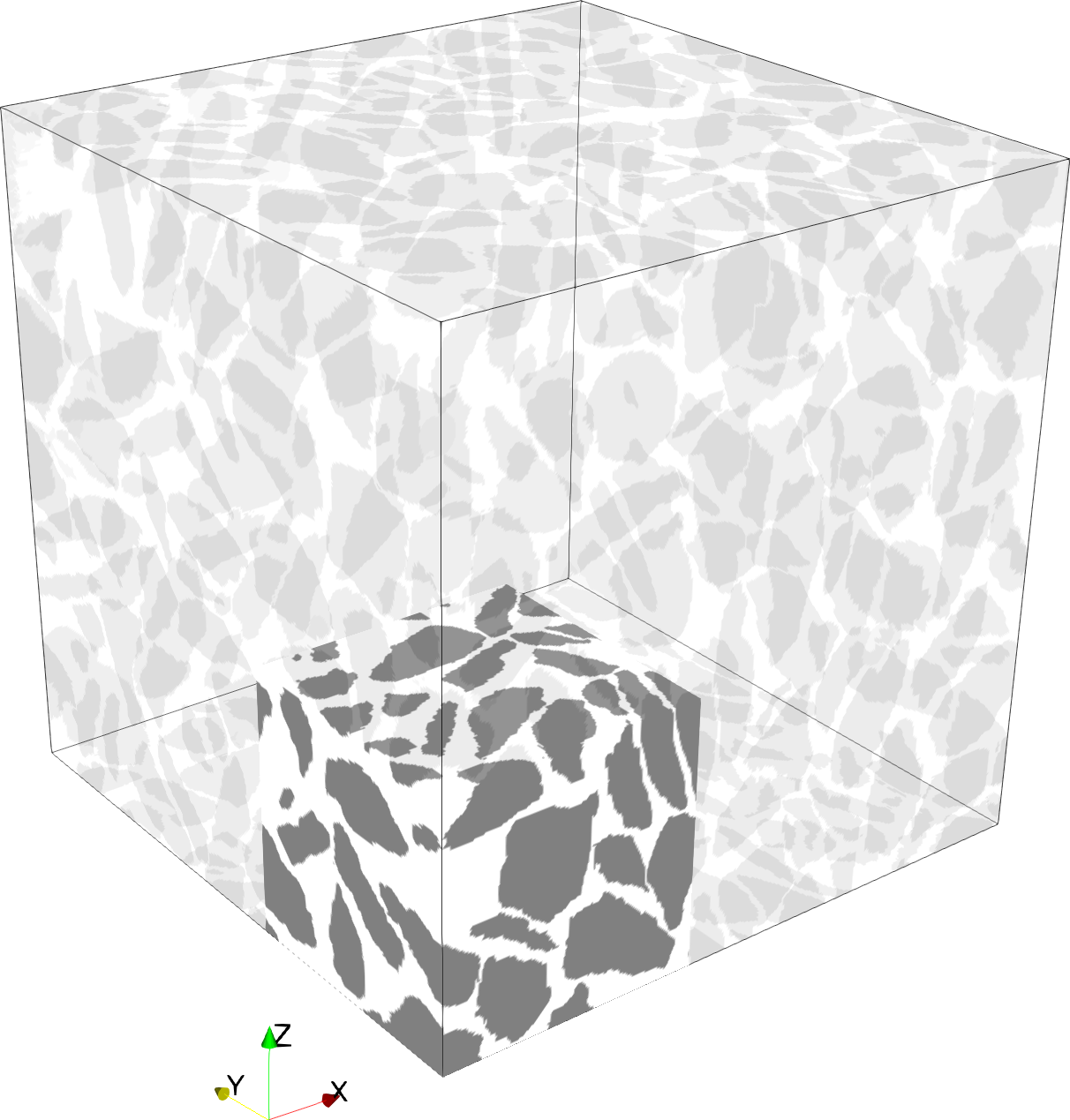}} \\
%{\includegraphics[height=2.5cm, angle=0]{Subvolume_16x16x16.png}}  
{\includegraphics[height=2.5cm, angle=0]{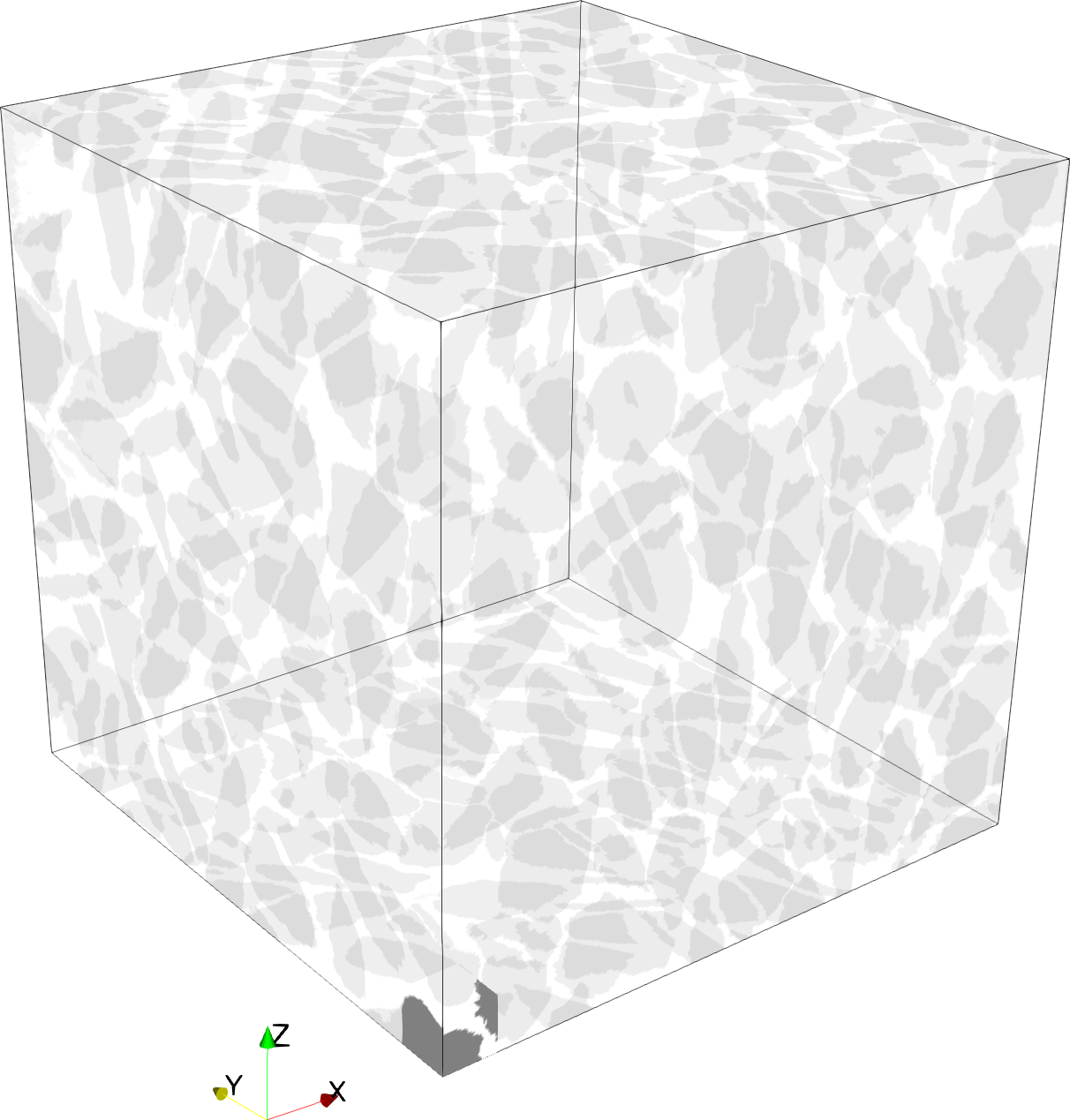}} 
{\includegraphics[height=2.5cm, angle=0]{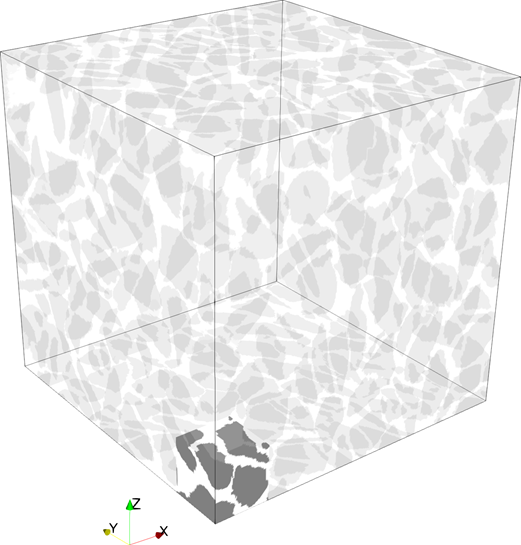}} 
{\includegraphics[height=2.5cm, angle=0]{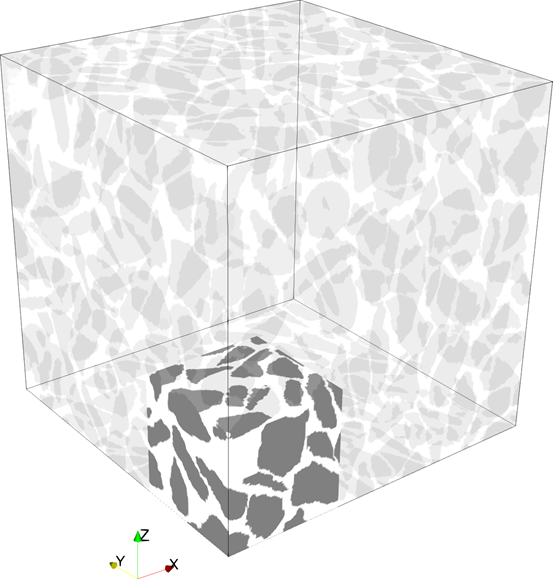}} 
\end{minipage}
\caption{\textbf{Size-dependent homogenized 3d elasticity tensor}. $\mathbb{C}_{ij}$ versus the specimen size and aggregate phase fraction, the corresponding numbers are given in the Supplement. In the right, a series of specimens from S32 to S371 are presented. %$S32,64,128,150,200,256,300,320,371.
}
\label{fig:component-elast-tensor}
\end{Figure}

\begin{Figure}[htbp]
	\centering
     \resizebox{0.99\columnwidth}{!}{% GNUPLOT: LaTeX picture with Postscript
\begingroup
  % Encoding inside the plot.  In the header of your document, this encoding
  % should to defined, e.g., by using
  % \usepackage[cp1252,<other encodings>]{inputenc}
  \inputencoding{cp1252}%
  \makeatletter
  \providecommand\color[2][]{%
    \GenericError{(gnuplot) \space\space\space\@spaces}{%
      Package color not loaded in conjunction with
      terminal option `colourtext'%
    }{See the gnuplot documentation for explanation.%
    }{Either use 'blacktext' in gnuplot or load the package
      color.sty in LaTeX.}%
    \renewcommand\color[2][]{}%
  }%
  \providecommand\includegraphics[2][]{%
    \GenericError{(gnuplot) \space\space\space\@spaces}{%
      Package graphicx or graphics not loaded%
    }{See the gnuplot documentation for explanation.%
    }{The gnuplot epslatex terminal needs graphicx.sty or graphics.sty.}%
    \renewcommand\includegraphics[2][]{}%
  }%
  \providecommand\rotatebox[2]{#2}%
  \@ifundefined{ifGPcolor}{%
    \newif\ifGPcolor
    \GPcolortrue
  }{}%
  \@ifundefined{ifGPblacktext}{%
    \newif\ifGPblacktext
    \GPblacktextfalse
  }{}%
  % define a \g@addto@macro without @ in the name:
  \let\gplgaddtomacro\g@addto@macro
  % define empty templates for all commands taking text:
  \gdef\gplbacktext{}%
  \gdef\gplfronttext{}%
  \makeatother
  \ifGPblacktext
    % no textcolor at all
    \def\colorrgb#1{}%
    \def\colorgray#1{}%
  \else
    % gray or color?
    \ifGPcolor
      \def\colorrgb#1{\color[rgb]{#1}}%
      \def\colorgray#1{\color[gray]{#1}}%
      \expandafter\def\csname LTw\endcsname{\color{white}}%
      \expandafter\def\csname LTb\endcsname{\color{black}}%
      \expandafter\def\csname LTa\endcsname{\color{black}}%
      \expandafter\def\csname LT0\endcsname{\color[rgb]{1,0,0}}%
      \expandafter\def\csname LT1\endcsname{\color[rgb]{0,1,0}}%
      \expandafter\def\csname LT2\endcsname{\color[rgb]{0,0,1}}%
      \expandafter\def\csname LT3\endcsname{\color[rgb]{1,0,1}}%
      \expandafter\def\csname LT4\endcsname{\color[rgb]{0,1,1}}%
      \expandafter\def\csname LT5\endcsname{\color[rgb]{1,1,0}}%
      \expandafter\def\csname LT6\endcsname{\color[rgb]{0,0,0}}%
      \expandafter\def\csname LT7\endcsname{\color[rgb]{1,0.3,0}}%
      \expandafter\def\csname LT8\endcsname{\color[rgb]{0.5,0.5,0.5}}%
    \else
      % gray
      \def\colorrgb#1{\color{black}}%
      \def\colorgray#1{\color[gray]{#1}}%
      \expandafter\def\csname LTw\endcsname{\color{white}}%
      \expandafter\def\csname LTb\endcsname{\color{black}}%
      \expandafter\def\csname LTa\endcsname{\color{black}}%
      \expandafter\def\csname LT0\endcsname{\color{black}}%
      \expandafter\def\csname LT1\endcsname{\color{black}}%
      \expandafter\def\csname LT2\endcsname{\color{black}}%
      \expandafter\def\csname LT3\endcsname{\color{black}}%
      \expandafter\def\csname LT4\endcsname{\color{black}}%
      \expandafter\def\csname LT5\endcsname{\color{black}}%
      \expandafter\def\csname LT6\endcsname{\color{black}}%
      \expandafter\def\csname LT7\endcsname{\color{black}}%
      \expandafter\def\csname LT8\endcsname{\color{black}}%
    \fi
  \fi
    \setlength{\unitlength}{0.0500bp}%
    \ifx\gptboxheight\undefined%
      \newlength{\gptboxheight}%
      \newlength{\gptboxwidth}%
      \newsavebox{\gptboxtext}%
    \fi%
    \setlength{\fboxrule}{0.5pt}%
    \setlength{\fboxsep}{1pt}%
\begin{picture}(11904.00,3684.00)%
    \gplgaddtomacro\gplbacktext{%
      \csname LTb\endcsname%%
      \put(1058,368){\makebox(0,0)[r]{\strut{}$-5$}}%
      \csname LTb\endcsname%%
      \put(1058,815){\makebox(0,0)[r]{\strut{}$0$}}%
      \csname LTb\endcsname%%
      \put(1058,1262){\makebox(0,0)[r]{\strut{}$5$}}%
      \csname LTb\endcsname%%
      \put(1058,1709){\makebox(0,0)[r]{\strut{}$10$}}%
      \csname LTb\endcsname%%
      \put(1058,2157){\makebox(0,0)[r]{\strut{}$15$}}%
      \csname LTb\endcsname%%
      \put(1058,2604){\makebox(0,0)[r]{\strut{}$20$}}%
      \csname LTb\endcsname%%
      \put(1058,3051){\makebox(0,0)[r]{\strut{}$25$}}%
      \csname LTb\endcsname%%
      \put(1058,3498){\makebox(0,0)[r]{\strut{}$30$}}%
      \csname LTb\endcsname%%
      \put(1190,148){\makebox(0,0){\strut{}$0$}}%
      \csname LTb\endcsname%%
      \put(1577,148){\makebox(0,0){\strut{}$50$}}%
      \csname LTb\endcsname%%
      \put(2351,148){\makebox(0,0){\strut{}$150$}}%
      \csname LTb\endcsname%%
      \put(3124,148){\makebox(0,0){\strut{}$250$}}%
      \csname LTb\endcsname%%
      \put(3898,148){\makebox(0,0){\strut{}$350$}}%
      \put(1809,3248){\makebox(0,0)[l]{\strut{}$\mathbb{C}_{11}$}}%
    }%
    \gplgaddtomacro\gplfronttext{%
      \csname LTb\endcsname%%
      \put(640,1933){\rotatebox{-270}{\makebox(0,0){\strut{}Deviation (in $\%$)}}}%
      \put(2737,-182){\makebox(0,0){\strut{}Size S}}%
      \csname LTb\endcsname%%
      \put(3694,3325){\makebox(0,0)[r]{\strut{}$\text{KUBC}$}}%
      \csname LTb\endcsname%%
      \put(3694,3105){\makebox(0,0)[r]{\strut{}$\text{PBC}$}}%
      \csname LTb\endcsname%%
      \put(3694,2885){\makebox(0,0)[r]{\strut{}$\text{SUBC}$}}%
    }%
    \gplgaddtomacro\gplbacktext{%
      \csname LTb\endcsname%%
      \put(4272,368){\makebox(0,0)[r]{\strut{} }}%
      \csname LTb\endcsname%%
      \put(4272,815){\makebox(0,0)[r]{\strut{} }}%
      \csname LTb\endcsname%%
      \put(4272,1262){\makebox(0,0)[r]{\strut{} }}%
      \csname LTb\endcsname%%
      \put(4272,1709){\makebox(0,0)[r]{\strut{} }}%
      \csname LTb\endcsname%%
      \put(4272,2157){\makebox(0,0)[r]{\strut{} }}%
      \csname LTb\endcsname%%
      \put(4272,2604){\makebox(0,0)[r]{\strut{} }}%
      \csname LTb\endcsname%%
      \put(4272,3051){\makebox(0,0)[r]{\strut{} }}%
      \csname LTb\endcsname%%
      \put(4791,148){\makebox(0,0){\strut{}$50$}}%
      \csname LTb\endcsname%%
      \put(5564,148){\makebox(0,0){\strut{}$150$}}%
      \csname LTb\endcsname%%
      \put(6338,148){\makebox(0,0){\strut{}$250$}}%
      \csname LTb\endcsname%%
      \put(7111,148){\makebox(0,0){\strut{}$350$}}%
      \put(5023,3248){\makebox(0,0)[l]{\strut{}$\mathbb{C}_{44}$}}%
    }%
    \gplgaddtomacro\gplfronttext{%
      \csname LTb\endcsname%%
      \put(5951,-182){\makebox(0,0){\strut{}Size S}}%
      \csname LTb\endcsname%%
      \put(6907,3325){\makebox(0,0)[r]{\strut{}$\text{KUBC}$}}%
      \csname LTb\endcsname%%
      \put(6907,3105){\makebox(0,0)[r]{\strut{}$\text{PBC}$}}%
      \csname LTb\endcsname%%
      \put(6907,2885){\makebox(0,0)[r]{\strut{}$\text{SUBC}$}}%
    }%
    \gplgaddtomacro\gplbacktext{%
      \csname LTb\endcsname%%
      \put(7486,368){\makebox(0,0)[r]{\strut{} }}%
      \csname LTb\endcsname%%
      \put(7486,815){\makebox(0,0)[r]{\strut{} }}%
      \csname LTb\endcsname%%
      \put(7486,1262){\makebox(0,0)[r]{\strut{} }}%
      \csname LTb\endcsname%%
      \put(7486,1709){\makebox(0,0)[r]{\strut{} }}%
      \csname LTb\endcsname%%
      \put(7486,2157){\makebox(0,0)[r]{\strut{} }}%
      \csname LTb\endcsname%%
      \put(7486,2604){\makebox(0,0)[r]{\strut{} }}%
      \csname LTb\endcsname%%
      \put(7486,3051){\makebox(0,0)[r]{\strut{} }}%
      \csname LTb\endcsname%%
      \put(8005,148){\makebox(0,0){\strut{}$50$}}%
      \csname LTb\endcsname%%
      \put(8778,148){\makebox(0,0){\strut{}$150$}}%
      \csname LTb\endcsname%%
      \put(9552,148){\makebox(0,0){\strut{}$250$}}%
      \csname LTb\endcsname%%
      \put(10325,148){\makebox(0,0){\strut{}$350$}}%
      \put(8237,3248){\makebox(0,0)[l]{\strut{}$\mathbb{C}_{12}$}}%
    }%
    \gplgaddtomacro\gplfronttext{%
      \csname LTb\endcsname%%
      \put(9165,-182){\makebox(0,0){\strut{}Size S}}%
      \csname LTb\endcsname%%
      \put(10121,3325){\makebox(0,0)[r]{\strut{}$\text{KUBC}$}}%
      \csname LTb\endcsname%%
      \put(10121,3105){\makebox(0,0)[r]{\strut{}$\text{PBC}$}}%
      \csname LTb\endcsname%%
      \put(10121,2885){\makebox(0,0)[r]{\strut{}$\text{SUBC}$}}%
    }%
    \gplbacktext
    \put(0,0){\includegraphics{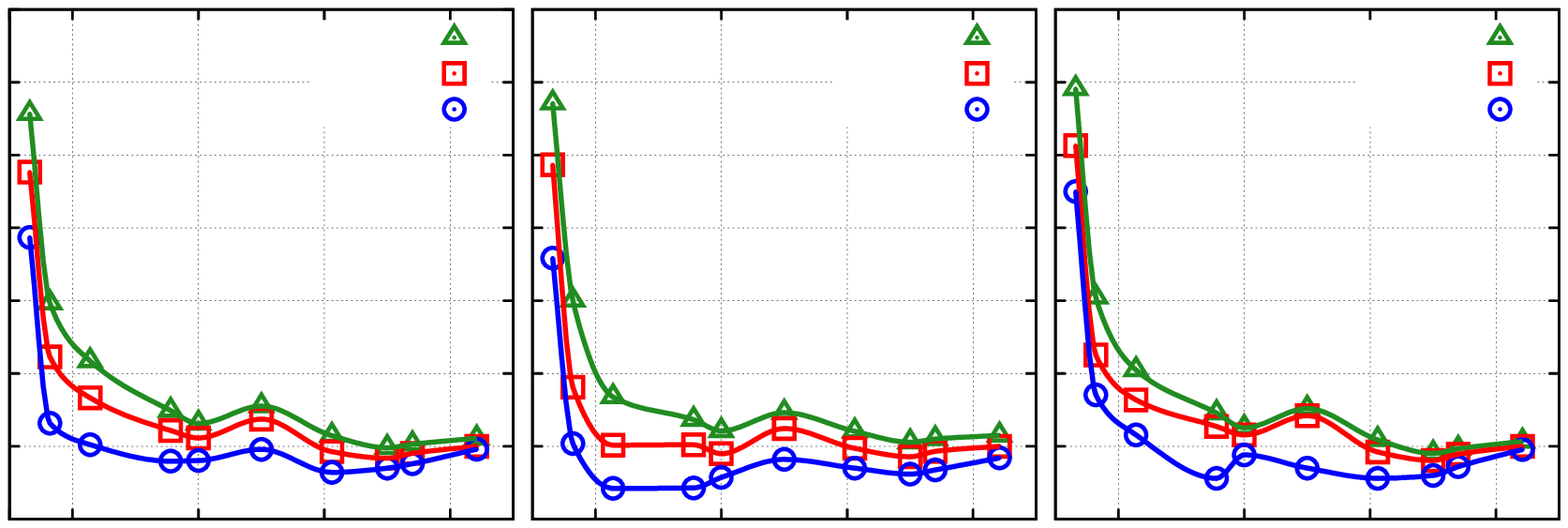}}%
    \gplfronttext
  \end{picture}%
\endgroup
}\\[2mm]
	\caption{\textbf{From apparent to effective properties in 3d}. Percentage deviation of the coefficients $\mathbb{C}_{11}$, $\mathbb{C}_{44}$ and $\mathbb{C}_{12}$ from the reference $\mathbb{C}_{ij}$(PBC,SRD371) for different microdomain sizes SRD$i$ subject to KUBC, PBC and SUBC.} 
	\label{fig:Apparent2Effective_RVE_size}
\end{Figure}

The diagrams in Fig.~\ref{fig:component-elast-tensor} indicate a direct dependence of $\mathbb{C}_{ij}$ on the phase fraction of the stiffest phase. Their larger values for small sample sizes is of course accidental since they depend on the selection in the domain of S371. Apart from that, the phase fraction ratio is almost ''converged'' at S64.

Figure~\ref{fig:Apparent2Effective_RVE_size} displays the percentage deviation of different coefficients $\mathbb{C}_{ij}$ for KUBC, SUBC and PBC, where the case of PBC(S371) serves as the reference. The type of convergence, for SUBC from below, for KUBC from above, indicate the transition from  apparent to effective properties. The order $\mathbb{C}$(KUBC) $\geq$ $\mathbb{C}$(PBC) $\geq$ $\mathbb{C}$(SUBC) is observed for all sizes. The characteristics show rapid convergence and are monotonic for small S, where S200 is an outlier already visible in Fig.~\ref{fig:component-elast-tensor}.

Since the concrete specimen is not periodic, its true deformation in the bulk of a structural element is unknown. For this reason the invariance of the calculated elasticity tensor to different BCs is of practical relevance in building materials and beyond, and similar transitions are observed in coupled problems, for e.g. in magneto-mechanics see Zabihyan et al. \cite{Zabihyan.2018}.

%---------------------------------------------------------------------------------------

\subsection{Validation of error estimation}

Two key questions are investigated now. Firstly, how does the estimated error match the actual one? Secondly, can the error estimation be validated by using smaller sample sizes where a reference solution is readily available? 
 
\begin{figure}[htbp]
	\centering
	\subfloat[estimated error $\bar e_{\text{mic}}$] 
	{\includegraphics[height=4.0cm, angle=0]{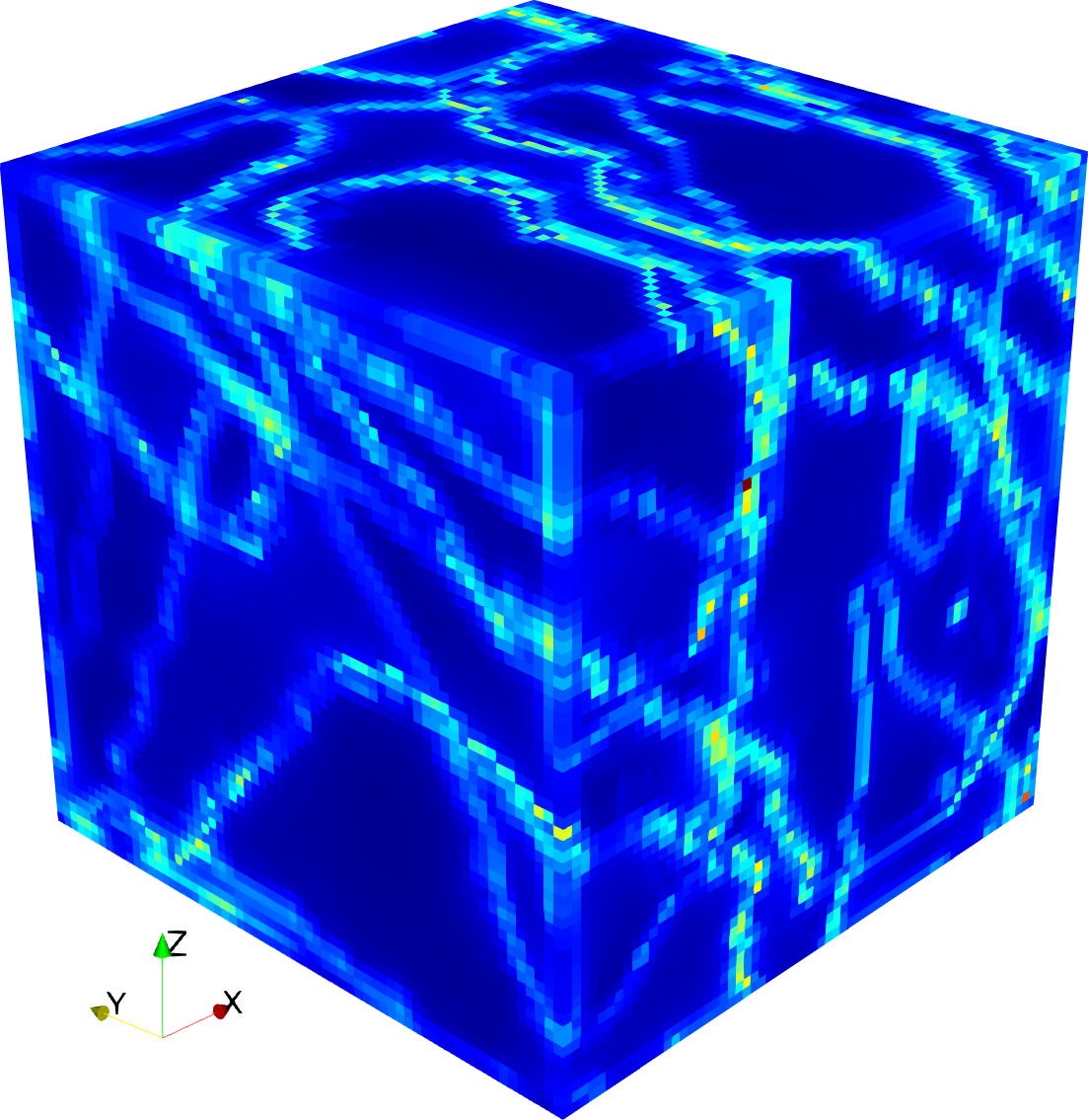}} 
	\hspace*{0.0\linewidth}
	\subfloat[$YZ$-plane]
	{\includegraphics[height=4.0cm, angle=0]{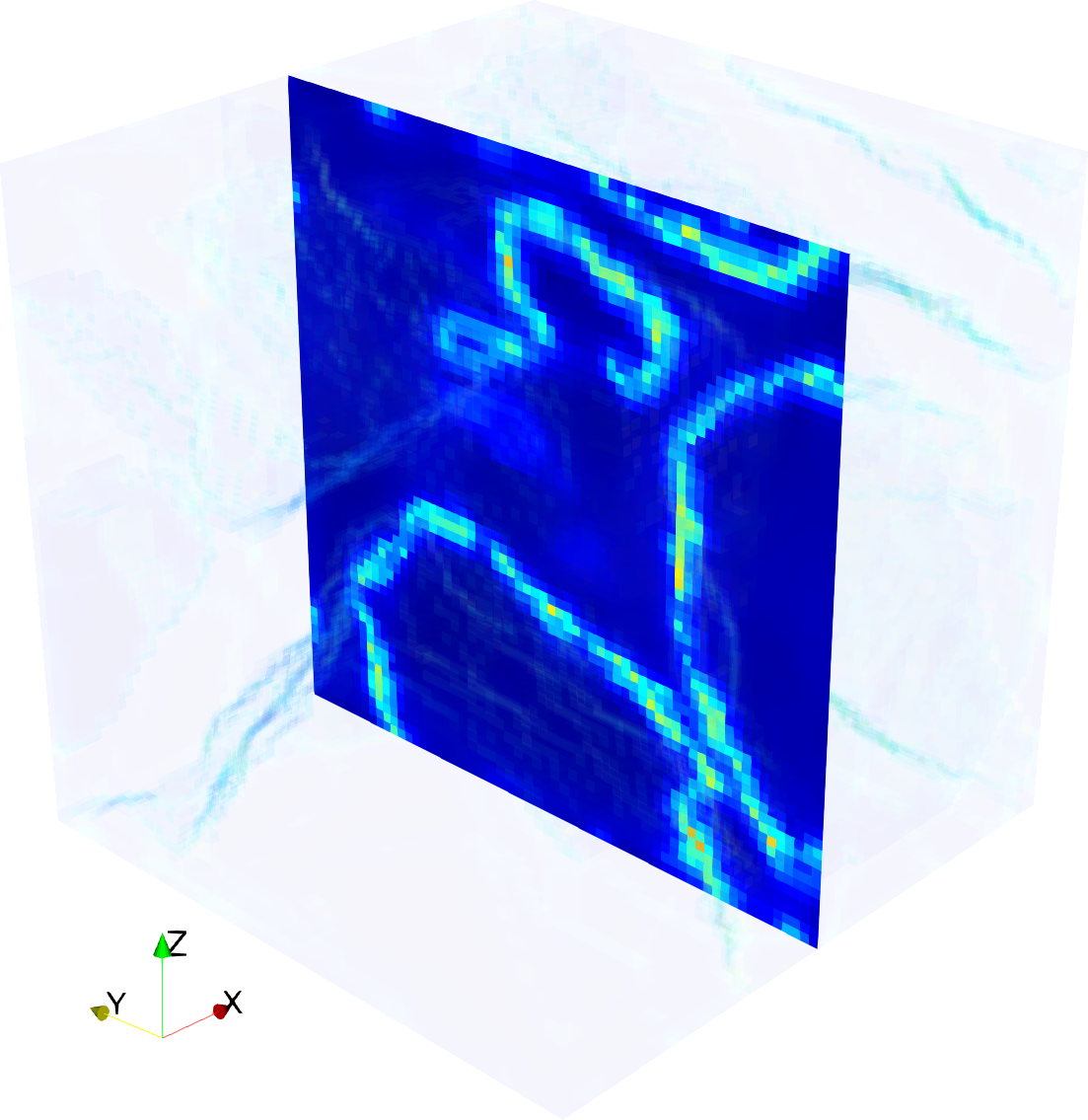}}
    \hspace*{0.0\linewidth}
	\subfloat[$XZ$-plane]
	{\includegraphics[height=4.0cm, angle=0]{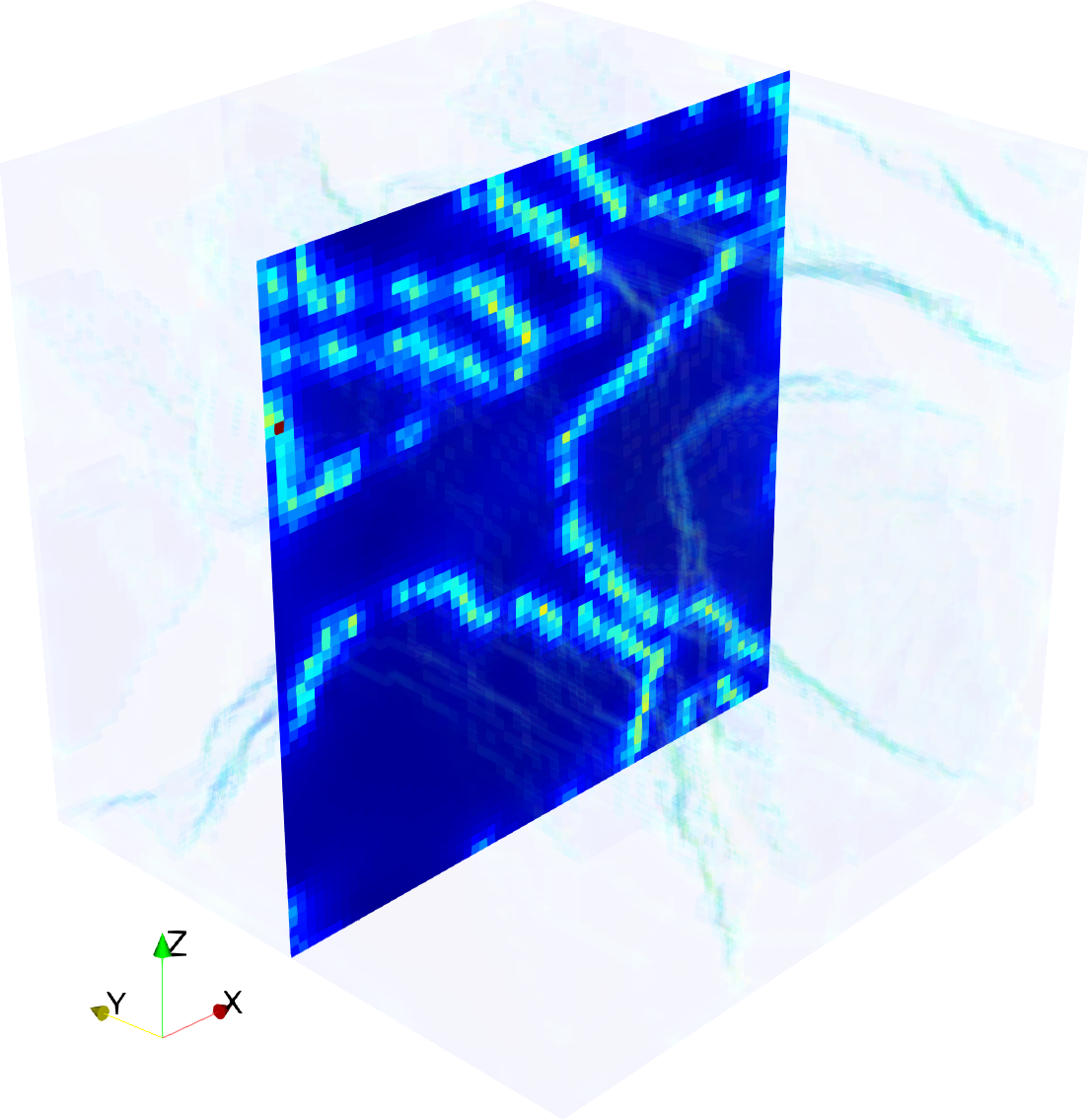}}
	\hspace*{0.0\linewidth}
    \subfloat[$XY$-plane]
	{\includegraphics[height=4.0cm, angle=0]{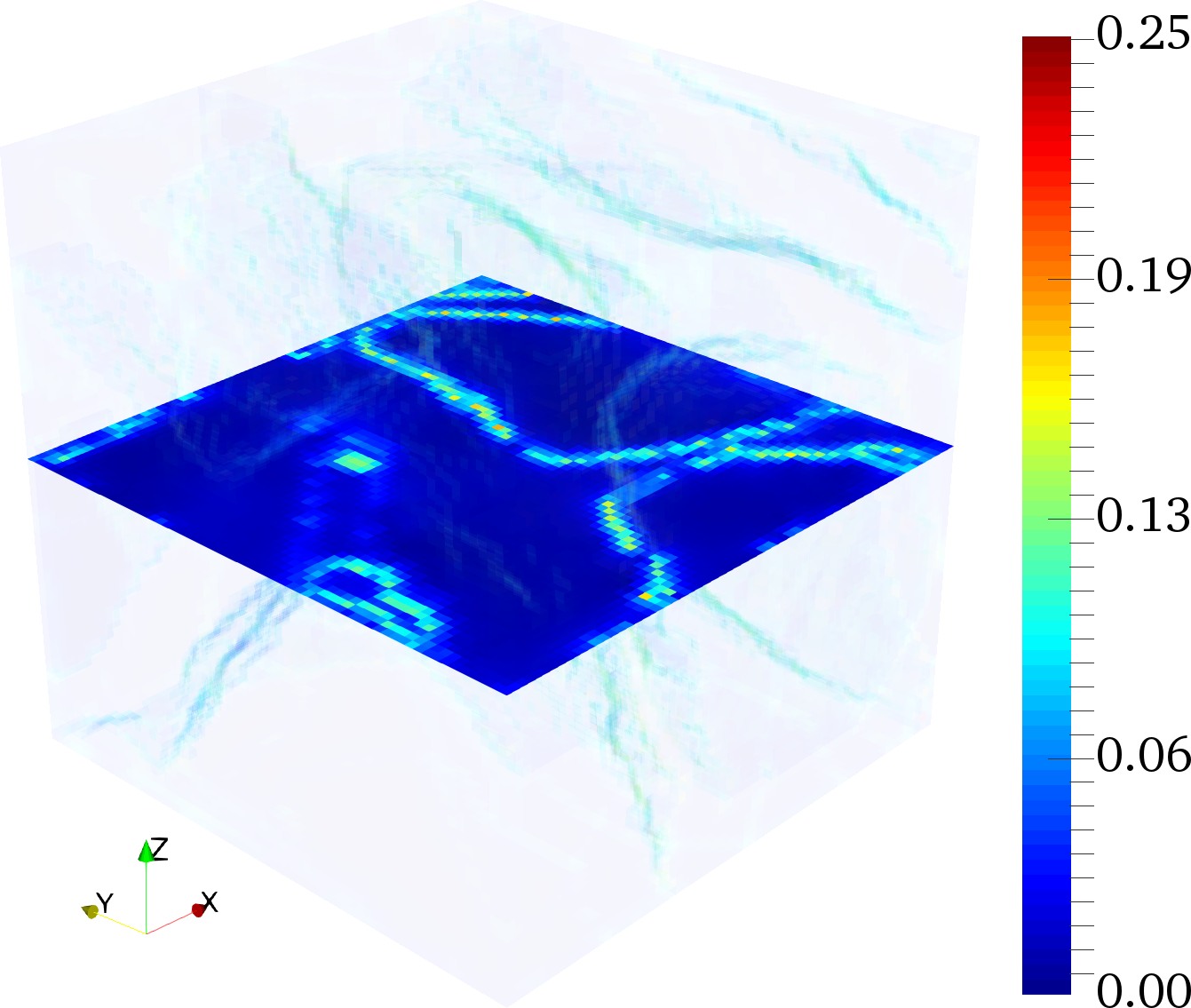}}	
	\\
	\subfloat[actual error $e_{\text{mic}}$]
	{\includegraphics[height=4.0cm, angle=0]{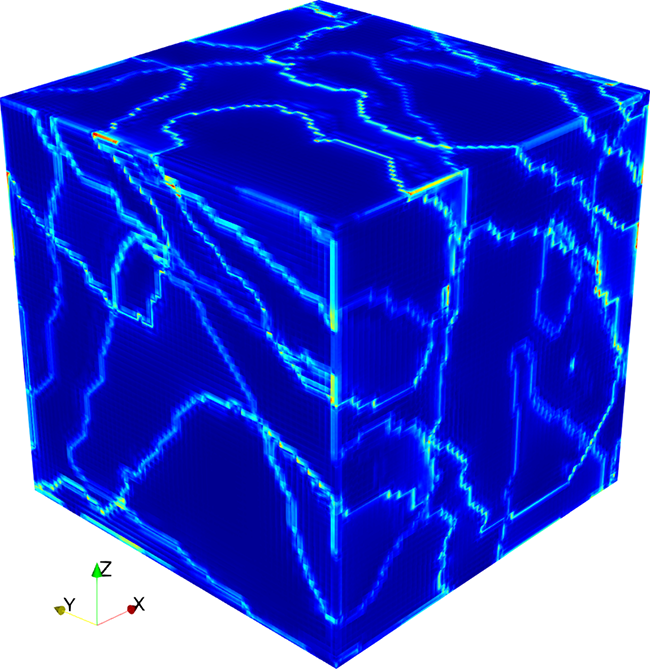}}
	\hspace*{0.0\linewidth}
	\subfloat[$YZ$-plane]
	{\includegraphics[height=4.0cm, angle=0]{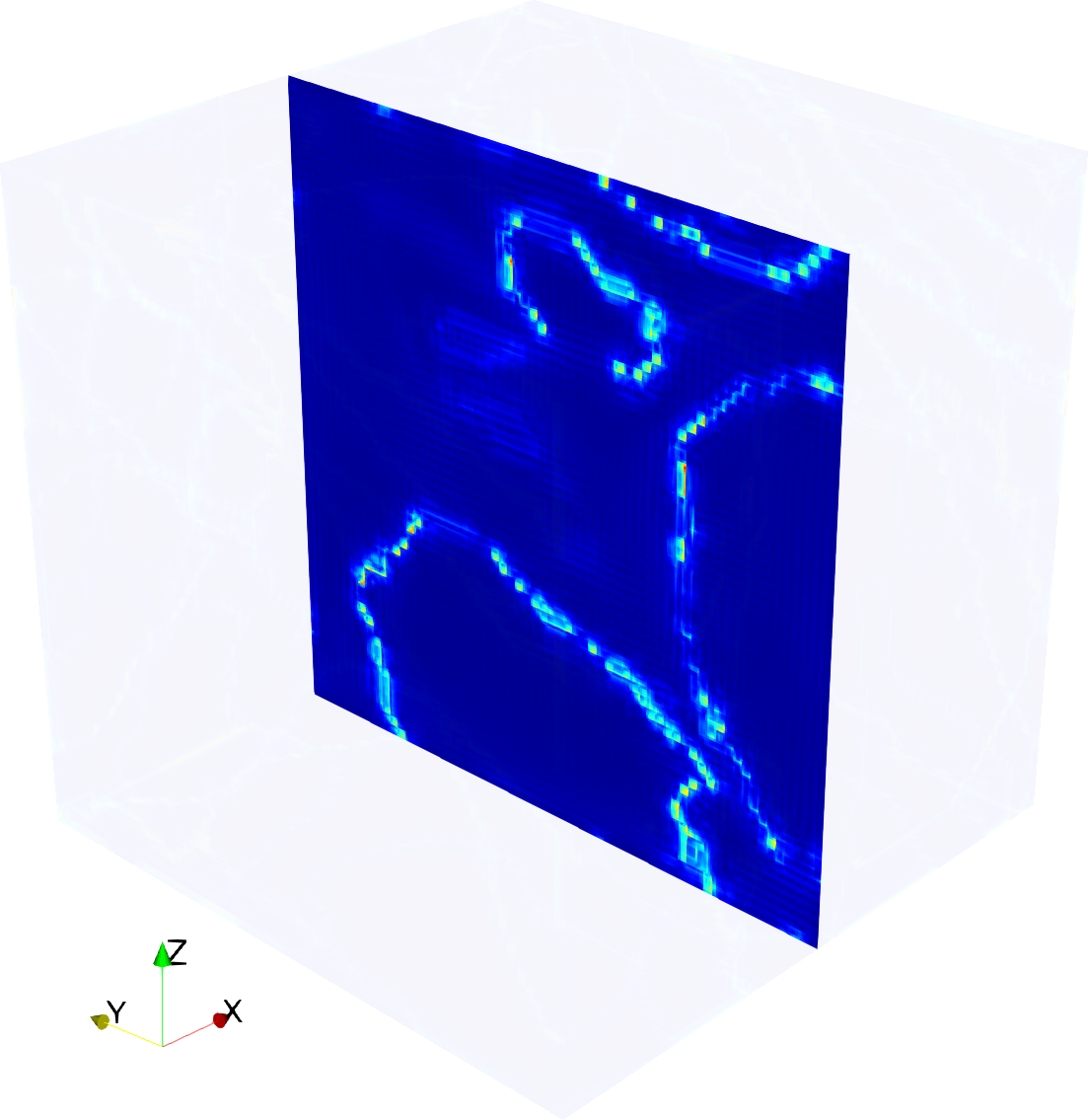}}
    \hspace*{0.0\linewidth}
	\subfloat[$XZ$-plane]
	{\includegraphics[height=4.0cm, angle=0]{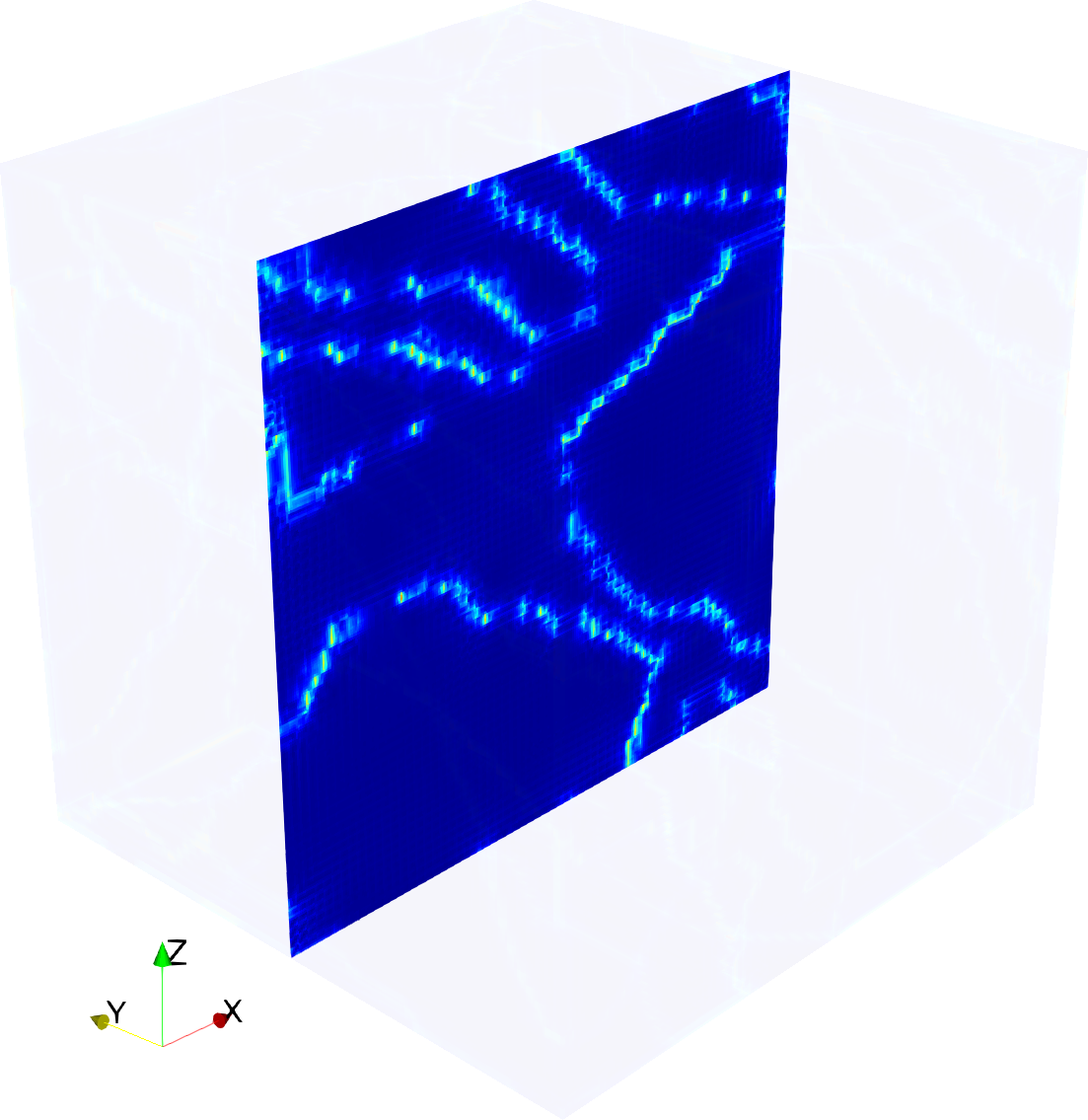}}
	\hspace*{0.0\linewidth}
    \subfloat[$XY$-plane]
	{\includegraphics[height=4.0cm, angle=0]{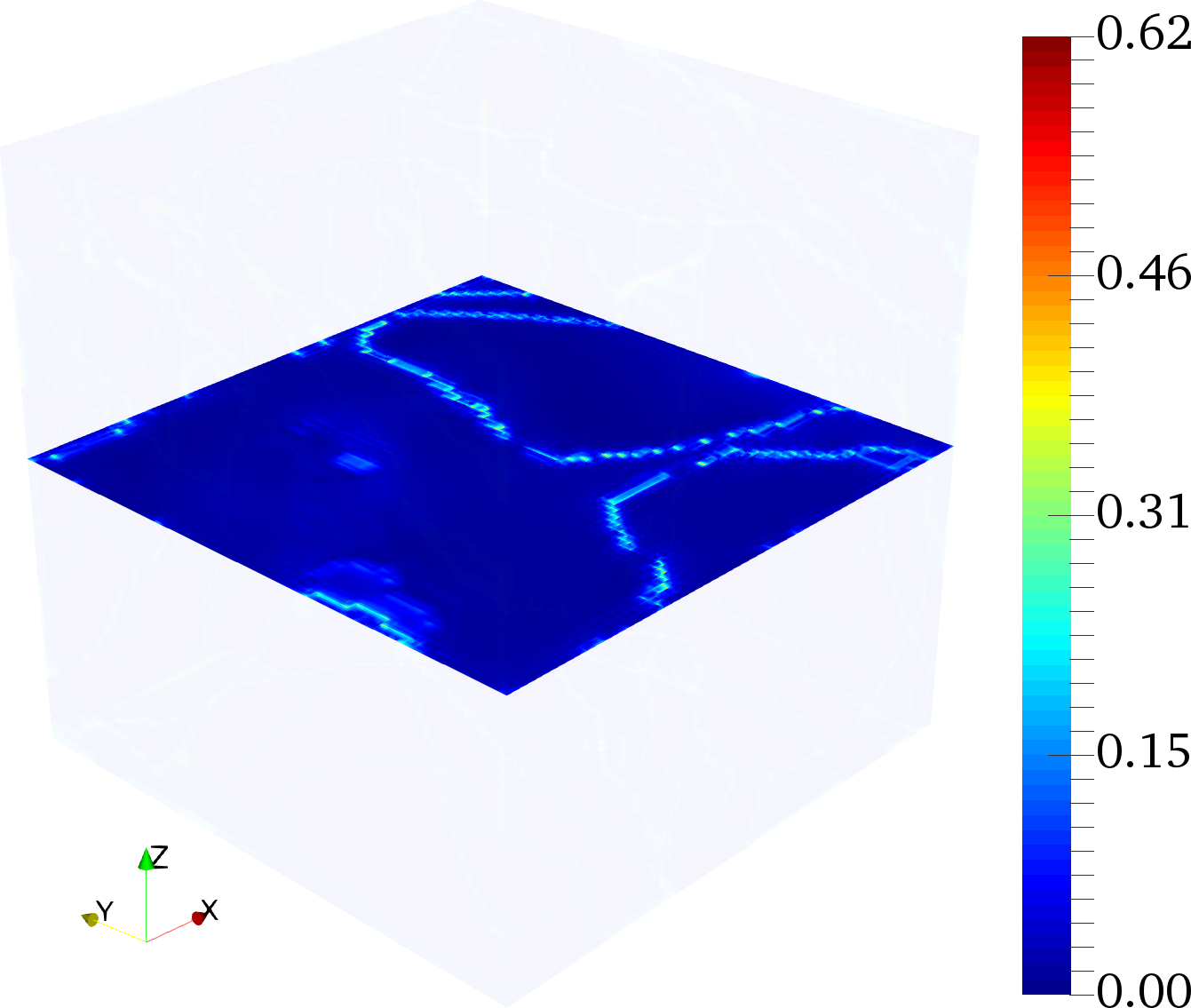}}
	\caption{\textbf{Error distributions for SRD64}. Estimated (a)--(d) versus actual (e)--(h) relative errors. Reference solutions for actual errors on SR64-D256.}
	\label{fig:Concrete_3d_32-64vx_Actual-vs-Estimted-Error}
\end{figure}

The results shown in Fig.~\ref{fig:Concrete_3d_32-64vx_Actual-vs-Estimted-Error} and listed in Tab.~\ref{tab:Concrete_error_3D} indicate the following error characteristics. 
\begin{itemize}
    \item {\bf Global micro errors}: The error estimation for the full microdomain is close to the actual error as indicated by effectivity indices $\Theta$ close to 0.8 at uniform meshes of sizes S32 and S64, for adaptively coarsened meshes it is throughout above 0.65. The fact that $\Theta$ is almost invariant for the two different samples, most notably at uniform discretization, suggests that the accuracy of error estimation can be transferred to larger sample sizes. In that case, the expensive computation of actual errors becomes dispensable. 
    \item {\bf Local micro error distributions}: The error distributions in the contour plots of 
    Fig.~\ref{fig:Concrete_3d_32-64vx_Actual-vs-Estimted-Error} indicate (i) the confinement of larger errors to interfaces and low errors in phase interiors. This behavior (ii) justifies the adaptive mesh-coarsening which preserves high resolution at interfaces but carries out coarse-graining  coarsening in the phase interior. In these characteristics, (iii) the error estimation agrees well with the error computation. The (iv) true errors are more confined to interfaces, less blurred and exhibit higher values. The reason is simply that the true errors are obtained with a much finer discretization, which localizes the stress and error concentrations to the interface.    
\end{itemize}

\begin{Table}[htbp]
	\begin{minipage}{15.5cm}  
		\centering
		\renewcommand{\arraystretch}{1.3} 
		\resizebox{0.55\columnwidth}{!}{%
		\begin{tabular}{r c c c c c }
			\hline
			       &   &     \multicolumn{4}{c}{Adaptive coarsening} \\  
		    Size   &   & $0$ & $1$ & $2$  & $3$ \\
            \hline
%		   	S16 & ndof      & $14\,739$  & $12\,144$ & -- & -- \\
%	         	   & factor  & $1.000$  & $0.8239$    & -- & --  \\
%	         	   \cmidrule{2-6} 
%				   & $e_{\text{mic}}$(D64)  & $1.8619$ & $2.1905$     & -- & --  \\
%		   	       & $\bar e_{\text{mic}}$(D16) & $1.3227$ & $1.4798$ & -- & --  \\
%				   & $\Theta$                & $0.7104$ & $0.6756$    & -- & --  \\
%	               & \multicolumn{5}{l}{ref. for $e_{\text{mic}}$ is D64.} \\	
%	        \hline
		   	S32 & ndof       & $107\,811$  & $62\,715$ & $61\,200$ & $61\,179$ \\
			       & Factor   & $1.000$     & $0.5817$  & $0.5677$  & $0.5675$ \\
			       \cmidrule{2-6} 
				   & $e_{\text{mic}}$(D128) & $4.6283$    & $5.3454$  & $6.2689$  & $6.6116$ \\
		   	       & $\bar e_{\text{mic}}$(D32) & $3.6734$    & $4.0361$  & $4.2783$  & $4.3090$ \\
				   & $\Theta$        & $0.7936$    & $0.7551$  & $0.6825$  & $0.6517$ \\
%                   & \multicolumn{5}{l}{ref. for $e_{\text{mic}}$ is D128.} \\    
		   	\hline
			S64 & ndof  & $823\,875$  & $436\,728$ & $424\,221$ & $424\,110$ \\
			       & Factor    & $1.000$     & $0.5301$   & $0.5149$   & $0.5148$ \\
			       \cmidrule{2-6} 
			       & $e_{\text{mic}}$(D256) & $11.2579$   & $12.2820$  & $12.9257$  & $13.0031$ \\
		   	       & $\bar e_{\text{mic}}$(D64) & $8.6605$    & $9.3428$   & $9.6798$   & $9.7025$ \\
				   & $\Theta$         & $0.7693$    & $0.7607$   & $0.7489$   & $0.7462$ \\
%		           & \multicolumn{5}{l}{ref. for $e_{\text{mic}}$ is D256.} \\
%			  	\hline
%			S128 & ndof                             & $6\,440\,067$  & $3\,521\,739$  & $3\,440\,115$ & $3\,439\,674$ \\
%	         	    & factor                           & $1.000$        & $0.5468$  & $0.5342$ & $0.5341$ \\
%		   	        & $\bar e^{\epsilon}_{\text{mic}}$ & $26.8484$      & $28.4945$  & $29.1221$ & $29.2352$  \\
%		    	\hline
%			S256  & ndof   & $50\,923\,779$  & $28\,904\,403$ & $28\,385\,649$ & $28\,384\,992$ \\
%	         	    & factor & $1.000$  & $0.5676$ & $0.5574$ & $0.5574$ \\
%		   	        & $\bar e^{\epsilon}_{\text{mic}}$ & $80.935$  & $y.zz$       & $y.zz$    & $y.zz$     \\
%			\hline
%			S320  & ndof   & $99\,228\,483$  & $xx\,...\,...$ & $..\,...\,...$ & $..\,...\,...$ \\
%	         	    & factor & $1.000$  & $0.cccc$ & $0.ffff$ & $0.dddd$ \\
%		   	        & $\bar e^{\epsilon}_{\text{mic}}$ & $xy.zz$  & $y.zz$       & $y.zz$    & $y.zz$     \\
			\hline
		\end{tabular} 
		}
	\end{minipage}
	\caption{\textbf{Specimen sizes SRD32, 64}. Number of unknowns (ndof), actual errors $e_{\text{mic}}$, estimated errors $\bar e_{\text{mic}}$, effectivity index $\Theta$ for the uniform mesh (referred to as 0) and the adaptively coarsened discretizations (1--3). The discretization for the reference solution in the error computation is $h^{\text{ref}}=h/4$. Errors in (Nmm).}
	\label{tab:Concrete_error_3D} 
\end{Table}

\subsection{Uniform coarsening of resolution}

The diagrams in Fig.~\ref{fig:component-elast-tensor_resolution-3d} display the changes in the homogenized elasticity tensor for both variants of the uniform resolution coarsening. For phase-preserving coarsening as shown in Fig.~\ref{fig:component-elast-tensor_resolution-3d}(b) the homogenized elastic constants follow the fraction of the stiffest phase during coarsening, which increase following the drop from RD256 to RD128. For the resolution coarsening taking the arithmetic mean of the elastic constants as shown in Fig.~\ref{fig:component-elast-tensor_resolution-3d}(a) the homogenized elastic constants continuously increase with the coarsening step\footnote{the volume average of the Young's modulus $<E>$ remains constant.}. In this behavior both variants are in good agreement with the 2d case. 
 
The dashed lines in Fig.~\ref{fig:component-elast-tensor_resolution-3d} refer to the case where resolution and discretization coincide, hence R=D, while the solid lines stand for the case where the discretization is kept constantly fine at D256. The comparison quantifies the influence of the discretization, which turns out to be relatively small compared to the impact of the resolution, for the phase preserving case somewhat stronger (see Fig.~\ref{fig:component-elast-tensor_resolution-3d}(b)) than for the case with new phases (see Fig.~\ref{fig:component-elast-tensor_resolution-3d}(a)). 

\begin{Figure}[htbp]
	\centering
	\subfloat[new phases, S256-RD\,$i$]
     {\resizebox{0.44\columnwidth}{!}{% GNUPLOT: LaTeX picture with Postscript
\begingroup
  % Encoding inside the plot.  In the header of your document, this encoding
  % should to defined, e.g., by using
  % \usepackage[cp1252,<other encodings>]{inputenc}
  \inputencoding{cp1252}%
  \makeatletter
  \providecommand\color[2][]{%
    \GenericError{(gnuplot) \space\space\space\@spaces}{%
      Package color not loaded in conjunction with
      terminal option `colourtext'%
    }{See the gnuplot documentation for explanation.%
    }{Either use 'blacktext' in gnuplot or load the package
      color.sty in LaTeX.}%
    \renewcommand\color[2][]{}%
  }%
  \providecommand\includegraphics[2][]{%
    \GenericError{(gnuplot) \space\space\space\@spaces}{%
      Package graphicx or graphics not loaded%
    }{See the gnuplot documentation for explanation.%
    }{The gnuplot epslatex terminal needs graphicx.sty or graphics.sty.}%
    \renewcommand\includegraphics[2][]{}%
  }%
  \providecommand\rotatebox[2]{#2}%
  \@ifundefined{ifGPcolor}{%
    \newif\ifGPcolor
    \GPcolortrue
  }{}%
  \@ifundefined{ifGPblacktext}{%
    \newif\ifGPblacktext
    \GPblacktextfalse
  }{}%
  % define a \g@addto@macro without @ in the name:
  \let\gplgaddtomacro\g@addto@macro
  % define empty templates for all commands taking text:
  \gdef\gplbacktext{}%
  \gdef\gplfronttext{}%
  \makeatother
  \ifGPblacktext
    % no textcolor at all
    \def\colorrgb#1{}%
    \def\colorgray#1{}%
  \else
    % gray or color?
    \ifGPcolor
      \def\colorrgb#1{\color[rgb]{#1}}%
      \def\colorgray#1{\color[gray]{#1}}%
      \expandafter\def\csname LTw\endcsname{\color{white}}%
      \expandafter\def\csname LTb\endcsname{\color{black}}%
      \expandafter\def\csname LTa\endcsname{\color{black}}%
      \expandafter\def\csname LT0\endcsname{\color[rgb]{1,0,0}}%
      \expandafter\def\csname LT1\endcsname{\color[rgb]{0,1,0}}%
      \expandafter\def\csname LT2\endcsname{\color[rgb]{0,0,1}}%
      \expandafter\def\csname LT3\endcsname{\color[rgb]{1,0,1}}%
      \expandafter\def\csname LT4\endcsname{\color[rgb]{0,1,1}}%
      \expandafter\def\csname LT5\endcsname{\color[rgb]{1,1,0}}%
      \expandafter\def\csname LT6\endcsname{\color[rgb]{0,0,0}}%
      \expandafter\def\csname LT7\endcsname{\color[rgb]{1,0.3,0}}%
      \expandafter\def\csname LT8\endcsname{\color[rgb]{0.5,0.5,0.5}}%
    \else
      % gray
      \def\colorrgb#1{\color{black}}%
      \def\colorgray#1{\color[gray]{#1}}%
      \expandafter\def\csname LTw\endcsname{\color{white}}%
      \expandafter\def\csname LTb\endcsname{\color{black}}%
      \expandafter\def\csname LTa\endcsname{\color{black}}%
      \expandafter\def\csname LT0\endcsname{\color{black}}%
      \expandafter\def\csname LT1\endcsname{\color{black}}%
      \expandafter\def\csname LT2\endcsname{\color{black}}%
      \expandafter\def\csname LT3\endcsname{\color{black}}%
      \expandafter\def\csname LT4\endcsname{\color{black}}%
      \expandafter\def\csname LT5\endcsname{\color{black}}%
      \expandafter\def\csname LT6\endcsname{\color{black}}%
      \expandafter\def\csname LT7\endcsname{\color{black}}%
      \expandafter\def\csname LT8\endcsname{\color{black}}%
    \fi
  \fi
    \setlength{\unitlength}{0.0500bp}%
    \ifx\gptboxheight\undefined%
      \newlength{\gptboxheight}%
      \newlength{\gptboxwidth}%
      \newsavebox{\gptboxtext}%
    \fi%
    \setlength{\fboxrule}{0.5pt}%
    \setlength{\fboxsep}{1pt}%
\begin{picture}(6802.00,5102.00)%
    \gplgaddtomacro\gplbacktext{%
      \csname LTb\endcsname%%
      \put(548,3979){\makebox(0,0)[r]{\strut{}$43$}}%
      \csname LTb\endcsname%%
      \put(548,4195){\makebox(0,0)[r]{\strut{}$44$}}%
      \csname LTb\endcsname%%
      \put(548,4412){\makebox(0,0)[r]{\strut{}$45$}}%
      \csname LTb\endcsname%%
      \put(548,4628){\makebox(0,0)[r]{\strut{}$46$}}%
      \csname LTb\endcsname%%
      \put(1768,3542){\makebox(0,0){\strut{} }}%
      \csname LTb\endcsname%%
      \put(2856,3542){\makebox(0,0){\strut{} }}%
      \csname LTb\endcsname%%
      \put(3944,3542){\makebox(0,0){\strut{} }}%
      \csname LTb\endcsname%%
      \put(5032,3542){\makebox(0,0){\strut{} }}%
    }%
    \gplgaddtomacro\gplfronttext{%
      \csname LTb\endcsname%%
      \put(5529,4672){\makebox(0,0)[r]{\strut{}$\mathbb{C}_{11}$}}%
      \csname LTb\endcsname%%
      \put(5529,4452){\makebox(0,0)[r]{\strut{}$\mathbb{C}_{22}$}}%
      \csname LTb\endcsname%%
      \put(5529,4232){\makebox(0,0)[r]{\strut{}$\mathbb{C}_{33}$}}%
    }%
    \gplgaddtomacro\gplbacktext{%
      \csname LTb\endcsname%%
      \put(548,2859){\makebox(0,0)[r]{\strut{}$17.5$}}%
      \csname LTb\endcsname%%
      \put(548,3220){\makebox(0,0)[r]{\strut{}$18.5$}}%
      \csname LTb\endcsname%%
      \put(548,3581){\makebox(0,0)[r]{\strut{}$19.5$}}%
      \csname LTb\endcsname%%
      \put(1768,2458){\makebox(0,0){\strut{} }}%
      \csname LTb\endcsname%%
      \put(2856,2458){\makebox(0,0){\strut{} }}%
      \csname LTb\endcsname%%
      \put(3944,2458){\makebox(0,0){\strut{} }}%
      \csname LTb\endcsname%%
      \put(5032,2458){\makebox(0,0){\strut{} }}%
    }%
    \gplgaddtomacro\gplfronttext{%
      \csname LTb\endcsname%%
      \put(-134,3219){\rotatebox{-270}{\makebox(0,0){\strut{}$\mathbb{C}_{ij}$ (in GPa)}}}%
      \csname LTb\endcsname%%
      \put(5529,3588){\makebox(0,0)[r]{\strut{}$\mathbb{C}_{12}$}}%
      \csname LTb\endcsname%%
      \put(5529,3368){\makebox(0,0)[r]{\strut{}$\mathbb{C}_{13}$}}%
      \csname LTb\endcsname%%
      \put(5529,3148){\makebox(0,0)[r]{\strut{}$\mathbb{C}_{23}$}}%
    }%
    \gplgaddtomacro\gplbacktext{%
      \csname LTb\endcsname%%
      \put(548,1865){\makebox(0,0)[r]{\strut{}$12.5$}}%
      \csname LTb\endcsname%%
      \put(548,2407){\makebox(0,0)[r]{\strut{}$13.5$}}%
      \csname LTb\endcsname%%
      \put(1768,1374){\makebox(0,0){\strut{} }}%
      \csname LTb\endcsname%%
      \put(2856,1374){\makebox(0,0){\strut{} }}%
      \csname LTb\endcsname%%
      \put(3944,1374){\makebox(0,0){\strut{} }}%
      \csname LTb\endcsname%%
      \put(5032,1374){\makebox(0,0){\strut{} }}%
    }%
    \gplgaddtomacro\gplfronttext{%
      \csname LTb\endcsname%%
      \put(5529,2505){\makebox(0,0)[r]{\strut{}$\mathbb{C}_{44}$}}%
      \csname LTb\endcsname%%
      \put(5529,2285){\makebox(0,0)[r]{\strut{}$\mathbb{C}_{55}$}}%
      \csname LTb\endcsname%%
      \put(5529,2065){\makebox(0,0)[r]{\strut{}$\mathbb{C}_{66}$}}%
    }%
    \gplgaddtomacro\gplbacktext{%
      \colorrgb{0.00,0.00,1.00}%%
      \put(548,781){\makebox(0,0)[r]{\strut{}$0.55$}}%
      \colorrgb{0.00,0.00,1.00}%%
      \put(548,1323){\makebox(0,0)[r]{\strut{}$0.57$}}%
      \csname LTb\endcsname%%
      \put(1768,290){\makebox(0,0){\strut{}256}}%
      \csname LTb\endcsname%%
      \put(2856,290){\makebox(0,0){\strut{}128}}%
      \csname LTb\endcsname%%
      \put(3944,290){\makebox(0,0){\strut{}64}}%
      \csname LTb\endcsname%%
      \put(5032,290){\makebox(0,0){\strut{}32}}%
    }%
    \gplgaddtomacro\gplfronttext{%
      \colorrgb{0.00,0.00,1.00}%%
      \put(-134,1052){\rotatebox{-270}{\makebox(0,0){\strut{}$\text{PF}$}}}%
      \csname LTb\endcsname%%
      \put(3400,-40){\makebox(0,0){\strut{}Resolution R}}%
      \csname LTb\endcsname%%
      \put(5529,1421){\makebox(0,0)[r]{\strut{}$\text{Aggregate phase fraction}$}}%
    }%
    \gplbacktext
    \put(0,0){\includegraphics{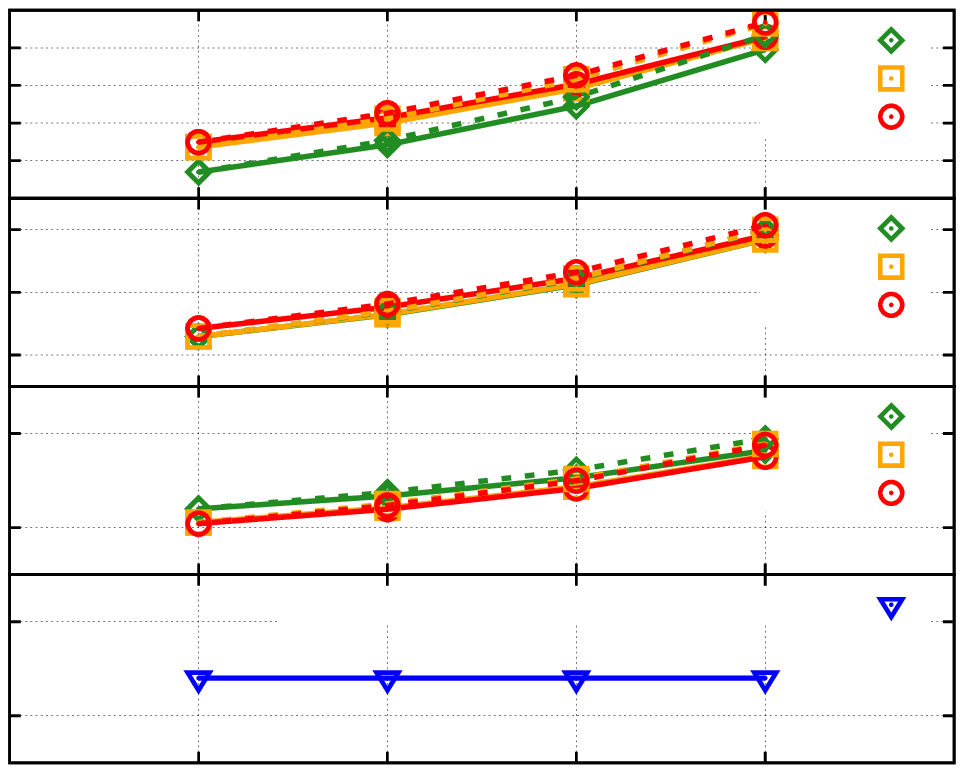}}%
    \gplfronttext
  \end{picture}%
\endgroup
}}
     \hspace*{4mm}
     \subfloat[phase preserving, S256-RD\,$i$]
	 {\resizebox{0.44\columnwidth}{!}{% GNUPLOT: LaTeX picture with Postscript
\begingroup
  % Encoding inside the plot.  In the header of your document, this encoding
  % should to defined, e.g., by using
  % \usepackage[cp1252,<other encodings>]{inputenc}
  \inputencoding{cp1252}%
  \makeatletter
  \providecommand\color[2][]{%
    \GenericError{(gnuplot) \space\space\space\@spaces}{%
      Package color not loaded in conjunction with
      terminal option `colourtext'%
    }{See the gnuplot documentation for explanation.%
    }{Either use 'blacktext' in gnuplot or load the package
      color.sty in LaTeX.}%
    \renewcommand\color[2][]{}%
  }%
  \providecommand\includegraphics[2][]{%
    \GenericError{(gnuplot) \space\space\space\@spaces}{%
      Package graphicx or graphics not loaded%
    }{See the gnuplot documentation for explanation.%
    }{The gnuplot epslatex terminal needs graphicx.sty or graphics.sty.}%
    \renewcommand\includegraphics[2][]{}%
  }%
  \providecommand\rotatebox[2]{#2}%
  \@ifundefined{ifGPcolor}{%
    \newif\ifGPcolor
    \GPcolortrue
  }{}%
  \@ifundefined{ifGPblacktext}{%
    \newif\ifGPblacktext
    \GPblacktextfalse
  }{}%
  % define a \g@addto@macro without @ in the name:
  \let\gplgaddtomacro\g@addto@macro
  % define empty templates for all commands taking text:
  \gdef\gplbacktext{}%
  \gdef\gplfronttext{}%
  \makeatother
  \ifGPblacktext
    % no textcolor at all
    \def\colorrgb#1{}%
    \def\colorgray#1{}%
  \else
    % gray or color?
    \ifGPcolor
      \def\colorrgb#1{\color[rgb]{#1}}%
      \def\colorgray#1{\color[gray]{#1}}%
      \expandafter\def\csname LTw\endcsname{\color{white}}%
      \expandafter\def\csname LTb\endcsname{\color{black}}%
      \expandafter\def\csname LTa\endcsname{\color{black}}%
      \expandafter\def\csname LT0\endcsname{\color[rgb]{1,0,0}}%
      \expandafter\def\csname LT1\endcsname{\color[rgb]{0,1,0}}%
      \expandafter\def\csname LT2\endcsname{\color[rgb]{0,0,1}}%
      \expandafter\def\csname LT3\endcsname{\color[rgb]{1,0,1}}%
      \expandafter\def\csname LT4\endcsname{\color[rgb]{0,1,1}}%
      \expandafter\def\csname LT5\endcsname{\color[rgb]{1,1,0}}%
      \expandafter\def\csname LT6\endcsname{\color[rgb]{0,0,0}}%
      \expandafter\def\csname LT7\endcsname{\color[rgb]{1,0.3,0}}%
      \expandafter\def\csname LT8\endcsname{\color[rgb]{0.5,0.5,0.5}}%
    \else
      % gray
      \def\colorrgb#1{\color{black}}%
      \def\colorgray#1{\color[gray]{#1}}%
      \expandafter\def\csname LTw\endcsname{\color{white}}%
      \expandafter\def\csname LTb\endcsname{\color{black}}%
      \expandafter\def\csname LTa\endcsname{\color{black}}%
      \expandafter\def\csname LT0\endcsname{\color{black}}%
      \expandafter\def\csname LT1\endcsname{\color{black}}%
      \expandafter\def\csname LT2\endcsname{\color{black}}%
      \expandafter\def\csname LT3\endcsname{\color{black}}%
      \expandafter\def\csname LT4\endcsname{\color{black}}%
      \expandafter\def\csname LT5\endcsname{\color{black}}%
      \expandafter\def\csname LT6\endcsname{\color{black}}%
      \expandafter\def\csname LT7\endcsname{\color{black}}%
      \expandafter\def\csname LT8\endcsname{\color{black}}%
    \fi
  \fi
    \setlength{\unitlength}{0.0500bp}%
    \ifx\gptboxheight\undefined%
      \newlength{\gptboxheight}%
      \newlength{\gptboxwidth}%
      \newsavebox{\gptboxtext}%
    \fi%
    \setlength{\fboxrule}{0.5pt}%
    \setlength{\fboxsep}{1pt}%
\begin{picture}(6802.00,5102.00)%
    \gplgaddtomacro\gplbacktext{%
      \csname LTb\endcsname%%
      \put(548,3979){\makebox(0,0)[r]{\strut{}$43$}}%
      \csname LTb\endcsname%%
      \put(548,4195){\makebox(0,0)[r]{\strut{}$44$}}%
      \csname LTb\endcsname%%
      \put(548,4412){\makebox(0,0)[r]{\strut{}$45$}}%
      \csname LTb\endcsname%%
      \put(548,4628){\makebox(0,0)[r]{\strut{}$46$}}%
      \csname LTb\endcsname%%
      \put(1768,3542){\makebox(0,0){\strut{} }}%
      \csname LTb\endcsname%%
      \put(2856,3542){\makebox(0,0){\strut{} }}%
      \csname LTb\endcsname%%
      \put(3944,3542){\makebox(0,0){\strut{} }}%
      \csname LTb\endcsname%%
      \put(5032,3542){\makebox(0,0){\strut{} }}%
    }%
    \gplgaddtomacro\gplfronttext{%
      \csname LTb\endcsname%%
      \put(5529,4672){\makebox(0,0)[r]{\strut{}$\mathbb{C}_{11}$}}%
      \csname LTb\endcsname%%
      \put(5529,4452){\makebox(0,0)[r]{\strut{}$\mathbb{C}_{22}$}}%
      \csname LTb\endcsname%%
      \put(5529,4232){\makebox(0,0)[r]{\strut{}$\mathbb{C}_{33}$}}%
    }%
    \gplgaddtomacro\gplbacktext{%
      \csname LTb\endcsname%%
      \put(548,2859){\makebox(0,0)[r]{\strut{}$17.5$}}%
      \csname LTb\endcsname%%
      \put(548,3220){\makebox(0,0)[r]{\strut{}$18.5$}}%
      \csname LTb\endcsname%%
      \put(548,3581){\makebox(0,0)[r]{\strut{}$19.5$}}%
      \csname LTb\endcsname%%
      \put(1768,2458){\makebox(0,0){\strut{} }}%
      \csname LTb\endcsname%%
      \put(2856,2458){\makebox(0,0){\strut{} }}%
      \csname LTb\endcsname%%
      \put(3944,2458){\makebox(0,0){\strut{} }}%
      \csname LTb\endcsname%%
      \put(5032,2458){\makebox(0,0){\strut{} }}%
    }%
    \gplgaddtomacro\gplfronttext{%
      \csname LTb\endcsname%%
      \put(-134,3219){\rotatebox{-270}{\makebox(0,0){\strut{}$\mathbb{C}_{ij}$ (in GPa)}}}%
      \csname LTb\endcsname%%
      \put(5529,3588){\makebox(0,0)[r]{\strut{}$\mathbb{C}_{12}$}}%
      \csname LTb\endcsname%%
      \put(5529,3368){\makebox(0,0)[r]{\strut{}$\mathbb{C}_{13}$}}%
      \csname LTb\endcsname%%
      \put(5529,3148){\makebox(0,0)[r]{\strut{}$\mathbb{C}_{23}$}}%
    }%
    \gplgaddtomacro\gplbacktext{%
      \csname LTb\endcsname%%
      \put(548,1865){\makebox(0,0)[r]{\strut{}$12.5$}}%
      \csname LTb\endcsname%%
      \put(548,2407){\makebox(0,0)[r]{\strut{}$13.5$}}%
      \csname LTb\endcsname%%
      \put(1768,1374){\makebox(0,0){\strut{} }}%
      \csname LTb\endcsname%%
      \put(2856,1374){\makebox(0,0){\strut{} }}%
      \csname LTb\endcsname%%
      \put(3944,1374){\makebox(0,0){\strut{} }}%
      \csname LTb\endcsname%%
      \put(5032,1374){\makebox(0,0){\strut{} }}%
    }%
    \gplgaddtomacro\gplfronttext{%
      \csname LTb\endcsname%%
      \put(5529,2505){\makebox(0,0)[r]{\strut{}$\mathbb{C}_{44}$}}%
      \csname LTb\endcsname%%
      \put(5529,2285){\makebox(0,0)[r]{\strut{}$\mathbb{C}_{55}$}}%
      \csname LTb\endcsname%%
      \put(5529,2065){\makebox(0,0)[r]{\strut{}$\mathbb{C}_{66}$}}%
    }%
    \gplgaddtomacro\gplbacktext{%
      \colorrgb{0.00,0.00,1.00}%%
      \put(548,618){\makebox(0,0)[r]{\strut{}$0.53$}}%
      \colorrgb{0.00,0.00,1.00}%%
      \put(548,835){\makebox(0,0)[r]{\strut{}$0.55$}}%
      \colorrgb{0.00,0.00,1.00}%%
      \put(548,1052){\makebox(0,0)[r]{\strut{}$0.57$}}%
      \colorrgb{0.00,0.00,1.00}%%
      \put(548,1269){\makebox(0,0)[r]{\strut{}$0.59$}}%
      \csname LTb\endcsname%%
      \put(1768,290){\makebox(0,0){\strut{}256}}%
      \csname LTb\endcsname%%
      \put(2856,290){\makebox(0,0){\strut{}128}}%
      \csname LTb\endcsname%%
      \put(3944,290){\makebox(0,0){\strut{}64}}%
      \csname LTb\endcsname%%
      \put(5032,290){\makebox(0,0){\strut{}32}}%
    }%
    \gplgaddtomacro\gplfronttext{%
      \colorrgb{0.00,0.00,1.00}%%
      \put(-134,1052){\rotatebox{-270}{\makebox(0,0){\strut{}$\text{PF}$}}}%
      \csname LTb\endcsname%%
      \put(3400,-40){\makebox(0,0){\strut{}Resolution R}}%
      \csname LTb\endcsname%%
      \put(5529,1421){\makebox(0,0)[r]{\strut{}$\text{Aggregate phase fraction}$}}%
    }%
    \gplbacktext
    \put(0,0){\includegraphics{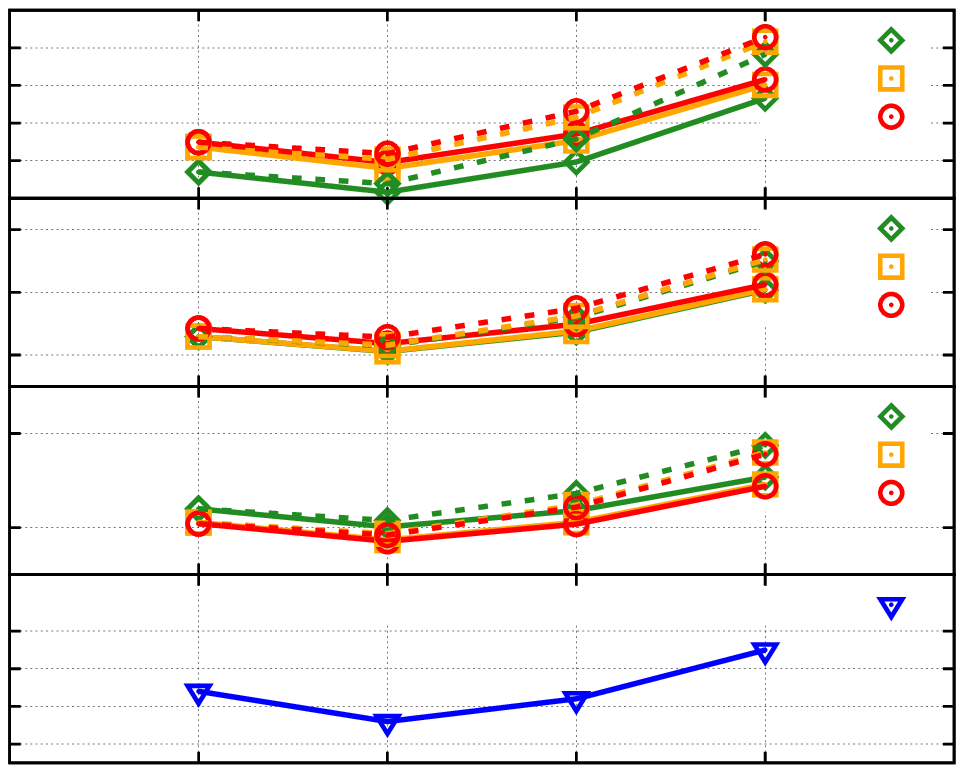}}%
    \gplfronttext
  \end{picture}%
\endgroup
}}
	\caption{\textbf{Resolution-dependent 3d elasticity}. Components $\mathbb{C}_{ij}$ of size S256 at different resolutions R. Dashed lines for the case that R and D coincide, hence S256-RD\,$i$, solid lines for the case of throughout constant D256. Corresponding numbers in the Supplement.} 
	\label{fig:component-elast-tensor_resolution-3d}
\end{Figure}

\begin{Table}[htbp]
\begin{minipage}{15.5cm}  
\centering
\renewcommand{\arraystretch}{1.3} 
\resizebox{0.65\columnwidth}{!}{%
	\begin{tabular}{r c c c c c }
	\hline
	& & \multicolumn{4}{c}{$\longrightarrow$ Uniform resolution coarsening $\longrightarrow$} \\
	& RD & $256$ & $128$ & $64$  & $32$ \\
	\hline
	S256 & ndof                               & $50\,923\,779$  & $6\,440\,067$ & $823\,875$ & $107\,811$ \\
 & Factor                                     & $1.000$ &  $0.1265$   &  $0.0162$   &  $0.0021$  \\
 & $\bar e_{\text{mic}}$                      & $80.935$  &  $110.536$  & $151.250$  & $191.304$  \\
 & $\bar e_{\text{mic}}/|| {\bm u} ||_{A(\mathcal{B}_{\epsilon})}$ &  $4.67\%$ &  $6.38\%$  & $8.73\%$  & $11.04\%$  \\
 & err[$\mathbb{C}_{11}$] & 0 \% & 0.72 \% & 2.06 \% & 7.33 \% \\
 & err[$\mathbb{C}_{12}$] & 0 \% & 0.82 \% & 1.71 \% & 6.78 \% \\
 & err[$\mathbb{C}_{44}$] & 0 \% & 0.96 \% & 1.29 \% & 5.32 \% \\
	\cmidrule{2-6} 
	& adap 1  & ndof                                       & $5\,261\,679$ &  $803\,556$ & $107\,778$ \\
	&         & Factor                                     &    $0.1033$   &  $0.0158$   & $0.0021$   \\
	&         & $\bar e_{\text{mic}}$                      &   $113.835$   &  $152.063$  & $191.308$  \\
	&         & $\bar e_{\text{mic}}/|| {\bm u} ||_{A(\mathcal{B}_{\epsilon})}$ &  $6.57\%$ &  $8.78\%$    & $11.04\%$  \\
	\cmidrule{3-6}
	& adap 2  & ndof                                       & $5\,259\,459$ & -- & -- \\
	&         & Factor                                     &   $0.1032$    & -- & -- \\
	&         & $\bar e_{\text{mic}}$                      &   $113.922$   & -- & -- \\
	&         & $\bar e_{\text{mic}}/|| {\bm u} ||_{A(\mathcal{B}_{\epsilon})}$ &   $6.58\%$   & -- & -- \\[1mm]
	\hline
	\end{tabular} 
}
\end{minipage}
\caption{\textbf{Resolution and mesh coarsening of SRD256}. Numbers of ndof reduction and increase of (relative) estimated errors for SRD256 its derived variants for phase-preserving resolution coarsening and adaptive mesh coarsening. Energies in (Nmm). SRD256 is the reference for $|| {\bm u} ||_{A(\mathcal{B}_{\epsilon})}$ and for err[$\mathbb{C}_{ij}$].}
\label{tab:Concrete_error_256_uni-adap} 
\end{Table}

The data for the ndof and the absolute and relative energy errors in Tab.~\ref{tab:Concrete_error_256_uni-adap} complement the information in Fig.~\ref{fig:component-elast-tensor_resolution-3d}. By resolution coarsening the ndof can be drastically reduced, while the relative energy errors remain small, even for S32 they are hardly above 11\%. Here and for error calculation in the components of the homogenized elasticity tensor, SRD256 is chosen as the reference. 
As an alternative route to consecutive resolution coarsening, two adaptive mesh coarsening steps are carried out for the resolution RD128, hence S256-RD128adap1/2 which further realizes an ndof reduction at very small error increase, for corresponding numbers see the second and third block of Tab. \ref{tab:Concrete_error_256_uni-adap}. In the same table the results for S256RD64/32 are displayed; in these two cases only one adaptive mesh coarsening step can be carried out. Generally, the ndof-reduction by adaptive mesh-coarsening is very limited for resolutions R128 and below.

\begin{figure}[htbp]
	\centering
	\subfloat[SRD256]
	{\includegraphics[height=4.0cm, angle=0]{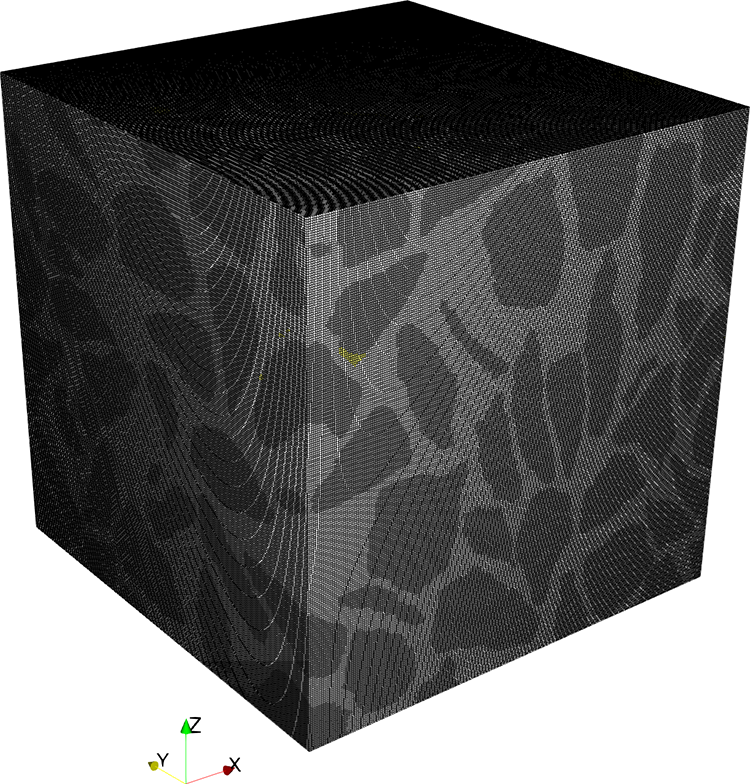}} 
	\hspace*{0.0\linewidth}
	\subfloat[S256-RD128]
	{\includegraphics[height=4.0cm, angle=0]{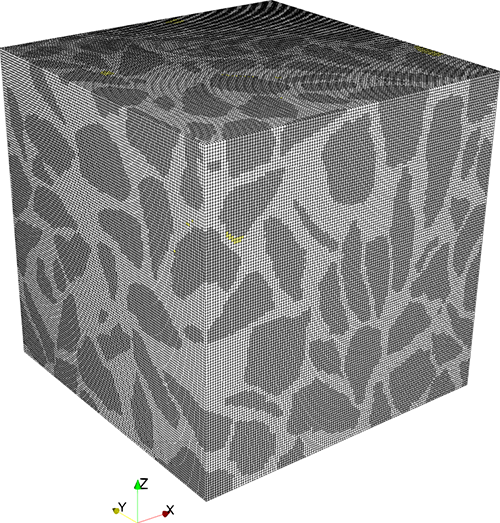}}
    \hspace*{0.0\linewidth}
	\subfloat[S256-RD64]
	{\includegraphics[height=4.0cm, angle=0]{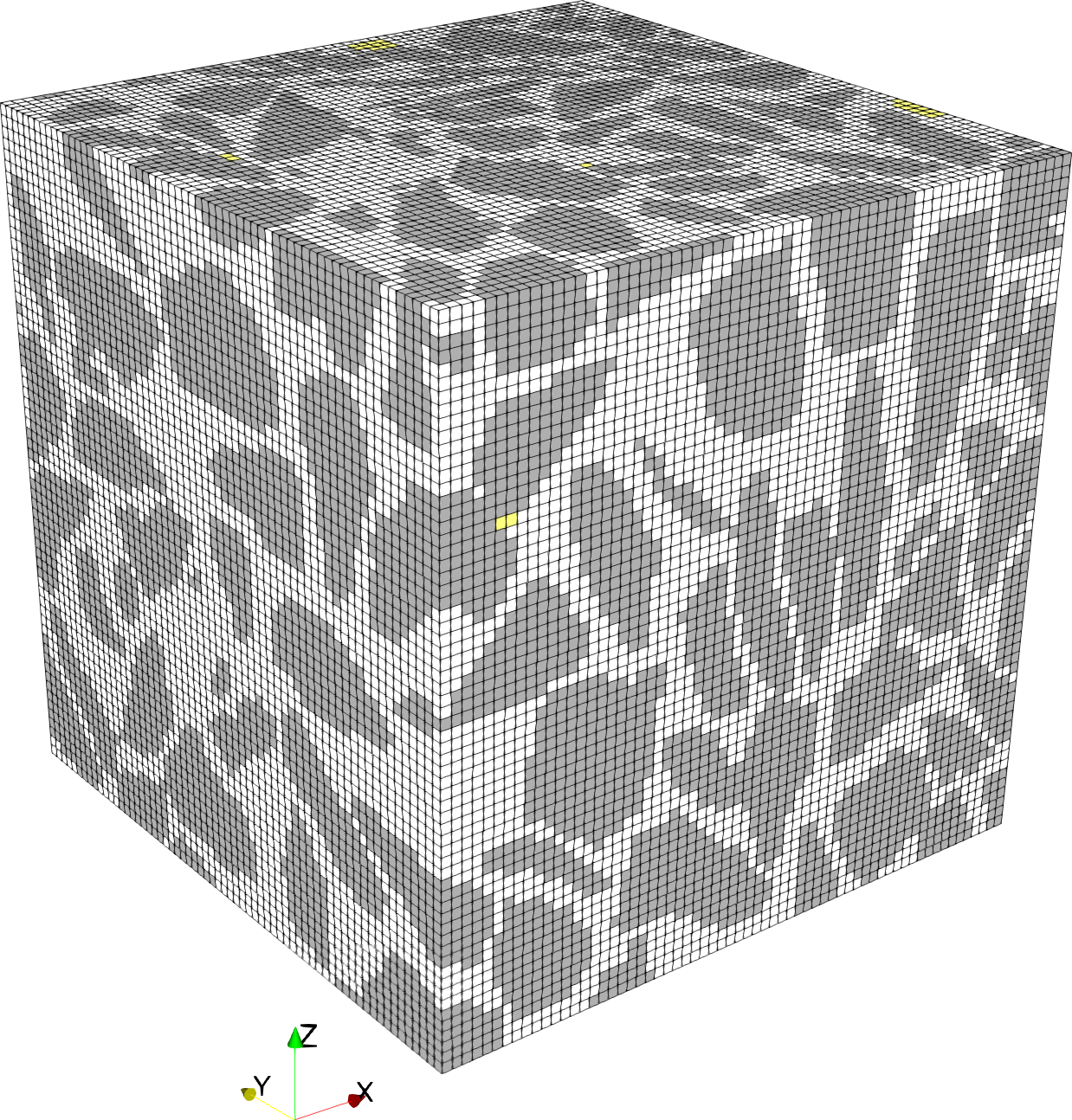}}
	\hspace*{0.0\linewidth}
    \subfloat[S256-RD32]
	{\includegraphics[height=4.0cm, angle=0]{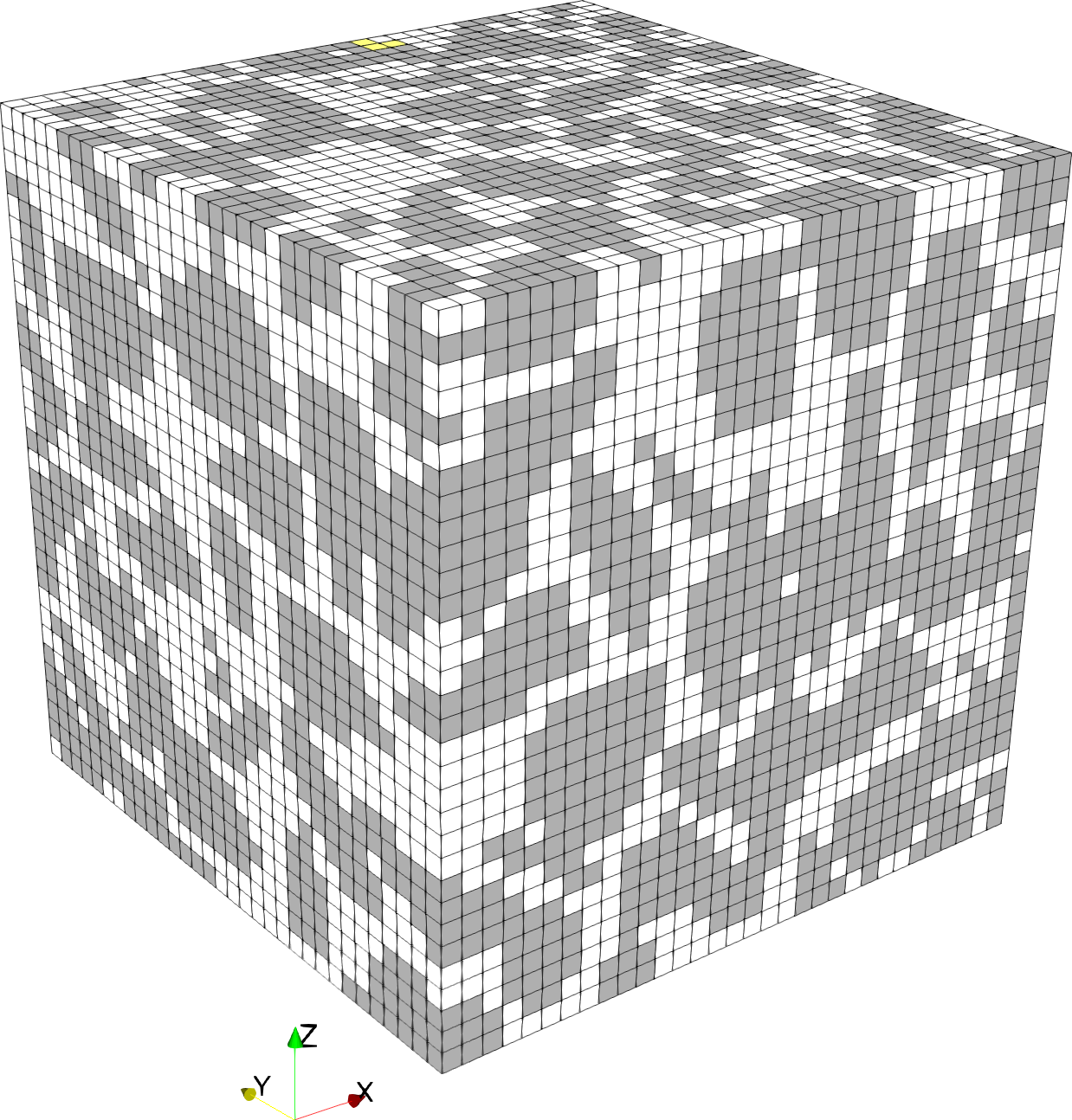}}	
	\\[2mm]
	\hspace*{80mm}
	\subfloat[S256-RD128adap1]
	{\includegraphics[height=4.0cm, angle=0]{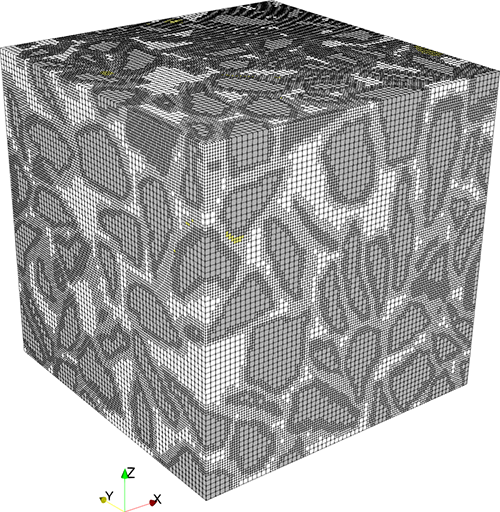}} \hspace*{0.0\linewidth}
	\subfloat[S256-RD128adap2]
	{\includegraphics[height=4.0cm, angle=0]{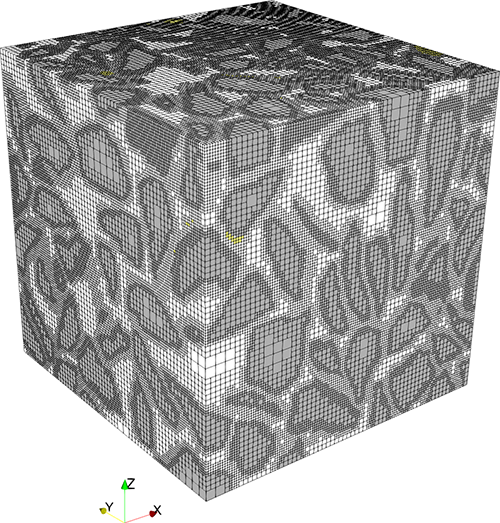}}
	\caption{\textbf{Coarsening 3d}. Resolution and discretization coarsening with coinciding RD for the phase-preserving case in (a)--(d). In (e), (f) two consecutive adaptive mesh coarsening steps are displayed starting from the phase-preserving, uniform case S256-RD128.}
	\label{fig:Concrete_256voxel_uniform_coarsening}
\end{figure}

Figures \ref{fig:Concrete_256voxel_uniform_coarsening}(a)--(d) show the outcome of the consecutive RD coarsening, the images (e) and (f) display the adaptive discretizations based on S256-RD128.
  
\subsection{Check of isotropy}

The lack of pronounced material directions in the two-phase composite suggests an isotropy. The specimen size S256 is chosen to measure its deviation from elastic isotropy, since it is sufficiently large and exhibits ''converged'' elastic properties. In the 3d case, the linear elastic constitutive law is given by    

\begin{equation}
\left[\begin{array}{c}
\varepsilon_{11} \\ \varepsilon_{22} \\ \varepsilon_{33} \\ 2\varepsilon_{12} \\ 2\varepsilon_{23} \\ 2\varepsilon_{13}
\end{array}\right]=
\displaystyle \underbrace{\left(\begin{array}{c c c c c c}
\phantom{-}\frac{1}{E} & -\frac{\nu}{E} &  -\frac{\nu}{E}  &  0  & 0 & 0   \\ 
-\frac{\nu}{E}  & \phantom{-}\frac{1}{E}  &  -\frac{\nu}{E}  & 0  & 0   &  0  \\ 
-\frac{\nu}{E}  &    -\frac{\nu}{E}       & \phantom{-}\frac{1}{E} &           0         &      0              &      0            \\ 
            &                          &                           & \frac{2(1+\nu)}{E} &        0            &     0             \\
            &                          &                           &                    & \frac{2(1+\nu)}{E} &      0            \\
   sym.         &                          &                           &                    &                    & \frac{2(1+\nu)}{E}   
\end{array}\right)}_{\substack{\text{Elastic compliance matrix \,} \mathbf{\mathbb S} \\ \text{for isotropy}}} 
\left[\begin{array}{c}
\sigma_{11} \\ \sigma_{22} \\ \sigma_{33} \\ \sigma_{12} \\ \sigma_{23} \\ \sigma_{13}
\end{array}\right]
\label{3D-isotropy}
\end{equation}

\begin{Table}[htbp]
	\begin{minipage}{15.5cm}  
		\centering
		\renewcommand{\arraystretch}{1.3} 
		\resizebox{1.00\columnwidth}{!}{
	\begin{tabular}{cc|cccccccc}
		\hline 
		 \multicolumn{2}{c}{Identification} & \multicolumn{8}{c}{Check of isotropy} \\
		\hline
		\rule{0pt}{25pt}
		 $E$ (MPa)  & $\nu$  & $\displaystyle \frac{\mathbb C_{11}-\mathbb C_{22}}{\mathbb C_{11}}$ & 
		$\displaystyle \frac{\mathbb C_{11}-\mathbb C_{33}}{\mathbb C_{11}}$ &
		$\displaystyle \frac{\mathbb C_{44}-G}{G}$ &
		$\displaystyle \frac{\mathbb C_{55}-G}{G}$ & 
		$\displaystyle \frac{\mathbb C_{66}-G}{G}$ &
		$\displaystyle \frac{\mathbb C_{14}}{\mathbb C_{11}}$ & 
		$\displaystyle \frac{\mathbb C_{24}}{\mathbb C_{11}}$ &
		$\displaystyle \frac{\mathbb C_{34}}{\mathbb C_{11}}$ \\

		\hline 
		
		 32374.7 & 0.29 &  1.56$\%$ 
		&  1.86$\%$
		&  1.26$\%$
		&  0.11$\%$
		&  0.02$\%$  
		&  0.00$\%$ 
		&  0.03$\%$ 
		&  0.02$\%$    \\
		\hline
	\end{tabular} 
}
\end{minipage}
	\caption{\textbf{Specimen SRD256}. Identified isotropic material parameters and the deviation [$\%$] of the remaining elastic coefficients from those of isotropy.} 	
	\label{tab:Concrete256vox-Tensor-Ident-and-Check}
\end{Table}
Isotropic elasticity is characterized by two independent material parameters $E$ and $\nu$ which are identified by the homogenized coefficients $\mathbb S_{11}$ and $\mathbb S_{12}$, the shear modulus follows through $G=E/(2(1+\nu))$. The remaining coefficients are used to validate the hypothesis of isotropy, which turns out to be justified by the slight deviations throughout below 2 \% in Tab. \ref{tab:Concrete256vox-Tensor-Ident-and-Check}.

\section{Results and Discussion}
On the search for an error-controlled reduction of the computational costs of the microstructure sample having 371$^3$ voxels, the present analysis has introduced and explored the SRD parameter space with the following results:
\begin{enumerate}
    \item {\bf Size S.} For the identification of a necessary specimen size for an RVE we have used three criteria, (i) phase fraction ratio, (ii) homogenized components of the elasticity tensor, (iii) invariance of the elastic response with respect to the applied BCs. The results for the homogenized components of the elasticity tensor in Fig.~\ref{fig:component-elast-tensor} indicate that the homogenized elastic constants correlate approximately with the phase fraction of the stiffest phase, the aggregate phase. From the sample size S64 the phase fractions and therefore the elasticities remain almost unaltered up to the size S371. This suggests that already the considered S64 can provide a representative volume size for characterizing the homogenized elastic behavior. {\color{black} The same is true with respect to criterion (iii) where we see in 3d that at the size S128 (with an outlier for S200) and above the maximal deviations between the different BCs are lower than 3\%. For deviations below 1\% it requires a S256 specimen.} For the case of S256 in 3d and S320 in 2d we can show that the effective elasticity law is virtually isotropic.  
    \item {\bf Resolution coarsening R.} It enables a considerable efficiency gain at controlled accuracy. The reduction of SRD256 to S256RD32 reduces the ndof from almost 51 millions to less than 0.11 million, while the relative error in the energy norm increases from 4.7\% to 11.1\% as shown in  Tab.~\ref{tab:Concrete_error_256_uni-adap}, and in the main components of the homogenized elasticity tensor  less than {\color{black} 7.5~\%} throughout, compare data in Fig.~  \ref{fig:component-elast-tensor_resolution-3d}.
    \item {\bf Discretization adapted to microstructure D.} For adaptive mesh coarsening of the samples S64 to S256, it is mainly the first step which realizes the best efficiency-accuracy trade-off, since it results in the strongest ndof-reduction of 53\%--57\% at rather moderate error increases.  
    \\
    The comparison of the corresponding error distributions in Fig.~\ref{fig:Concrete_3d_32-64vx_Actual-vs-Estimted-Error} reveals that the discretization errors are confined to aggregate-mortar interfaces but remain small in the interior of the phases, which justifies the microstructure-guided coarsening strategy. In the error distributions, the estimation is in its qualitative picture similar to actual errors, although local maximal values are underestimated.  
    Combinations of the RD coarsening fully unfolds the potential of the ndof reduction. 
    \item {\bf Quality of error estimation? Best practice of assessment?} The accuracy of error estimation is assessed for the sample sizes of 32 and 64 voxels per edge. The reference solutions are obtained on the corresponding meshes with $h^{\text{ref}}=h/4$. The effectivity index in the range of 0.7 to 0.8 indicates a sufficient and almost size-independent accuracy of the reconstruction-based error estimate as shown in Tab.~\ref{tab:Concrete_error_3D}. The results suggest that the validation of error estimation at smaller subvolumes is transferable to larger sample sizes, if the phase fractions and interface structure are approximately representative for the smaller sample sizes. 
    \item {\bf Transfers from 2d analyses to 3d valid?} 
    Since a single 2d slice can not be statistically representative for the full 3d specimen, conclusions from 2d to 3d with respect to the sample size, the required resolution and discretization are generally questionable. 
    A new finding and very useful for applications is that the accuracy of error estimation for 2d samples is very close to that for the 3d case, and the corresponding effectivity indices are in the range of 0.7 to 0.8. Hence, the transferability from 2d error analysis to 3d is possible, which implies a drastic reduction of the computational costs in 3d, since the true error computation based on very fine discretizations can be replaced. 
    \item {\bf Relevance in practice.} Real concrete volumes in the bulk of structural elements are hardly periodic and therefore do hardly deform according to PBC which is equally the case for direct numerical simulations (DNS). Hence, the particular BCs (KUBC, PBC, SUBC) replace in homogenization analyses, what is truly unknown by an idealized, artificial setup. For this reason the demonstrated invariance of the effective elastic response of an RVE to the applied BCs is of considerable relevance in practice; it brings the results back from the cyber into the physical world.
\end{enumerate}
 
%, where effective properties in the bulk are hardly accessible by experiments.  
\section{Conclusions and Outlook}

The present work proposes a concept for error-controlled reduction of the computational complexity for the numerical homogenization analysis of image-based multiphase composite microstructures. Reduction is achieved in the so-called SRD parameter space of size, resolution and discretization, and combinations thereof. This concept is applied to the highly resolved voxel-representation of a concrete specimen. Key criteria for reductions in the elastic analysis of homogenization are the (i) phase fraction ratio, (ii) homogenized elasticity tensor components, (iii) invariance of the elastic response with respect to the applied BCs, and (iv) total errors as well as error distributions in the energy norm.

For the elastic analyses of the concrete sample the best practice turns out to be a combination of coarsening the voxel resolution along with an additional adaptive coarsening of the finite element mesh. For the latter, adaptivity preserves the high resolution at mortar-aggregate interfaces and carries out mesh coarsening only in the interior of both phases. The achieved computational savings are considerable and monitored by validated error estimation.  

The present work considers the elastic properties of the analyzed microstructure not only at a few spots but rather for a broad range in the SRD parameter space. Therefore it is a first, with respect to elastic properties, a comprehensive step towards a digital twin of a representative concrete specimen in homogenization analysis. 
\\
In databases currently under construction or rapidly increasing in numbers and volume, digital twins obtain their properties typically from both physical and cyber worlds. Which properties are relevant and reliable? Largest errors and deviations among different choices of R and D are observed not in the homogenized quantities, but locally in the energy errors at interfaces (see e.g. Fig.\ref{fig:Concrete_3d_32-64vx_Actual-vs-Estimted-Error}). Energy/stress concentrations are typically the sources of inelastic mechanisms up to failure. Hence R and D are important for discretization methods and therefore shall be incorporated as object properties of a material's digital twin in databases. Corresponding ontologies\footnote{e.g. the European Materials Modelling Ontology EMMO (https://emmc.info/emmo-info/)}, which bring a structure in databases and enable a fast finding, should equally account for that.  

Future work shall analyze how the coarsening in voxel resolution and finite element discretization does influence the material response taking into account the inelastic behavior. Then it will become clear, whether and to which extent the presented elastic pre-analyses are guiding for the inelastic case. Since for concrete the weakest link occurs at the mortar-aggregate interfaces, the nucleation and propagation of cracks is expected in these regions. For this case the high resolution preserved at interfaces through the present adaptive mesh-coarsening is a valuable feature for the sake of accuracy.  

Notwithstanding, and it is also a limitation of the present work, that the spatial variations in the microstructure of composites may have a minor influence on the elastic properties but a strong impact on the overall inelastic behavior \cite{Geers.2010}. Modeling the inelastic behavior of concrete must account for effects of combined damage-plasticity, cracking, phase separation and decohesion, scatter in properties, see e.g. Unger and Eckardt \cite{Unger.2011}, Snozzi et al. \cite{Snozzi.2011}, Nguyen et al. \cite{Nguyen.2012}, Grassl et al. \cite{Grassl.2012}, Huang et al. \cite{Huang.2015}, Wang et al. \cite{Wang.2015}, Oliver et al. \cite{Oliver.2017}, Liu et al. \cite{Liu.2018}, Thilakarathna et al. \cite{Thilakarathna.2020}, Aldakheel \cite{Aldakheel.2020} to name but a few of many previous relevant works.

Beyond the particular material system of concrete considered here, the present work describes a generally applicable concept for the systematic and error-controlled reduction of the computational complexity of numerical homogenization analysis of image-based, multiphase composite microstructures. 

%Finally it shall be pointed out, how datasets of voxelized microstructures from tomography, which are made available, can boost research, here even in different disciplines. The dataset \cite{Huang.2015} has not only boosted concrete research but also method development in the field of octree-based adaptive remeshing, the Scaled Boundary Finite Element Method (SBFEM) and the Finite Cell Method (FCM). This indicates the demand for and benefits of databases with microstructure information made freely available.

\bigskip

{\bf Acknowledgements.} Bernhard Eidel acknowledges support by the Deutsche Forschungsgemeinschaft (DFG) within the Heisenberg program (grant no. EI 453/5-1). Simulations were in parts performed with computing resources granted by RWTH Aachen University under project ID BUND0005, in parts with resources at the University of Siegen. The authors are very grateful to Professor Zhenjun Yang (Wuhan University, China) for providing the XCT image data used in \cite{Huang.2015}.
  
\bigskip

{\bf Declaration of Interest.} None. 

\bibliographystyle{abbrv}  
\bibliography{concrete,octree}

%=================================================================

\begin{appendix}
%
%\addcontentsline{toc}{section}{Appendix}
%\renewcommand{\thesubsection}{\Alph{section}.\arabic{subsection}}
%\renewcommand{\theequation}{\Alph{section}.\arabic{equation}}
%\renewcommand{\thefigure}{\Alph{section}.\arabic{figure}}
%\renewcommand{\thetable}{\Alph{section}.\arabic{table}}
%\newcommand {\ssectapp}{
%                        \setcounter{equation}{0}
%                        \setcounter{figure}{0}
%                        \setcounter{table}{0}
%		                \subsection
%                        }
%
%\setcounter{equation}{0}
%
%%========================================================================================
%
%%\vfill
%%\newpage
%
%\input{appendix_Miscellaneous}
%\input{appendix_Misc-to-Surplus}

\end{appendix}

\end{document}